# Potterian economics

Daniel Levy[1,2,3,4,*] and Avichai Snir[5]

[1]Department of Economics, Bar-Ilan University, Ramat Gan 5290002, Israel
[2]Department of Economics, Emory University, Atlanta, GA 30322, USA
[3]International School of Economics at Tbilisi State University, 0108 Tbilisi, Georgia
[4]ICEAand RCEA
[5]Department of Banking and Finance, Netanya Academic College, Netanya 42365, Israel
*Correspondence address. Department of Economics, Bar-Ilan University, Ramat Gan 5290002, Israel. E-mail: Daniel.Levy@biu.ac.il

**Abstract**

Recent studies in psychology and neuroscience offer systematic evidence that fictional works exert a surprisingly strong influence on readers and have the power to shape their opinions and worldviews. Building on these findings, we study 'Potterian economics', the economic ideas, insights and structure, found in Harry Potter books, to assess how the books might affect economic literacy. A conservative estimate suggests that more than 7.3% of the world's population has read the Harry Potter books, and millions more have seen their movie adaptations. These extraordinary figures underscore the importance of the messages the books convey. We explore the Potterian economic model and compare it to professional economic models to assess the consistency of the Potterian economic principles with the existing economic models. We find that some of the principles of Potterian economics are consistent with economists' models. Many other principles, however, are distorted and contain numerous inaccuracies, contradicting professional economists' views and insights. We conclude that Potterian economics can teach us about the formation and dissemination of folk economics—the intuitive notions of naïve individuals who see market transactions as a zero-sum game, who care about distribution but fail to understand incentives and efficiency and who think of prices as allocating wealth but not resources or their efficient use.

'I think the writers [of popular literature] are not particularly sympathetic to or don't understand how a market works. It's not easy to convey that to a child. It's not always easy to convey it to grown-ups.'

**Gary Becker** (*New York Times*, August 21, 2011, p. SR5).

'With all due respect to Richard Posner, Cass Sunstein, or Peter Schuck [reference to the books these scholars published in 2005], no book released in 2005 will have more influence on what kids and adults around the world think about government than [Rowling's] *The Half-Blood Prince*.'

**Benjamin Barton** (*Michigan Law Review*, 2006, p. 1525).

'As economic theorists, we organize our thoughts using what we call models. The word "model" sounds more scientific than "fable" or "fairy tale" although I do not see much difference between them. The author of a fable draws a parallel to a situation in real life. He has some moral he wishes to impart to the reader. The fable is an imaginary situation that is somewhere between fantasy and reality. Any fable can be dismissed as being unrealistic or simplistic, but this is also the fable's advantage. Being something between fantasy and reality, a fable is free of extraneous details and annoying diversions. In this unencumbered state, we can clearly discern what cannot always be seen in the real world. On our return to reality, we are in possession of some sound advice or a relevant argument that can be used in the real world."

**Ariel Rubinstein** (*Econometrica*, 2006, p. 881).

'An investigation of novels and [economic] models ... may help us better understand how the public thinks about economic issues.'

**Tyler Cowen** (*The Street Porter and the Philosopher: Conversations on Analytical Egalitarianism*, 2008, p. 321).

## INTRODUCTION

According to Caplan (2007), ' ... modern economic theories ... begin by assuming that the typical citizen understands economics and votes accordingly.' Empirical evidence, however, suggests that the economic literacy of the public is low. For example, OECD (2014, p. 1) reports that only ' ... 10 percent of students can analyze complex financial products ... while 15 percent can, at best, make simple decisions about everyday spending.' Similarly, to assess financial literacy, Tang et al. (2015) surveyed American young adults, asking them three simple intuitive questions, and find that the participants answered correctly only 1.8 questions, on average.

Similar findings were reported in 2014 in Israel: 82.2% of the Israeli adults could not identify the interest rate that the Bank of Israel sets monthly. This is despite the attention the media has paid to it with the rate cut to 0.25%, the lowest level ever (Source: www.ynet.co.il/articles/0,7340,L-4568826,00.html (in Hebrew), accessed June 6, 2022.). Consistent with these findings, Nelson and Sheffrin (1991, p. 157) state that 'The level of economic knowledge among most high school students is shocking.'

There is evidence that better economic literacy improves economic decision-making, and therefore, the low level of economic literacy is worrisome as it may lead to poor financial planning (Note: See Bernheim et al. (2001), Boisclair et al. (2014), Gerardi et al. (2013), Brown et al. (2014), Grohmann et al. (2015), Lusardi & Mitchell (2014), Wiedrich et al. (2014) and Rubin (2018).). It may also lessen the effectiveness of economic policy (Bernanke, 2006). There is also evidence that the general public has biases and misconceptions about the economy, which influence and shape its views about the role of public policy (Rubin, 2003; Caplan, 2007; Hillman, 2010). These observations raise questions about how people acquire their knowledge about the economy.

It is widely accepted that literature, even fictional, is a mirror of culture and society (Albrecht, 1954). Recent studies in psychology and neuroscience, however, find that causality goes the other way as well, offering systematic evidence of fictional stories' effects on the human mind (Hakemulder, 2000; Mar, 2004; Appel, 2008).

We focus on the potential influence of literature on people's views and opinions about the economy and its functioning, by studying the economic principles that Harry Potter books convey. We chose Harry Potter books because studies that focus specifically on these books find that they not only relate to and reflect the readers' views but also have a powerful influence on them.

Although Harry Potter books belong to the fantasy genre, Harry Potter lives not only in a social world but also in an economic one. Indeed, the Potterian world is rich with economic institutions and ideas, including monopolies, inefficient government, limited social mobility, trade restrictions and other regulations, insufficient social capital, commodity money, prices, banking, etc. Not surprisingly, these economic ideas are unnoticed by most readers. But perhaps surprisingly, they are often unnoticed also by professional economists. Indeed, when we offered a colleague a list of the economic ideas we identified in the books, his response was, 'Well, you are right, I completely missed that.'

Studying what we term 'Potterian economics' is interesting for several reasons. First, as Lucas Jr. et al. (2002), Blinder & Krueger (2004), Giovannini & Malgarini (2012), and Cruijsen et al. (2010) note, the public acquires much of its knowledge about economics through popular intermediaries (books/newspapers, etc.), but the role of these intermediaries in shaping the public's views and opinions about economics has received little attention.

Second, among all such intermediaries, Harry Potter books are of particular importance because they are among the most popular books of our times. The books are popular among children and adults, men and women, irrespective of income and education. In sales, Harry Potter books rank fifth after *The Holy Quran* with over 3 billion copies sold, *The King James Bible* with over 2.5 billion copies sold, *The Quotations of Chairman Mao Zedong* with 800 million copies sold, and *Don Quixote* with 500 million copies sold (Source: www.stylist.co.uk/books/the-all-time-most-popular-books-in-the-world-revealed/127306, accessed June 6, 2022.). Rowling (2005) sold almost 7 million copies in the US and over two million copies in the UK just on the first weekend of its release (Source: news.bbc.co.uk/2/hi/entertainment/4692093.stm, accessed June 6, 2022.). In total, over 450 million copies were sold in over 200 countries. For comparison, Tolkien's *The Hobbit* was in print for more than 60 years and sold over 40 million copies, while Tolkien's *The Lord of the Rings* was in print for almost 50 years and sold over 50 million copies (Shippey, 2002, p. xxiv).

Harry Potter books have been translated into 73 languages, including Afrikaans, Albanian and Arabic, all the way to Vietnamese, Welsh and Zulu. The books have even been translated into two dead languages, Latin ('Harrius Potter') and Ancient Greek ('Αρειοσ Ποτηρ'), the latter translation being the longest work in the language since the novels of Heliodorus of Emesa in the third century AD (Source: https://en.wikipedia.org/wiki/Harry_Potter_in_translation#List_of_translations_by_language, accessed June 6, 2022.) In some countries, the books have been translated into several local languages. There are also pirate and unofficial translations as well as counterfeit versions. For example, according to the July 13, 2007 NPR's Morning Edition, 'Faking Harry Potter books has become a cottage industry in China.' Examples include 'Harry Potter and the Golden Vase,' 'Rich Dad, Poor Dad and Harry Potter, etc........ Don't worry if you've never heard of these books. They're totally made up with no resemblance to the real thing.' (Source: www.npr.org/templates/story/story.php?storyId=11945354, accessed June 6, 2022.).

By our conservative estimates, more than 7.3% of the world's population of 6.08 billion has read the books (Source of population figures: www.infoplease.com/ipa/A0762181.html, accessed June 6, 2022.). The estimate is conservative, and thus the actual share likely exceeds it because (1) the sales figure doesn't include unauthorized translations and their sales, and (2) the actual number of readers likely exceeds the number of copies sold because a single copy is often read by many, e.g. by an entire family. Hundreds of millions have also seen their movie adaptations. In the US alone, Harry Potter movies sold 350 million tickets (Source: https://mrob.com/pub/film-video/topadj.html, accessed June 6, 2022.). Nine thousand FedEx trucks were used in the US, to deliver the initial release of *The Goblet of Fire* alone. The last four books in the Harry Potter series were the fastest-selling books in history, with the final book selling roughly 11 million copies in the US within 24 hours of its release.

A recent study by Facebook's computer scientists reported that in a meme about 'What Books Stayed with You,' 130 000 participants (their average age was 37) ranked the *Harry Potter* books as the number 1. *The Bible* ranked sixth (Source: 'Harry Potter Tops List of Facebook Users' Favorite Books,' *The Telegraph*, September 10, 2014. For details about the survey, see the Facebook post of the authors of the study: https://www.facebook.com/notes/facebook-data-science/books-that-have-stayed-with-us/10152511240328859/, accessed June 6, 2022.). These extraordinary figures underscore the importance of the messages the books convey.

In that context, compare the Harry Potter books to the best-selling economics textbooks that were authored by Samuelson, selling over 4 million copies in 40 languages, and Mankiw, selling over one million copies in 17 languages (Source: en.wikipedia.org/wiki/Paul_Samuelson, and en.wikipedia.org/wiki/Greg_Mankiw, accessed June 6, 2022.). These figures, although impressive, are dwarfed by the sales figures of Harry Potter. Clearly, it is not fair to compare a popular fantasy book to academic textbooks, no matter how successful the latter are. The point, however, is that to the extent that Harry Potter books teach millions of people of all ages, principles of economics, identifying these principles is informative and important.

Third, Potterian economics can teach us about the formation and dissemination of folk economics (Rubin, 2003)—the intuitive notions of naive individuals who care about distribution but fail to understand incentives and efficiency. Cowen (2008), for example, argues that literary works can help us understand what people think about economic issues. Following these arguments, our thesis is that Potterian economics may reveal some of the ideas of folk economics. Indeed, as far as we know, J.K. Rowling is not an economist (she majored in French and Classics), and therefore Potterian economic model may be viewed as a layman's model and thus, it might reflect on the general public's attitudes and understanding of the economy.

Fourth, if fiction's influence is particularly strong on adolescents, and if a significant proportion of the readers of Harry Potter are children and teenagers, then understanding Potterian economics may also shed light on the sources of illiteracy that studies have documented among young adults (Tang et al., 2015) (Note: Lunt & Furnham (1996) contain essays on children's and adolescents' knowledge of economics and economic matters.) Indeed, many colleges and universities, including some Ivy League schools, recognize the influence of Harry Potter, and try to take advantage of the Harry Potter bandwagon in their sales' pitch, in their efforts to make their institutions more attractive to the candidates of Harry Potter generation. According to Lauren Edelson, during her visits to various university campuses, she heard from numerous host tour guides, how similar their school was to the Hogwarts. For example, 'During a Harvard information session, the admissions officer compared the intramural sports competitions there to the Hogwarts House Cup. The tour guide told me that I wouldn't be able to see the university's huge freshman dining hall as it was closed for the day, but to just imagine Hogwarts's Great Hall in its place.' (Source: Lauren Edelson, 'Taking the Magic Out of College,' *New York Times* (New York Edition), December 6, 2009, Op-Ed Contribution, p. WK12.).

Literary scholars have also emphasized the unique importance of Harry Potter books. For example, according to John Pazdziora, speaking at Britain's first academic conference on Harry Potter in 2012, 'We cannot avoid the fact that Harry Potter is the main narrative experience of an entire generation, the children who quite literally grew up with Harry Potter. The Harry Potter novels are simply the most important and influential children's books of the late 20[th] and early 21[st] centuries. For very many people, this is their first experience of literature, and of literary art … These are the most important, seminal texts for an entire generation of readers … In 100, 200 years' time, when scholars want to understand the early 21[st] century, when they want to understand the ethos and culture of the generation that's just breaking into adulthood, it is a safe bet that they'll be looking at the Harry Potter novels.' (See: A. Cramb, 'Harry Potter and the Philosophers Conference at St Andrews University,' *The Telegraph*, May 17, 2012, and A. Flood, 'Harry Potter and the Order of the 60 Scholars Gets Mixed Initial Reception,' *The Guardian*, May 18, 2012.)

The importance of Harry Potter books has been recognized by scholars in other fields besides literature. For example, they have been used to assess the completeness of tort theories (Hershovitz, 2010), and to shed light on the Anglo-American interpretation of rule of law (Liston, 2009; Thomas and Snyder, 2010). Barton (2006) studies governments' legitimacy. Sheftell et al. (2007) identify the headache episodes found in the books as symptomatic of migraine. Woeste (2010) revisits the debate over free will and determinism. As of June 6, 2022, Amazon listed over 40 000 items with 'Harry Potter' in the title, over 20 000 of them under books. The titles include: *The Psychology of Harry Potter*, *Harry Potter and History*, *Harry Potter and International Relations*, *Harry Potter and Philosophy*, *Ethics in the Bible and the World of Harry Potter*, *The Law and Harry Potter*, and *The Sociology of Harry Potter*. Other Harry Potter titles (along with comparable economics titles) include, *Who Killed Albus Dumbledore?* by Wendy Harte 2006 (compare it to *Who Killed John Maynard Keynes?* by Carl Biven 1989), *Rowling Revisited* by James Thomas 2010 (compare it to *Revisiting Keynes* by Lorenzo Pecchi and Gustavo Piga 2008), *The Complete Idiot's Guide to the World of Harry Potter* by Tere Stouffer 2007 (compare it to *The Complete Idiot's Guide to Economic Indicators* by Mark Rogers 2009), *Harry Potter and the Classical World* by Richard Spencer 2016 (compare it to *Keynes and the Classics Reconsidered* by James Ahiakpor 1998), etc.

In addition to books, there are also hundreds of papers, published and unpublished. For example, *SSRN* lists 63 Harry Potter studies in law, administrative science, political science, philosophy, history, marketing, etc. *SciVerse-Scopus* lists 46 items in engineering, neuroscience, psychology, etc. (Source: estrip.org/articles/read/tiny pliny/45318/Harry_Potter_in_Scientific_Literature.html, accessed June 6, 2022.). We also found 51 articles in *PubMed*. Finally, the Jstor database includes 159 papers, with 'Harry Potter' in their title (accessed June 6, 2022).

In addition, numerous academic conferences were and are still devoted to these studies. For example, in 2012, the University of California, San Diego, and the National Institute of Health, held a joint lecture series and an exhibit on 'Harry Potter's World: Renaissance Science, Magic, and Medicine.' The lecture series featured UCSD scholars from medicine, mathematics, engineering and literature. According to the UCSD News center, the interest of the UCSD medical scholars in Harry Potter stems from the fact that ' … the magic depicted in the popular Harry Potter novels … can be traced to Renaissance traditions that played a pivotal role in the development of modern science and medicine.' (Source: http://ucsdnews.ucsd.edu/pressrelease/harry_potters_world_renaissance_science_magic_and_medicine/, accessed June 6, 2022.) Chestnut Hill College in Philadelphia, PA, holds an annual conference on Harry Potter. The interdisciplinary conference includes several parallel sessions, on such topics as young adult literature, textual analysis, politics and justice, education and science, literature and education, psychology and philosophy, character studies, etc. (Source: https://www.chc.edu/events/10th-anniversary-harry-potter-academic-conference-hpac-x, accessed June 6, 2022.).

However, economic studies of Harry Potter are scarce (Note: Economic analysis of literary works includes Watts (2002, 2003), Rockoff (1990) and Bookman & Bookman (2008).). Exceptions include Gouvin (2010) and Schooner (2010), who study the role of the Potterians' only bank, the Gringotts, and Snir & Levy (2010), who try to understand the reasons for the lack of economic growth in the Potterian economy. Podemska-Mikluch & Deyo (2014) and Podemska-Mikluch et al. (2016) focus on the benefits of using Harry Potter series for undergraduate economics teaching. Salonikov et al. (2022) find evidence that the Harry Potter books affect readers' perceptions of various economic issues including banks, bureaucracy and money.

In this paper, we investigate the Potterian economy by analyzing its full structure. We find that it combines ingredients from various economic models but is not fully consistent with any particular model. Some features of the Potterian economy are in line with Marxist views, while others fit the public choice perspective. Prices in the Potterian economy are rigid in the Keynesian spirit, yet Potterians enjoy full employment as in the Classical model.

We conclude that the Potterian model reflects folk economics. As such, although it is sometimes consistent with economists' views, many of its deeper aspects of economics are distorted, containing numerous inaccuracies, which might potentially be absorbed by the public, and contribute to its lack of literacy on economic matters. This is particularly true for the young readers, who figure prominently among Harry Potter readers.

The paper is organized as follows. In section 2, we review the economic literacy literature. In section 3, we discuss fiction's influence. In section 4, we describe the setting. In section 5, we study money, credit, and banking. In section 6, we consider prices and their properties. In section 7, we focus on the foreign exchange rate and the commodity value of the Galleon. In section 8, we look at the Potterian government. In section 9, we discuss the law and order. In section 10, we focus on monopolies, oligarchies, and other pathologies. In section 11, we study income distribution in the Potterian economy. In section 12, we study international trade and migration. In section 13, we study war economics. In section 14, we study technological progress. In section 15, we discuss the Potterian education system and human capital. We conclude in section 16 by summarizing the implications of our findings, addressing possible objections to our interpretation, and discussing some caveats (Note: The online supplementary appendix contains a detailed reference to all economic themes, topics and issues we have identified in the Harry Potter books, along with the quotations of the relevant texts from the books, and their exact locations.).

## ECONOMIC LITERACY

The topic of economic literacy is not new. Newcomb (1893, p. 395) wrote over 120 years ago about the need to educate the public because of the gap between 'well-established economic conclusions on the one hand and the ideas of the public on the other.'

In 1948, at the First Workshop on Economic Education, Ernest Melby stated that economic education was a key to the survival of democracies: 'Democracy will live if it works and … die if it does not…if it fails in the economic front it will … go down to defeat … [for] survival of our way of life, there is no kind of education more important than that which seeks to make the average American intelligent about our economic system' (Troelstrup, 1954, p. 238).

The importance of economic education was also emphasized by the US Fed. For example, the Fed Governor Ben Bernanke (2006) stated: 'The Federal Reserve's mission of conducting monetary policy and maintaining a stable financial system depends upon the participation and support of an educated public.' Bernanke (2011) further emphasizes the individual benefits of economic literacy: 'Well-informed consumers … are one of the best lines of defense against the proliferation of financial products … that are unsuitable, unnecessarily costly, or abusive.'

The American Economic Association (AEA) has been involved in economic education since 1885. Hinshaw & Siegfried (1991) describe the AEA's efforts ' … to educate public … about economic questions and economic literature' (p. 373). The AEA's Committee on Economic Education has been active since the mid-1940s. The AEA routinely holds panels about teaching economics at its annual meetings. The establishment of the *Journal of Economic Education* in 1969 is also noteworthy. The US Fed has also been engaged in economic education for decades. For example, it offers teaching resources for K-12 grades on finance, banking, monetary policy, etc. OECD is also involved in these efforts (Atkinson and Messy, 2011) (Note: Gleason & Van Scyoc (1995) offer evidence on economic literacy in the US. Jappelli (2010) offers international evidence. See also Nelson & Sheffrin (1991) and the studies cited therein. Lusardi & Mitchell (2014) offer a survey.).

Despite these projects and efforts, the public rarely interprets economic ideas the same way as economists do (Alston et al. (1992) and Blendon et al. (1997).). Caplan (2007) argues that the public has various biases. (Note: Many biases of today's public are remarkably similar to the biases noted more than a century ago by Newcomb (1893).), which Rubin (2003, pp. 157–158) calls folk economics, and which ' … can explain the beliefs of naïve individuals regarding [economic] matters … [Folk economics] is the economics of wealth allocation, not production. Naive people … think of prices as allocating wealth but not … resources … The world of folk economics is zero-sum…if one person gets a job someone else must lose a job … Economists would do a better job of persuading others … if we paid explicit attention to folk economics.' Krugman (1996) lists some popular books, and notes two features they share: 'They all offer a view … of international trade as … "win-lose" competition … And they all contain little or nothing of what economists think … about international trade' (p. viii). Referring to Lunt & Furnham (1996), Rubin (2003, p. 158) notes that … 'the index to the book contains no entry for efficiency or productivity.' Paldam & Nannestad (2000) find that Danish voters are myopic and have prediction biases.

Blinder & Krueger (2004) and Caplan (2001) find that educated people tend to think like economists. Jappelli (2010) finds that the overall economic literacy level is low, although he documents a positive correlation between education and economic literacy (Blinder et al., 2008). Education by itself, however, does not always guarantee economic literacy. For example, according to Blank (2002, p. 476), 'A high share of … Congressional staff (never mind their bosses) do not understand basic economic principles. … in most meetings, my main role … was to lean forward and convincingly apply one of the following three concepts to the problem at hand: supply and demand, opportunity cost, or scarcity.'

To learn economics, one could take college economics courses. However, 'only about 10% of adults ever take college economics.' It would be better to increase ' … the quality and quantity of discussion of economics in the mass media. When a TV show like "West Wing" considers the benefits and costs of free trade, it probably has more impact on the economic literacy … than all freshmen economic courses combined' (Krueger, 2002, pp. 475–476). Indeed, some studies document the effect of TV shows on the political attitudes of the audience. See, for example, French & van Hoorn (1986), Lenart & McGraw (1989), and Mutz & Nir (2010). Mutz (2016) surveys some of the studies that find that fictional stories can influence political opinions, and offers evidence-based on the most recent data from the 2017 US presidential elections.

Confirming Krueger's (2002) assessments, Blinder & Krueger (2004) find that for US adults, print and electronic mass media serve as the primary source of information about economic issues. Similarly, Cruijsen et al. (2010) find that the Dutch public learns about ECB's monetary policy through the media. Caplan (2007) suggests that people are gullible and might believe what they read. A good analogy is offered by Gottschall (2012) in analyzing the stubborn persistence of superstitions, which it turns out, are not limited to the uneducated. Indeed, according to Park (2008) and Aaronovitch (2010), many conspiracy theories originate and circulate among the educated. Many superstitions have survived long periods of time, and an incredible number of people seem to believe them.

According to Gottschall (2012, p. 106), 'Many conspiracy theories would be funny except for the fact that stories … have consequences. For example, in Africa, many believe that AIDS is a racist hoax designed to … perpetrate a bloodless genocide. Believing this gets a lot of Africans killed.' On July 21, 2014, Christiane Amanpour ended her CNN show with a discussion of a denial of moon landing: 'One of man's greatest achievements [is] still … denied. 45 years ago … Armstrong and Aldrin walked on the moon … But almost from the start, there were those who said … that the whole thing had been staged … the deniers persist.' (Source: edition.cnn.com/TRANSCRIPTS/1407/21/ampr.01.html, accessed June 6, 2022.).

## HOW FICTION REFLECTS AND AFFECTS POPULAR VIEWS AND PERCEPTIONS

Literary works serve as a mirror of society. According to Bloch (1961, p. 102), for example, 'In every literature, a society contemplates its own image.' Similarly, Tiemensma (2010, p. 3) states that 'Stories are the structural coding of social values, beliefs, and goals that underlie human interaction.' Literature, even fictional, reflects the life, the views, the norms and the beliefs of the society. Indeed, texts are often used in social sciences to learn how people of different societies and different times view the world (Dickstein, 2005).

The causal relationship between fiction and society, however, goes the other direction as well. The recognition that literature can influence society is not new. Consider, for example, the list of the books and writings that were banned throughout the history precisely because of the conviction that they had a strong and deep influence on the human mind. According to Newth (2010), the practice of censorship of books can be traced to the Romans, as far back as 443 BC, with the establishment of the Office of Censor: 'In Rome, as in the ancient Greek communities, the ideal of good governance included shaping the character of the people. Hence, censorship was regarded as an honorable task. In China, the first censorship law was introduced in 300 AD.'

The invention of the printing press in the 15th century increased the need for censorship further, primarily because of the emerging tensions between the Catholic Church and the Protestant movement, which led to the introduction of the first *Index Librorum Prohibitorum* (Index of Prohibited Books) by Pope Paul IV in 1559. According to Newth (2010), the Index was re-issued 20 times since then by different popes, and the last update was published in 1948.

However, book censorship was not limited to religious authorities, and throughout the history, it served other goals and motives, primarily political. USSR and its allied countries took book censorship to new levels (recall the books of Solzhenitsyn, Sakharov, etc.). Perhaps surprisingly, libraries played an important role in supporting censorship: Public libraries were expected to act as the benevolent guardians of literature, particularly books for young readers. Consequently, this gave teachers and librarians license to censor a wide range of books in libraries, under the pretext of protecting readers from morally destructive and offensive literature.

Even in liberal-minded countries such as Sweden and Norway, which boasts the earliest press freedom laws, surveillance of public and school libraries remained a concern to authors and publishers even through the latter part of the 20th century. Not less surprising is the die-hard tradition of surveillance of books in schools and libraries in the United States. For example, the Library of Congress is holding a multi-year exhibition featuring *The Books that Shaped America* (https://www.loc.gov/exhibits/books-that-shaped-america/overview.html, accessed June 6, 2022.). According to the curators, 'Some of the titles on display have been the source of great controversy, even derision, yet they nevertheless shaped Americans' views of their world and often the world's view of the United States.' Many of the books displayed at the exhibit were banned in the US (Haight, 1978).

The American Library Association lists on its page the top 100 banned books during the years 2000–2009. Harry Potter books top the list, which were banned in the US by some religious authorities as well as by some schools (Scheffer, 2010) (Source: http://www.ala.org/advocacy/bbooks/top-100-bannedchallenged-books-2000-2009, accessed June 6, 2022.). This, however, was not limited to the US. For example, Cardinal Joseph Ratzinger, who later became Pope Benedict 16, in 2003 wrote a letter to the author of the book, *Harry Potter: Good or Evil*: 'It is good that you enlighten us on the Harry Potter matter, for these are subtle seductions … barely noticeable, and precisely because of that have a deep effect and corrupt the Christian faith in souls even before it could properly grow.' (Source: Pope Benedict XVI, March 2003 Letter, London Times, July 13, 2005.)

Confirming this view, the Vatican newspaper *L'Osservatore Romano* published an article on January 15, 2008, which stated that Cardinal Ratzinger was right to worry: 'Despite the values that we come across in the narration, at the base of this story, witchcraft is proposed as a positive ideal. The violent manipulation of things and people, comes thanks to knowledge of the occult. The ends justify the means because the knowledgeable, the chosen ones, the intellectuals know how to control the dark powers and turn them into good. This a grave and deep lie, because it is the old Gnostic temptation of confusing salvation and truth with a secret knowledge. The characterization of common men who do not know magic as "muggles" who know nothing other than bad and wicked things is a truly diabolical attitude.' (Source: M. Moore and N. Reynolds, 'JK Rowling's Harry Potter condemned in Vatican newspaper,' *The Telegraph*, January 15, 2008.) Perhaps ironically, however, Amazon-UK ' … revealed it had received advance orders for *Harry Potter and the Half-Blood Prince* from … the Vatican.' (Source: 'Pope Criticizes Harry Potter,' R. Blakely, *The Times*, July 13, 2005.)

The perceived dangers that Harry Potter books pose, have been noted by other religious authorities as well. For example, in orthodox Judaism, some public figures, newspaper editorials, etc. ridiculed the fascination of so many people, young and old, with the imaginary tales of Harry Potter and his friends, arguing that it is indicative of the '… emptiness of the Western cultures' (Source: Meir Shalev, 'Holy Harry Potter,' Yedioth Ahronoth (daily newspaper published in Israel in Hebrew), August 4, 2000). Some Rabbinical authorities, however, were more open minded. See, for example, Yvette Alt Miller, 'Harry Potter and Jewish Values,' December 21, 2014 (Source: https://aish.com/harry-potter-and-jewish-values/, accessed June 24, 2022), '5 Quintessentially Jewish Concepts in the Harry Potter Saga,' Forward, (Source: https://forward.com/opinion/375984/5-quintessentially-jewish-concepts-in-the-harry-potter-saga/, accessed June 24, 2022), and Judy Siegel-Itzkovich, 'Researchers: Harry Potter's Wizardry OK with Rambam,' Jerusalem Post, May 17, 2006 (Source: https://www.jpost.com/jewish-world/judaism/researchers-harry-potters-wizardry-ok-with-rambam, accessed June 24, 2022).

Similarly, Harry Potter books have been banned in some Islamic countries, but not everywhere. For example, the books were banned in the United Arab Emirates (Source: 'Emirates Ban Potter Book,' BBC News, February 12, 2002, http://news.bbc.co.uk/2/hi/

entertainment/1816012.stm, accessed June 24, 2022). According to some Islamic scholars, the books themes conflict with Islamic teaching because 'Islam prohibits both pointless entertainment (lahw) and sorcery' (Source: Khalid Baig, 'Harry Potter: Facts about Fiction', June 21, 2003, http://www.albalagh.net/current_affairs/harry_potter.shtml, accessed June 24, 2022). In Iran, in contrast, the books were published by the Ministry of Culture and Islamic Guidance (Source: https://www.hogwartsprofessor.com/ayatollah-condemns-harry-potter/, accessed June 24, 2022).

Peter Smith of the UK Teachers Association warns against the supernatural: 'Children who had enjoyed the magic and wizardry of the stories should be careful about extending their interest in the occult.' (Source news.bbc.co.uk/2/hi/uk_news/education/1638887.stm, June 6, 2022.) Some schools have banned the books because '… they go against the Bible's teaching.' (Source: news.bbc.co.uk/2/hi/uk_news/education/693779.stm, June 6, 2022.) Some British toy shops have even refused to stock Potterian merchandise fearing it will attract children to the occult (Source: news.bbc.co.uk/2/hi/entertainment/1560335.stm, accessed June 6, 2022.). According to the American Library Association, in 2005 '… there were 26 challenges to remove the Harry Potter books from bookshelves in 16 states.' (Source: www.educationworld.com/a_admin/admin/admin157.shtml, accessed June 6, 2022.)

A recent episode in Israel suggests that even People of the Book might be afraid of books and find them dangerous. In 2015, the Israeli Education Ministry has decided not to include 'Border Life,' a novel by Dorit Rabinyan, in its list of required readings for the matriculation exams, because it includes a love story between a Jewish Israeli woman and a Muslim Palestinian man. Many Israeli school directors and educators fought the Ministry over this decision and insisted on including the book in their school's required readings lists.

Authors also believe that their work can affect the readers. According to Gottschall (2012), Tolstoy believed that an artist's job is to 'infect' his audience with his own ideas and emotions—'the stronger the infection, the better is the art as art' (p. 134). Similarly, 'In his book "The Act of Reading," Wolfgang Iser writes that ideally, a book should transform a reader by "disconfirming" his habitual notions and perceptions and thus forcing him or her to a new understanding of them' (Tuck, 2015). Indeed, there are many fictional works with a long-lasting impact on popular views and opinions. These include Uncle Tom's Cabin, Black Beauty, The Birth of a Nation, Jaws, 1984, Darkness at Noon, Roots, etc.

The belief that books influence the readers has been corroborated by recent studies in psychology that offer systematic evidence of fictional stories' effects on the human mind and attitudes. These studies find that when we read fiction, ' … we allow ourselves to be invaded by the teller (Note: In social psychology, the term 'attitude' refers to personal and social norms, prejudices, and stereotypes. We use the term similarly to describe the way people perceive and think about the world in which they live, operate and make decisions.). The story maker penetrates our skulls and seizes control of our brains … fiction subtly shapes our beliefs, behaviors, ethics … ' (Gottschall, 2012, pp. xvi–xvii).

Hakemulder (2000) reviews dozens of studies in psychology that demonstrate that fiction can have a profound effect on the readers' thinking. Gottschall (2012, p. 133–134) argues that 'Fiction does mold our minds … influences our moral logic … alters our behavior…shaping our minds without our knowledge … Most of us believe that we know how to separate fantasy and reality … this is not always the case. In the same mental bin, we mix information gleaned from both fiction and nonfiction.' Vezzali et al. (2012) report that after reading books where characters with different cultural backgrounds had positive interactions with one another, Italian teenagers displayed more positive and less stereotypical attitudes toward immigrants.

Mar (2004, p. 1414), based on evidence from neuroimaging, argues that ' … reader attitudes shift to become more congruent with the ideas expressed in a narrative after exposure to fiction.' Appel & Richter (2007) and Appel (2008) find that fiction shapes readers' views on fairness and justice (Note: The July 29, 2014 edition of 'Room for Debate' of the NY Times asked, 'Will Fiction Influence How We React to Climate Change?') Vezzali et al. (2015) survey psychology literature, which shows that novels can have positive social impact. Green et al. (2004) find that fictional worlds alter the way we process information, and that the deeper we are immersed in a story, the more influential the story is. According to Gottschall (2012, p. 135), 'Fiction readers who reported a high level of absorption tended to have their beliefs changed in a more "story-consistent" way... [and] detected significantly fewer "false notes" in stories—inaccuracies, infelicities … When we read nonfiction, we read with our shields up. We are critical and skeptical. But when we are absorbed in a story, we drop our intellectual guard. We are moved emotionally, and this seems to leave us defenseless'. See also Sklar (2009).

We focus on the Harry Potter books because of the evidence that these particular books affect popular views. Indeed, studies suggest that the attitudes of the readers of the Harry Potter books are influenced by the messages these books convey. Hallett (2005) notes that Harry Potter books influence culture. Brown & Patterson (2009) study brand assessment by focusing on 'Pottermania.' Brown & Patterson (2010) study consumers' treatment of the Harry Potter brand. See also the opening quote from Barton (2006).

Some studies offer even more direct evidence on the effect of Harry Potter books on the readers. Vezzali et al. (2015) report that Harry Potter books make the readers' attitudes toward stigmatized groups such as immigrants, homosexuals, refugees, etc. more positive and sympa-thetic. Gierzynski & Seger (2011) find that the books influence the readers' acceptance of difference, tolerance, equality and opposition to violence and corruption. Salonikov et al. (2022) offer evidence suggesting that the Harry Potter books affect the readers' attitudes towards economic issues, including banks, the public sector, money and monopolies.

Some studies have even documented that the Harry Potter books affected the readers' health. Gwilym et al. (2005), for example, find a drop in the number of children visiting the hospital emergency departments on the weekends that Harry Potter books are released, whereas Bennett (2003) reports that some of the young readers of the Harry Potter series have suffered from headaches because of their insistence to read the books cover-to-cover without taking a break.

Neuroscientists offer further evidence on the effect of the Potterian adventures on the human mind. Recently, for example, in a series of neuroimaging studies on the influence of Harry Potter books on their readers, Hsu et al. (2014, 2015a, 2015b) provide fMRI evidence for the fiction feeling hypothesis, which states that narratives with emotional content, in contrast to stories with neutral content, cause readers to empathize to a stronger degree with the protagonists, thus engaging the affective empathy network of the brain. See also Wehbe et al. (2014) and Lehne et al. (2015).

In sum, the existing evidence in psychology, sociology and neuroscience suggests that fictional literature in general, and Harry Potter books in particular, can have a subtle yet powerful influence on the readers' views, opinions and attitudes. Similar to these influences, and consistent with Krueger (2002) and Cowen (2008), we propose that the readers of the Harry Potter books, consciously and/or subconsciously, absorb the 'economic' ideas the books implicitly or explicitly convey. These ideas, we argue, can potentially shape the public's opinion on economics and economic issues. This is consistent with the argument of Cowen (2008) who describes this as a 'knowledge-generation process.' Novels and models, he argues, both are mechanisms for learning that complement each other. 'We should recognize the power of stories. Many models, especially the most relevant models, are embedded in stories, further illustrating the complementarities between novels and models     Both novelists and model builders have tacit knowledge about how the real world works, and they try to articulate that knowledge in the form of either a story or equations' (pp. 333–334).

The success of Harry Potter books suggests that in addition to telling a story that appeals to a broad and diverse audience, the author may have also been able to capture the readers' popular beliefs. Given the books' universal appeal, therefore, studying the Potterian economy and comparing it to standard economic models can shed light on the beliefs and views of millions of people with diverse cultures and norms about the economy and economic ideas.

Thomas and Snyder (2010, p. vii) state along these lines: 'Part of the appeal of [Harry Potter] is that the depictions resonate with readers … This may suggest that the depictions are consistent with readers' and viewers' values or opinions. Alternatively, if the depictions are not reflective, they may influence the development of values or opinions … any influence would be … subtle.'

A comparison between economic models and literary texts is possible because they share several key characteristics (Cowen, 2008; McCloskey, 1998, 2000; Thomson, 2001, Rubinstein 2006) (Note: According to McCloskey (1998, p. xiv): 'Economists are poets, but don't know it. Economists are storytellers.' Similarly, Cowen (2008, p. 15) states that 'Novels are more like models than is commonly believed' and recommends studying them to better understand how people think about economics and economic matters.). Both offer imaginary tales that are abstractions of reality. Both are composed of a set of actors—characters in stories, decision-makers in models and a set of assumptions—rules in stories and constraints and assumptions in models. In addition, the characters in a story are linked by initial relationships, the same way as decision-makers and variables in a model are linked by initial conditions. In both stories and models, the initial relationships evolve. In models, these follow the assumptions and optimal decisions and strategies. In stories, they follow the characters' attributes and the actions they can and cannot take. Eventually, both economic models and stories conclude in a final state (Rockoff, 1990; Watts, 2002).

Consider, for example, Cowen's (2008, pp. 325–326) description of the similarities between fictional stories and economic models: 'Science fiction stories … embody model-like thinking. The author writes down a description of some new technologies … The author then traces through the effects of these technologies and outlines how things would work or outlines an equilibrium in economic terminology. That equilibrium is then "disturbed" by some new change, such as an alien invasion or a new technology. The bulk of the novel then traces through the effects of the change, performing a kind of comparative statics exercise … these novels … use a stylized setting to show how one set of causes leads to particular effects, working through a mechanism of some generality. The mechanism is not always spelled out explicitly…They are like the models from earlier in the history of economics. Before the mathematization of the economics profession … models without explicit mathematical forms… It is no accident that contemporary model builders sometimes refer to earlier, non-formal economists as "telling stories".'

Another similarity between models and stories is that stories must maintain face validity, the same way as models need to maintain internal validity. Well-defined rules of math and logic ensure models' internal validity. Maintaining face validity in literary texts depends on the genre, on adhering to the readers' norms/expectations and on preserving the story's internal logic (Derrida, 1993). For example, if magic can solve all problems, then the resulting story, even fictional, is unlikely to be interesting. The Harry Potter books' success and near-universal popularity suggest that they satisfy this condition, and therefore, we believe that Potterian economics can offer lessons about the economic principles the books teach and convey.

## GENERAL BACKGROUND
## (SPOILER WARNING: SKIP IF YOU HAVE NOT READ THE HARRY POTTER BOOKS, BUT PLAN TO READ THEM!)

The Harry Potter story, an imaginary tale of a boy with extraordinary wizardry powers, is a series of seven books (Note: See Rowling (1998, 1999a, 1999b, 2000, 2003, 2005 and 2007). The discussion in this section is based on Anandane's (2011) descriptive summary of the seven volumes.). Most of Harry's adventures take place in and around Hogwarts, a boarding school of witchcraft and wizardry located in northern England, where Harry spends 7 years. The books follow Harry and his friends, as they grow up and mature, and the 7 volumes describe correspondingly the 7 years Harry spends at Hogwarts.

The story begins in 1991 with *Harry Potter and the Philosopher's Stone*, where Harry is a shy orphan living with his aunt in a suburb of London. At age 11, Harry learns that his parents were wizards and that they were killed by the most powerful wizard of the era, 'Lord Voldemort', whose reputation is so fearsome that his name is not mentioned and instead is referred to as 'You-Know-Who' or 'He Who Must Not Be Named.' Harry also learns that although he grew up among non-wizards ('*muggles*'), he has the power to become a wizard (Note: According to the Oxford dictionary, 'muggle' is now an official word, and it means 'a person who is not conversant with a particular activity or skill.'). Furthermore, he discovers that Lord Voldemort tried to kill him after killing his parents, which left a scar on his forehead. Harry survived the attack, which gives him a special role in the wizards' society. The school headmaster, Dumbledore, who knows about this incident, leaves Harry with his aunt until he is 11, when he is ready to enroll in Hogwarts. Following the Hogwarts' invitation, which is delivered by an owl, Harry enters the school, where he and his new friends, Hermione Granger and Ron Weasley, explore the world of magic, and slowly discover the powers of witchcraft and Wizardry. The first year ends when they recover the Philosopher stone, which can be used to brew an elixir that can make the drinker immortal.

In the second year at Hogwarts, *Harry Potter and the Chamber of Secrets*, Ron's younger sister Ginny, discovers Voldemort's old notebook, in which she reads about a 'Chamber of Secrets,' which, it turns out, leads to a monster. In this volume, Harry and his friends learn about the history of Hogwarts. Harry also discovers that he has some special skills, such as the ability to communicate with snakes (rather rare dark art). He also discovers some secrets about Voldemort. The book ends with Harry and his friends saving Ginny while fighting the monster. In the process, they inadvertently destroy a part of Voldemort's soul ('Horcrux').

In the third volume, *Harry Potter and the Prisoner of Azkaban*, Harry learns about Remus Lupin and Sirius Black, who were his father's friends. Lupin is a teacher of defensive measures against dark magic. Black, it turns out, is a murderer believed to have helped Voldemort in killing Harry's parents. In volume 4, *Harry Potter and the Goblet of Fire*, Harry is pressed by a Voldemort supporter, Barty Crouch (disguised as Professor Alastor Moody), to participate in a dangerous Tri-Wizard Tournament. Harry luckily escapes Crouch's plans, and at the same time he forces Voldemort to reenter the wizarding world as a mortal.

In volume 5, *Harry Potter and the Order of the Phoenix*, Harry learns about a secret society called the 'Order of the Phoenix', which is re-activated to protect Harry and his friends from Voldemort and his supporters. The school's new Headmaster Dolores Umbridge ('the High Inquisitor of Hogwarts') does not permit the students to learn defense against dark magic. Harry, therefore, forms a secret study group called 'Dumbledore's Army', where he teaches his friends how to fight and defeat dark arts. Eventually, Harry foresees Voldemort's actions, and thus manages to save Hogwarts from Voldemort's supporters, the Death Eaters.

By the sixth year, in *Harry Potter and the Half-Blood Prince*, the 17-year-old Harry, who is dating Ginny Weasley, incidentally, comes across an old potions textbook filled with annotations signed by an anonymous individual, the Half-Blood Prince. He also discovers that Voldemort's soul is split into a series of horcruxes (evil-enchanted items hidden in various locations). Draco Malfoy, Harry's foe, attempts to attack Dumbledore several times. Eventually, Dumbledore is killed by Professor Snape, another Harry adversary.

In the last volume, *Harry Potter and the Deathly Hallows*, Lord Voldemort takes control of the Ministry of Magic. Harry and his friends quit school and go on a mission of finding and destroying the remaining horcruxes of Voldemort. Harry discovers, however, that he is one of the horcruxes, and thus surrenders to Voldemort. It turns out, however, that the horcrux inside Harry has been destroyed when Voldemort tried to regain his full strength. In the end, the Order of the Phoenix, along with Harry and Harry's friends, defeat Voldemort and his supporters, and thus save the world of Witchcraft and Wizardry.

## MONETARY SYSTEM: POTTERIAN MONEY, CREDIT AND BANKING
### Money

Wizards use commodity money. They have three types of coins: gold Galleons, silver Sickles and bronze Knuts, where one Galleon equals 17 Sickles, and one Sickle equals 29 Knuts (Rowling, 1998, p. 49). The wizards' monetary system is therefore similar to the old English monetary system that existed from medieval times until 1971. In that system, one pound was worth 20 shillings and one shilling was worth 12 pennies. The similarity is also in the types of metals used in minting the coins. The pound (also known as Guinea) was originally made of gold, the Shilling (originally Scilling) was made of silver, and the penny (after 1796) was made of copper.

However, despite their superficial similarity, the wizards' monetary system differs greatly from its real-world counterpart. First, whereas in the old English system the value of the pound relative to the shilling and the penny fluctuated freely as a result of changes in the relative prices of gold, silver and copper, the relative values of the Galleons, Sickles and Knuts are fixed. Furthermore, whereas the values of the English coins depended on the amount of the precious metal minted in the coins, the values of the Potterian coins are independent of their physical size and weight. For example, when wizards from several countries gather, it seems that the value of all gold coins is the same, even though the foreign Galleons are 'the size of hubcaps' (Rowling, 2000, p. 50), suggesting that the Potterian money is not homogenous, a key property that money should have (Levy and Bergen, 1993). The Potterians however, seem not to care about this.

Thus, although the wizards' money is a type of commodity money, it behaves as if it was fiduciary money. For example, in the muggle (non-wizard) world, the ratio of the price of gold to the price of silver, over the period 1991–1998, when Harry Potter's adventures take place, varied between 48.2 to 91.6 (Source: https://sdbullion.com/, accessed June 6, 2022.). Yet the conversion rates between the Galleon, Sickle and Knut remained fixed.

Indeed, just like fiduciary money, the values of all types of Potterian coins decrease simultaneously when there is inflation as a result of a war that disrupts the supply (Rowling, 2005, p. 43). The values of the three types of coin denominations, however, do not change relative to each other, although we are told explicitly that at the same time as the inflation bout, the demand for silver increases because silver is used for making charms and apparatus that are in demand (Rowling, 2005, p. 73).

Thus, in the Potterian economy, commodity value of the money is distinct from the exchange value of money. This is counter to the models of commodity money where the value of the money is pegged to the value of the commodity it is made of (Sargent and Wallace, 1983; Rockoff, 1990) (Note: In *The Lord of the Rings*, for example, the value of gold as a medium of exchange is determined by weighing it. Thus, Tolkien seems to have understood this aspect of the difference between commodity money and fiduciary money.).

Another shortcoming of the Potterian monetary system is that, unlike monetary economy models, Potterian money lacks some key features that would facilitate trade. One advantage the old English system had over the modern decimal system is in the number of ways a pound could be divided into combinations of shillings and pennies. This divisibility allowed the minting of many coins that were combinations of pennies, shillings and pounds. For example, common English coins included 2 shillings, 2 shillings 6 pence, 5 shillings and many more.

In the case of Potterian money, however, the exchange rate between the Galleons and the Sickles and between the Sickles and the Knuts, are both prime numbers. It is, therefore, less useful to mint coins that are multiples of these basic coins.

Whatever the reason for using commodity money, its use has real effects beyond the opportunity cost of using the precious metals for making jewelry or silver daggers (Rowling, 1998, p. 8, Rowling, 2000, p. 413, Rowling, 2003, p. 45). Carrying too many coins is risky and cumbersome. Therefore, Wizards store most of their money in bank vaults. Withdrawals and deposits, however, are time-consuming, as they require a lengthy bureaucratic procedure, implying a high transaction cost (Rowling, 1998, pp. 47–49). Consequently, wizards make infrequent withdrawals from the bank (Rowling, 2000, pp. 61, 352).

In sum, the Potterian money lacks some of the basic properties that economists believe money should have, for it to serve its functions efficiently: homogeneity, portability, divisibility and storability (Levy and Bergen, 1993). The Potterian money is not homogenous, it is not easy to transport, it is not easily divisible and it is difficult to store. In any efficient economy, the Potterian money would have therefore been replaced by more efficient money, which is more divisible and less cumbersome.

It is possible that the Potterians use precious metals as money because it is harder to counterfeit money that is minted of precious metals than of other materials. However, it is possible to counterfeit the wizard's money. It is also possible that the Potterians use commodity money because they believe that pegging the value of money to commodities might be an effective way of controlling inflation (Woodford, 2003). However, it turns out that when the supply chain is disrupted, the outcome is inflation, as in an economy that uses fiduciary money.

Another important shortcoming of the Potterian monetary system is the lack of paper checks that could be used to transfer funds without the need to carry cumbersome coins, making transactions far more efficient. Medieval merchants and bankers took advantage of this benefit of paper checks (Quinn and Roberds, 2008), yet it seems that Potterians have yet to discover this technology.

## Banks, interest and credit

Potterians have only one bank, *Gringotts,* making it a textbook version of a perfect monopoly. The bank serves mostly wizards, although the bank's owners and employees are Goblins (Rowling, 1998, p. 41)—greedy, gold loving, selfish and unfriendly humanoids (Rowling, 2000, p. 81).

Gringotts offers several services. First, it is in charge of minting money and preventing its counterfeiting. There are, however, several ways to counterfeit money, and even schoolchildren can do it (Rowling, 2003, p. 297). The amount of counterfeit money in circulation is low, nevertheless.

Second, wizards take advantage of the Gringotts' money-storage and safekeeping services for storing their gold and other valuables (Rowling, 1998, pp. 48–49). Third, the bank offers currency exchange services, which include exchanging wizard money for precious stones, pieces of art and for muggle money (Rowling, 1998, p. 47, Rowling, 1999a, p. 37). The amount of money exchanged, however, is limited. Wizards use muggle money only when they run errands in the non-wizard parts of England. This happens rarely, and when it does, they face significant difficulties in using muggle banknotes (Rowling, 1998, p. 43, Rowling, 2000, p. 50). Muggles also have limited opportunities to exchange muggle money for wizards money because most muggles are not aware of the wizards' existence (Rowling, 1998, p. 42, Rowling, 1999b, p. 80, Rowling, 2000, p. 50).

Gringotts, however, is not an ordinary bank, as it does not offer lending and borrowing services. In fact, Potterians do not have a financial institution that offers such services. That is, unlike modern banks, the Potterian bank does not operate based on fractional reserves. For example, we do not find a single case of someone borrowing money from Gringotts, which suggests that Gringotts is not offering loans. This is the interpretation of other Harry Potter scholars as well. For example, consider Schooner's (2010, p. 265) interpretation: 'Most important, however, is the question of lending. If Gringotts is a true bank, then it not only takes deposits, but it also lends out a portion of those deposits to other customers. The image of Harry and his vault full of money suggests that all of Harry's money remains in the vault at all times. This would mean that Gringotts does not operate on the basis of fractional reserves, i.e. it does not lend out a percentage of the money deposited by its customers.' Wizards that want to borrow, must therefore borrow from a friend or from illegal usurers (Rowling, 2005, p. 78, Rowling, 2000, p. 471).

Indeed, we find numerous episodes where wizards borrow money from friends, private usurers, or some other wealthy individuals, but not from Gringotts. For example, to open their joke shop, the 'Weasleys' Wizard Wheezes,' Fred and George Weasley borrow the necessary start-up money from Harry Potter (Rowling, 2003, p. 79, Rowling, 2005, p. 78). As another example, Ludo Bagman, a senior public official who is in debt after running a failing private enterprise, borrows 'loads of gold' from Goblins (Rowling, 2000, p. 471).

Moreover, we find cases where wizards in need of money resort to gambling with the hope of winning the necessary funds. For example, the Weasley twins, before discovering that they can borrow money from Harry, consider gambling as a means of obtaining the funds needed to open their shop (Rowling, 2000, p. 57, Rowling, 2005, p. 78). Because there are no financial markets, the government also cannot issue debt and, therefore, it often depends on donations from wealthy individuals for funding public goods (Rowling, 2000, p. 66, 456).

It appears that it is not because of a lack of willingness that Gringotts does not lend money. Indeed, Gringotts' employees sometimes offer private usury services (Rowling, 2000, p. 471). There is no shortage of demand for loans either. On the contrary, the lack of borrowing options is a significant constraint. For example, wizards that make windfall gains spend them immediately, and entrepreneurs without capital cannot open businesses, suggesting that both consumers and businessmen face credit constraints (Rowling, 1999b, p. 5).

The books do not explain the reasons for the lack of financial intermediaries. However, it seems that the Potterians view financial service providers as immoral, the same way as prisoners of war were suspicious of those who offered financial services at the POW camp according to the account of Radford (1945) (Notes: See Rowling (1998, p. 41), Rowling (2000, pp. 81, 287, 471), Carlton (1995), and Hillman (2010). Portraying bankers in fictional works as evil has been common throughout history. Examples include Shylock in *Merchant of Venice*, Harpagon in *Miser*, Danglars in *The Count of Monte Christo*, Mr. Merdle in *Little Dorrit*, Mr. Banks in *Mary Poppins*, the hero in *American Psycho*, and Le Chiffre in *Casino Royale*.). The negative image of bankers is so strong that wizards shy away from such jobs. Consequently, the banking jobs are taken by Goblins, an inferior social group. (Note: Gringotts apparently employs some humans. Fleur Delacour (Rowling, 2003, p. 53) and Bill Weasley (Rowling, 2007, p. 196) are two such examples.).

Thus, in the Potterian economy, economic incentives are overridden by the power of socially stigmatizing financial service providers. In Europe, such processes led to the prosecution and stigmatization of minority groups (Carlton, 1995; Hillman, 2013a). Some have noted that in the movie adaptation of the Harry Potter books, the goblin bankers are depicted with aquiline noses and greedy-looking faces, similar to the cartoons that were used to depict stereotypical bankers and financiers in Europe in the late 19th century (Source: momentmagazine.wordpress.com/2011/07/14/debunking-the-harry-potter-anti-semitism-myth/, accessed June 6, 2022.) A reader has suggested that the goblins are merely a joke on the theme of *Gnomes of Zurich*. That, however, does not explain why they are depicted in the movies with aquiline noses. On the other hand, it could be argued that even if goblin characters were historically used to depict financial service providers negatively, currently they are a part of Western literary characters. (Note: A eugenics view could also be suggested. See Peart & Levy (2005) on eugenics in the post-classical 19th century economics.). Similarly, in the Potterian economy, the stigmatization of financial intermediation leads to a tension between wizards and Goblins, with Goblins being treated as inferiors even though they provide cheap, efficient, and economically beneficial services. It also inhibits most forms of interaction between wizards and Goblins (Rowling, 2007, pp. 323, 342).

Such negative attitudes still prevail. For example, in a 2005 Roper poll, 'Only 9% … had full trust in financial services institutions, down from 14% last year' (Source: 'New Surveys Show that Big Business Has a P.R. Problem', by C. Deutsch, *New York Times*, December 9, 2005, Late Edition-Final, Sec. C, Column 2). The 2008 financial markets' troubles have likely deepened the distrust between the public and bankers even further. For example, in July 2015, Argentines' President, Cristina Fernández de Kirchner, according to *Wall Street Journal*, told Buenos Aires schoolchildren, after learning that they were reading *Romeo and Juliet*, that they should ' … read "The Merchant of Venice" to understand the vulture funds', alluding to a decade-long financial battle to collect $1.5 billion from Argentina on defaulted foreign bonds held by a US hedge fund. In a follow up tweet, Kirchner added: 'No, don't laugh. Usury and bloodsuckers have been immortalized in the greatest literature for centuries.' As the author notes, 'The persistence of anti-Semitism over time and across cultures is one of mankind's darkest puzzles. So is the hatred of capitalism, property rights and freedom.' (Source: 'The Bigot Defense: The Oldest Prejudice Reappears in Attacks on American Capitalists', *Wall Street Journal*, July 10, 2015.

Thus, the Potterians' 'ban' on lending, leads to financial transactions being handled by goblins, which further strengthens their stereotypical image as greedy. These observations are striking given that the books are often viewed and interpreted as opposing all types of stereotypical biases (Gierzynski and Seger, 2011; Vezzali et al., 2015).

In summary, the ban on lending leads to an outcome that is opposite to the values that the books are believed to promote. Further, these credit constraints reduce welfare, increase corruption because it limits the ability of would-be-entrepreneurs to open new businesses and thus increase the influence of wealthy wizards (Paldam, 2002; Fudenberg and Levine, 2006).

In sum, in the Potterian economy, the role of financial intermediation is completely orthogonal to its role in economic models. Economists view financial intermediation as a key mechanism for promoting investment through efficient channeling of funds from savers to investors. The Potterians, whose bank enjoys monopoly power, in contrast, view financial intermediation as means to impoverish the borrowers and enrich the lenders.

## PRICES

The difficulties in carrying and transporting money, together with the lack of credit market, have real effects on the price level and on pricing strategies. Since bank withdrawals and deposits are time-consuming, wizards make infrequent cash withdrawals, as predicted by the Baumol–Tobin inventory model of money holding. This, in addition to the limited (or no) supply of credit in the Potterian economy, leads to wizards facing a cash (i.e. Clower) constraint.

The cash constraint forces retailers to set low prices for basic goods and services, as otherwise wizards would be caught cashless (Rowling, 1998, p. 41). For example, Harry Potter withdraws money only once every year, before going to Hogwarts, and the sum he withdraws has to last him through the entire school year (Rowling, 1999a, 50).

However, beyond their effect on the price level, the cumbersomeness of the Potterian money also affects the pricing strategies and possibly even the extent of price rigidity (Dutta et al., 2002). Many prices are set in round numbers, such as 5 Knuts, 10 Sickles, 30 Galleons, etc. Examples include Goblin-made armor for 500 Galleons (Rowling, 2005, p. 286), Acromantula venom for 100 Galleons/pint (Rowling, 2005, p. 316), Deflagration Deluxe for 20 Galleons (Rowling, 2003, p. 472), Omnioculars for 10 Galleons (Rowling, 2000, p. 60), weekly pay of 10 Galleons to house-elf (Rowling, 2000, p. 244), Metamorph-Medals for changing appearance for 10 Galleons (Rowling, 2005, p. 58), Basic Blaze box for 5 Galleons (Rowling, 2003, p. 472), Glittery-Black Beetle Eyes for 5 Knuts/scoop (Rowling, 1998, p. 52), etc.

In cases where prices are not round, then they are still denominated in units of a single type of coin, and thus they can be paid using one type of coin. Examples of such convenient prices include a new copy of *Advanced Potion-Making* for 9 Galleons (Rowling, 2005, p. 144), a ticket for a night bus to London for 11 Sickles (Rowling, 1999b, p. 22), hot chocolate on a night bus for 2 Sickles (Rowling, 1999b, p. 22), membership in the Society for the Promotion of Elfish Welfare for 2 Sickles (Rowling, 2000, pp. 144–145), etc. Such price-setting rules, however, have a disadvantage because they lead to a type of grid pricing, which is inefficient. For example, a Galleon is composed of $17 \times 29 = 493$ Knuts, but it seems that prices that are set in Galleons only change by Galleons and, consequently, retailers cannot change them by smaller amounts in response to small shocks. It is not surprising, therefore, that once there are disturbances to supply, prices go 'sky high' (Rowling, 2005, p. 65).

Thus, whereas retailers in real economies tend to set many prices at psychological price points such as 9-ending prices (Levy et al., 2011, Levy et al., 2020), which studies find affect the quantity demanded (Snir and Levy, 2021; Chen et al., 2022), Potterian retailers stick to convenient prices. The only product with a 9-ending price we were able to find is a new copy of *Advanced Potion-Making* from Flourish and Blotts, with a price tag of exactly 9 Galleons (Rowling, 2005, p. 144). Since the Potterian currency system is not decimal, it could be argued that the Potterian parallels of the Muggles' 9-ending prices are the prices that end with either 16 Sickles or 28 Knuts (i.e. 'just below prices'), which would be equivalent to prices like 11 pennies or 19 shillings that were documented by Gabor and Granger (1966) in the UK under the pre-decimal currency system. However, we find only one such good, dragon liver, with a price of 16 Sickles per ounce (Rowling, 1998, pp. 55–56). Thus, there are almost no 'just below' prices, despite the widespread recognition that they increase sales (Anderson and Simester 2003) and despite the fact that they were common in the old English monetary system (Gabor and Granger 1966). Thus, the inefficiency of the monetary system leads to inefficiencies in sellers' price setting practices.

Thus, unlike modern marketing practitioners and scholars, the Potterian retailers do not use psychological price points. It might be that unlike Muggle retailers (Anderson and Simester, 2003), Potterian sellers do not believe that psychological price points affect consumer demand. It might also be that the computational efforts needed in the triple coinage system are cognitively demanding and, consequently, the retailers use prices that reduce the cognitive load (Dehaene, 1997; Knotek II, 2008, 2011; Snir et al., 2022). This seems likely because it is clear that many wizards find it difficult to calculate the value of goods denominated in foreign currency units (Rowling, 1998, p. 43, Rowling, 2000, p. 50), similar to the difficulties reported in the EU during the transition to the Euro (Ehrmann, 2011). Thus, it is possible that wizards also find that converting prices between Galleons, Sickles and Knuts is a cognitively difficult task especially given that the exchange rates are prime numbers.

It could be that the complicated exchange rate between the Knut, Sickle and Galleon may have confused even the author and/or her book's editor. In the first edition of Rowling (1998), the price of one ounce of dragon liver is 17 sickles, that is, one galleon. But that would be like saying that an item costs 100¢, instead of $1. After the mistake was noticed, the price in the later editions was changed to 16 sickles (Source: http://harrypotter.wikia.com/wiki/Dragon_liver, accessed June 6, 2022, and www.hp-lexicon.org/edits-changes-text-ps/, accessed June 6, 2022.).

The abundance of convenient prices seems to affect price rigidity (Snir et al., 2021). Prices in the Potterian economy change only infrequently. For example, the price of Floo-powder remained the same for over 100 years, two Sickles a scoop (Source: pottermorehead.tumblr.com/post/102175802190/floo-powder, accessed June 6, 2022.). For comparison, the longest spell of price rigidity that economists have documented belongs to the 6.5 ounces Coca-Cola, whose price remained 5 cents between 1886 and 1960, over 74 years (Levy and Young, 2004). Thus, price rigidity in the Potterian economy is at least as prevalent as in the US, in line with the findings reported in some recent New Keynesian studies.

One reason for the price rigidity might be that retailers that set prices to maximize transaction convenience are reluctant to change them, as changing the price might, like in modern markets, make a transaction less convenient because of the increase in the number of coins used in the transaction, or because round prices are cognitively more accessible (Dehaene, 1997; Knotek 2008 and 2011; Snir et al., 2022, Fisher & Konieczny, 2006) (Note: This price rigidity cannot be explained by menu cost (Levy et al., 1997 and 1998, Dutta et al., 1999, Zbaracki et al., 2004, Fisher & Konieczny, 2006) because most prices in the Potterian economy are not posted.). In that case, round prices will be adjusted only if the round endings can be preserved. Coca-Cola faced a similar constraint in 1950s. It was unable to adjust the price because increasing it while allowing people to use a single coin to buy it, meant doubling the price from 5¢ (a nickel) to 10¢ (a dime) (See Levy & Young (2004, 2021) and Young & Levy (2014).). In modern markets, unlike the Potterian markets, however, round prices are usually limited to particular types of goods and settings (Snir et al., 2022). Thus, the Potterians' currency structure and the resulting pricing practices lead to prices that are set at a few convenient price points, which hinder price changes.

Another aspect of the behavior of prices that we should note is that there is very little price dispersion, even across highly heterogeneous goods, suggesting that the prices in the Potterian economy are not set in perfectly competitive markets. For example, the Omnioculars (Rowling, 2000, p. 60), Metamorph-Medals for changing one's appearance, one week's work of a house-elf (Rowling, 2005, p. 58), Slytherin's Locket (Rowling, 2005, p. 171) and Unicorn hair (Rowling, 2005, p. 320) all cost 10 Galleons. This type of pricing resembles somewhat the pricing of modern dollar stores, where highly heterogeneous products are all sold for $1.

Moreover, the Potterian economy is not growing, implying that there is no upward trend in the aggregate demand (Snir and Levy, 2010). Nor do the Potterian retailers experience major changes in their cost structure. These factors, along with the currency structure decrease the need and the willingness of the Potterian retailers to adjust prices. Simply, there are not sufficient changes in the market conditions that would warrant such price adjustments, except when a war breaks out (see section 13).

In sum, the price rigidity in the Potterian economy seems to be due to a combination of three factors. First, thanks to the monopolistic structure of the market, the Potterian retailers have relatively high profit margins, which allows them not to change the prices in response to small cost changes, and still remain profitable. Second, the lack of growth of the Potterian economy makes large changes in the cost and demand rare. Third, the prevalent use of convenient prices suggests that retailers believe that such prices increase the quantities sold. Retailers therefore change prices only when the cost or demand shock is large enough to merit a change to another convenient price. Given the low price level, however, this means a large price change (in percentage terms), which is unlikely because the Potterian economy is not growing (Snir and Levy, 2010).

## FOREIGN EXCHANGE RATE AND THE COMMODITY VALUE OF THE GALLEON

The Potterian monetary system has another important flaw: there is an unreasonably large gap between the commodity value of a gold Galleon and its exchange value. We draw this conclusion based on the Galleon–Dollar exchange rate on the one hand, and the estimated weight of the Gold Galleon, on the other.

Consider first the Galleon–Dollar exchange rate (i.e. the exchange rate between the muggles' money and the wizards' money), which is not mentioned explicitly in the original 7-volume Harry Potter books, but we know from the books that the Gringotts handles such exchanges (Rowling, 1999a, p. 50). Based on information from three sources, we estimate that the Galleon–Dollar exchange rate is about $7.30/Galleon.

First, J.K. Rowling authored two additional, although less known, Harry Potter books as a charity for the UK Comic Relief: (a) *Fantastic Beasts & Where to Find Them*, and (b) *Quidditch through the Ages*. *Fantastic Beasts & Where to Find Them* is included in the list of the things the first-year students are required to have at the Hogwarts School of Witchcraft and Wizardry (Rowling, 1998, p. 43). *Quidditch through the Ages* is mentioned in Rowling (1998, p. 93), where we are told that Hermione Granger borrows the book from the Hogwarts library, and then later we are told that she lends the book to Harry to help him better prepare for Quidditch practice (Rowling, 1998, p. 117). The foreword to both books is 'written' by Albus Dumbledore himself. In the foreword to *Fantastic Beasts & Where to Find Them*, Dumbledore writes that 'Comic Relief U.K. has raised over 250 million dollars since 1985 (that's also 174 million pounds, or thirty-four million, eight hundred and seventy-two Galleons, fourteen Sickles and seven Knuts)' (Rowling, 2001a, pp. vii–viii). In the foreword to *Quidditch through the Ages*, Dumbledore writes that 'Comic Relief U.K. uses laughter to fight poverty, injustice, and disaster. Widespread amusement is converted into large quantities of money (over 250 million dollars since they started in 1985—which is the equivalent of over 174 million pounds or thirty-four million Galleons)' (Rowling, 2001b, p. vii). These figures imply that one Galleon is equivalent to £4.95 or $7.30, and one Sickle is equivalent to £0.30 or $0.45.

Second, we compare the prices of the products consumed by wizards to the prices of similar products consumed by ordinary muggles in the same period, using the above exchange rate, i.e. one Galleon equals £4.95 or $7.30. For example, Hermione, Ron and Harry pay six Sickles for three butter beers (Rowling, 2003, p. 251), i.e. two Sickles per butter beer. We can reasonably assume that butter beer is not a real beer with significant alcoholic content, because otherwise, it would imply that Harry Potter and his friends, all of them underage wizards, are consuming alcoholic beverages, which is unlikely. In other words, we assume that butter beer is more like Muggles' root beer or cream soda. In that case, the price of a single bottle of butter beer, two Sickles, which is equivalent to about £0.60 or $0.90, seems reasonable for a price paid by school kids for a soft drink. Hot chocolate on a night bus also costs two Sickles (Rowling, 1999b, p. 22), i.e. £0.60 or $0.90, also

reasonable. As another example, Arthur Weasley, the Head the Office for the Detection and Confiscation of Counterfeit Defensive Spells and Protective Objects, is fined 50 Galleons for bewitching a Muggle car (Rowling, 1999a, p. 142), which is equivalent to about $375.00. This is comparable to the monetary fines assessed for various types of traffic violations in the real Muggle world, such as in California (Source: catrafficticket.com/ca-traffic-ticket-fines/, accessed on June 6, 2022.). Other examples include the cost of traveling on a night bus to London, 11 Sickles or about $4.95 (Rowling, 1999b, p. 22), birthday present of 10 Galleons, which is equivalent to about $73 (Rowling, 1999b, p. 36), etc. All these prices seem reasonable (Note: Using the above exchange rate, $7.5/Galleon, CNN has even published the Wizard Calculator, offering conversion services to/from the US dollars to/from the Potterian currency units. Source: money.cnn.com/2001/10/23/saving/wizard_calc/index.htm, accessed on June 6, 2022.).

Third, in an interview on March 12, 2001, when asked by Rebecca Boswell, 'What is the approximate value of a galleon?' J.K. Rowling's reply was 'About five pounds, though the exchange rate varies!' (Source: https://www.hp-lexicon.org/2007/02/04/wizard-money/, accessed June 6, 2022.) We conclude therefore that the Galleon–Dollar exchange rate is about $7.30/Galleon.

Next, consider the commodity value of a gold Galleon. To assess it, we need to know the weight of the gold it contains. Galleon's weight, however, is not mentioned in the Harry Potter books, although we know that it is not trivial. For example, in Rowling (2000, p. 61), Harry Potter finds that his money bag is 'considerably lighter' after paying 20–30 Galleons for some goods he bought.

To estimate the Galleon's weight, we take the average weight of three gold coins that were issued in the Kingdom of England. These are (1) the Noble, weighing 7.74 gr, which was issued during King Edward III, (2) the Ryal, weighing 7.58 gr, which was issued during King Edward IV and (3) the Angel, weighing 5.20 gr, which was issued during King Henry VII (Deutsche Bundesbank, 1983). These figures yield an average weight of 6.84 gr. Alternatively, we could consider the gold Guinea, which was used in England until 1814, and weighed up to 8.50 gr. (Source: https://www.royalmint.com/our-coins/ranges/guinea/, accessed June 6, 2022.). We thus estimate the weight of a gold Galleon to be in the range of 6.84–8.50 gr., which we consider an upper bound of the Galleon's true weight.

The actual price of 1 gr. gold in the years the Harry Potter story takes place, 1991–1998, varied between $10.14–$13.80 (Source: www.kitco.com/charts/historical gold.html, accessed June 6, 2022.), which implies that the commodity value of a gold Galleon is in the range of $69.36–$117.30. Thus, the commodity value of a gold Galleon is between 9 and 16 times the Galleon–Dollar exchange rate. As a lower bound of the Galleon's weight, we consider the estimate of Generalov (2006), who assess the weight of a gold Galleon and reaches a conclusion that a Galleon weighs about 2 gr. This is a more conservative estimate, yielding a commodity value of gold Galleon in the range of $20.28−$27.60, which is 3−4 times the Galleon–Dollar exchange rate. Because we know that wizards value gold, using gold for making coins with such a low purchasing value relative to its commodity value seems to be a misallocation: instead of using gold as a commodity, it is used as a medium of exchange. In this setting, gold is not directed to its most beneficial use, which is inefficient (Friedman, 2002).

Furthermore, the ridiculously low exchange rate points at an arbitrage opportunity. Wizards could melt the gold, sell it to muggles as a commodity and then exchange the muggle money for Galleons at Gringotts. However, wizards do not seem to take advantage of this arbitrage opportunity, although it promises immense profits and no risk. Not even the bankers, including the greediest ones, seem to notice it. Nor do rich businessmen who have interest in undermining the Potterians' political stability. Clearly, in any commodity money model, at least some of those who stand to make a profit would have exploited such profitable arbitrage opportunities leading to their eventual disappearance. Thus, the foreign exchange valuation of the gold Galleon is a textbook example of an inefficient market outcome.

Indeed, if gold in the Potterian economy is so cheap for the bank officials that they value it at $7.30 while muggles' valuation is as much as $69.36–$117.30, wizards could melt the gold and sell it to muggles in return for other things they value, such as other precious metals, gems, etc. They could buy, for example, silver and copper, because the commodity values of silver and copper in the Potterian economy, as can be deduced from the value of the Sickle and the Knut, are closer to the values of silver and copper in the muggle economy, in comparison to the value of gold. If they were to do so, the price differences between the two economies should have become smaller, or disappeared altogether, following the law of one price.

Some possible arguments could be made to confront the arbitrage opportunity puzzle (See: Generalov (2006), gilletts.com.au/jewellery-metal-information-i-39.html, www.goldpriceoz.com/gold-karat.html, taxfreegold.co.uk/goldcoinsbriefhistory.html, en.wikipedia.org/wiki/Gold_coin, and https://harrypotter.fandom.com/wiki/Wizarding_currency, all accessed June 6, 2022.). First, it could be that gold Galleons are not made of real gold but rather they are only gold colored, although there is nothing in the books that indicates this. Also, if wizards were to use coins that do not contain precious metals, then they could just as well use paper money, whose commodity value would be zero, or close to zero, like colored coins, but more convenient. Further, there are fake Galleons, made of Leprechaun gold, which look like the real gold Galleons, but they vanish within hours (Rowling, 2000, p. 350). The mere existence of 'fake' Galleons, however, suggests that the 'real' Galleons are 'real,' i.e. they are made of real gold, unlike the 'fake' ones. Also, it seems that wizards can fake Galleons relatively easily—Hermione does that when she was 16, in Rowling (2003) (Note: One of the readers of this paper, when he was young, was bewildered why wizards and witches did not fake 10 Euro from the muggle world with the appropriate charm so that they could at least live well in the muggle world.). Bank officials can probably identify counterfeit coins. Most wizards, however, cannot. Therefore, the only way to prevent wizards from counterfeiting Galleons on a large scale is if its cost is prohibitively high—i.e. if the coins are made of gold, because the only magical way to produce gold is by using the extremely rare philosopher's stone. This is essential for the stability of the Potterian currency as otherwise the Potterian economy would be flooded with fake gold coins, which would quickly make the Potterian money worthless.

Second, it could be that Galleons are made of real gold, but there are some limitations on the wizards' ability to take advantage of it for arbitrage purposes. For example, there may be some spell, prohibition or some other kind of restriction that prevents the wizards, even the greediest ones, from engaging in arbitrage. We, however, do not find any explicit mention of such restrictions in the books. Even if wizards are prohibited from melting the gold and selling it to muggles, the goblins that mint the coins are not. The goblins, however, also do not take advantage of this opportunity, although they trade with muggles regularly.

Third, it could be that gold Galleons are not made of pure, 100% gold. Indeed, gold is a soft substance and therefore 24 k gold is rarely used for making coins or jewelry. Historically, actual gold coins in circulation in the UK were made from 22 k (91.6%, known as 'Crown Gold') or 23 k (95.83%) gold. For example, the old Pound Sterling coin that was issued in 1489 under King Henry VII was made of 23 k gold. In 1526, under King Henry VIII, and in 1549, under King Edward VI, lower grade gold, 22 k, was used to make the coins. The Crown gold became the standard for making English gold coins from 1526 onward. In the US, 21.6 k (90% gold) became the standard for making American coins for circulation from 1837 onward.

However, even if the Galleon is made of 21 k–23 k gold like the Muggle gold coins, the puzzle remains because the price difference between 21 k gold and 24 k gold is not that significant and thus the Galleon's commodity value as gold is still high.

## POTTERIAN PUBLIC SECTOR: THE GOVERNMENT

The Potterian government is large, corrupt and inefficient. The government controls and regulates the production of every major good and service produced in the economy, including health, law, education, etc. Thus, to a large extent, the Potterian economy is planned. The government determines which products will be produced or imported and which will not. The government policies also ensure economic stability, as in the Marxian model, in the sense that Potterians do not experience business cycle fluctuations and, consequently, all workers are employed at all times.

Yet there is a private bank, and there is some private enterprise. The regulations, however, restrict competition in the private sector (Shughart II and Thomas, 2015). Firms in the private sector, therefore, do not have to fire workers or reorganize. Thus, Potterians have both stable prices and stable output in an economy that is not growing.

The public, however, knows very little about their government's decision-making process because of its lack of transparency, which makes it easier for corrupt public officials and interest groups to engage in rent-seeking activities (Tullock, 1967, 1989).

The lack of transparency is partly because in the Potterian economy there is a monopoly on information. The only important information source is a daily newspaper, the *Daily Prophet*, and its editors are on good terms with senior public officials and wealthy individuals. The latter often fund public goods and thus have a significant influence on the officials and their policy (Rowling, 2003, p. 116). The newspaper's reporters, therefore, publish information that favors the officials, who reciprocate by making decisions favoring the reporters and wealthy wizards (Rowling, 2003, p. 423, Strömberg, 2004, Gentzkow and Shapiro, 2006).

Power concentration and lack of transparency help senior officials in seeking rent and in obtaining other benefits such as bribes (Rowling, 2007, pp. 144, 172). Wealthy wizards that fund their office expenditures are reciprocated by getting access to the officials, and swaying their decisions (Rowling, 1999a, pp. 33–34, Rowling, 2003, p. 116). Although not identical, this process resembles political benefits gained through campaign donations (Ursprung, 1990).

Consider the following exchange that reads like a textbook case of rent-seeking, as surveyed for example, in Hillman (2013b): 'What are you doing here, anyway?' Harry asked Lucius Malfoy. 'I don't think private matters between myself and the Minister [of Magic, a position that parallels the position of Prime Minister in the muggle-world] are any concern of yours, Potter,' said Malfoy … Harry distinctly heard the gentle clinking of what sounded like a full pocket of gold. ' … shall we go up to your office, then, Minister?' 'Certainly,' said Fudge … 'This way, Lucius' … 'What private business have they got together, anyway?' 'Gold, I expect,' said Mr. [Arthur] Weasley angrily. 'Malfoy's been giving generously to all sorts of things for years … gets him in with the right people … then he can ask favors … delay laws he doesn't want passed … oh, he's very well-connected, Lucius Malfoy' (Rowling, 2003, pp. 115–116).

Nepotism is common and family members and associates of senior officials receive perks and benefits not offered to others. For example, Arthur Weasley and two of his sons, Percy and Ron, are all employed by the Ministry of Magic. Arthur Weasley works as the head of the Office for the Detection and Confiscation of Counterfeit Defensive Spells and Protective Objects (Rowling, 2005, p. 56), after promotion from his former position as the Head of the Office for the Misuse of Muggle Artifacts. Percy Weasley worked in the Department of International Magical Cooperation under Barty Crouch, Sr., until he became a Junior Assistant to the Minister of Magic (Rowling, 2000, p. 268, Rowling, 2003, p. 229). In Rowling (2003), Ron Weasley is about to become an Auror in the Ministry of Magic (Rowling, 2003, p. 170, Rowling, 2000, p. 392).

Junior public officials engage in rent-seeking (Hillman, 2013b) by putting effort into pleasing their superiors rather than doing their jobs, thereby increasing their prospects for promotion and for gaining higher status (Kahana and Liu, 2010). This, however, further deepens the inefficiency and corruption (Rowling, 2000, pp. 56, 58–59, 273). The tenure system makes it impossible to fire even the most inefficient workers, leading to hidden unemployment. In addition, undeserving promotions create underemployment. As a result, many offices are overstaffed with low productivity workers. For example, Bertha Jorkins' disappearance from her office is unnoticed for several weeks (Rowling, 2000, p. 40). These inefficiencies persist because the budget of each department depends primarily on the influence its head has in the Ministry or with the media. The resulting contest for contacts with powerful politicians and influence on their policy decisions over budgetary allocations leads to wasteful social welfare loss (Katz and Rosenberg, 1989).

However, junior public employees seem to be doing their best at carrying out their jobs, which contradicts the internal, bureaucratic rent-seeking noted above. Even extremely inefficient workers like Bertha Jorkins seem to be doing their best, and it is only their lack of talent that makes them inefficient, not their lack of effort (Rowling, 2000, p. 40). Thus, as in the Marxian model, the workers contribute according to their ability, and it is their work efforts, more than the pecuniary compensation that gives them satisfaction.

However, unlike the workers, those high in the social hierarchy behave as predicted by public choice theory. Senior public officials use their powers and the lack of transparency to advance their own goals, by taking advantage of the rent-seeking opportunities the Potterian government offers, from promoting their associates to releasing their relatives from jail. They also reciprocate by benefiting wealthy landowners, granting special benefits to those who donate to causes of their interest.

Senior officials also do not hesitate to use fear and violence, if needed, to achieve their goals or hide their mistakes (Levin and Satarov, 2015). When Harry and his colleagues express doubts about the threat of war, the press is used to intimidate and silence them, and some of those that raise doubts like the headmaster of the school lose their jobs. Others, including Harry Potter, are threatened (Rowling, 2005, pp. 341–348). At least on one occasion, a senior public official orders a 'neutralization' of a man to silence him (Rowling, 2000, pp. 453–454).

Thus, consistent with public choice theory, when public officials amass power over the entire economy and transparency is low, they use their power to advance their private goals and procure more power. The officials that comprise the upper layers of the public sector share similar values because only those with similar backgrounds are promoted.

Furthermore, it seems that one of the incentives to climb the ranks of the civil service is ego rents provided by status and power (Olson, 1965). The school headmaster, Dumbledore, for example, does not want to become a minister of magic because he is not as power hungry as those who apply for the job (Rowling, 2003, p. 89). The urge to satisfy his ego rent ambition for power even drives the Minister of Magic Pius Thicknesse to fight on the side of He-Who-Must-Not-Be-Named, the evil Voldemort (Rowling, 2007, p. 636). Hayek (2006) predicts that in a society in which power is concentrated in the hands of a few, the ones to get to the top are those that are the most power and status hungry, not the most competent or the most benevolent. Many characters that hold high positions in the Potterian economy fit this characterization.

Despite these inefficiencies, the government has considerable public support. Some of this support might be due to the public being unaware of the inefficiencies, due to the low transparency. It might also be because of the government's size (Alesina & Fuchs-Schuendeln, 2007). The public sector is the largest employer, and wizards depend on it for both work and services. Furthermore, the size of the government allows for full employment, with even the most unproductive workers finding a job in the public sector. Indeed, it seems that for most wizards, a job in the public sector is the default, and many of them do not even consider a job in the private sector. Therefore, further rent-seeking may be taking place through the benefits of a majority from the government (Paldam, 2015).

Thus, the support for the government seems to be partly because the public is unaware of much of the inefficiencies, and partly because those that are aware ignore them and look the other way, to protect their private benefits such as job security.

## LAW AND ORDER

Consider the following: 'What would you think of a government that engaged in this list of tyrannical activities: tortured children for lying; designed its prison specifically to suck all life and hope out of the inmates; placed citizens in that prison without a hearing; ordered the death penalty without a trial; allowed the powerful, rich or famous to control policy; selectively prosecuted crimes (the powerful go unpunished and the unpopular face trumped-up charges); conducted criminal trials without defense counsel; used truth serum to force confessions; maintained constant surveillance over all citizens; offered no elections and no democratic law-making process, and controlled the press? You might assume that the above list is the work of some despotic central African nation, but it is actually the product of the Ministry of Magic' (Barton, 2006, pp. 1523–1524).

The Potterians' legal institutions are rather limited in both breadth and scope. There are no lawyers, no independent court system or other muggle-type judicial entities. The Ministry of Magic seems to have all the regulatory, legislative and judicial powers, implying that the Potterians are not familiar with the notion of separation of powers. For example, the Ministry of Magic is not subject to any practical limits on its powers since there are no laws or regulations that impose any restrictions on what it can or cannot do. The Ministry operates the court, and the Minister of Magic himself serves as one of the 'inquisitors,' the Potterian equivalent of a prosecutor, as well as one of the judges.

Even the laws that are in place cannot be trusted because according to the Minister of Magic, there is no prohibition on their retroactive revision. On the contrary, the Ministry changes and manipulates the laws at will depending on its interests and circumstances. For example, the Ministry decides to change the time of Harry's court hearing but informs him of this change at the last minute, causing Harry to be late for the hearing (Rowling, 2003, p. 103). Moreover, the presiding judge (the 'Chief Inquisitor') in the hearing, the Minister of Magic himself, states during the court proceedings that 'laws can be changed.' For example, During Harry's trial for using underage magic, which the Ministry's laws prohibit, Dumbledore (who attends the court hearing as a witness) debates the Minister of Magic Cornelius Fudge about the applicability of the relevant laws (Rowling, 2003, p. 112): 'The Ministry does not have the power to expel Hogwarts students, Cornelius, as I reminded you on the night of the second of August,' said Dumbledore. 'Nor does it have the right to confiscate wands until charges have been successfully proven … In your admirable haste to ensure that the law is upheld, you appear, inadvertently I am sure, to have overlooked a few laws yourself.' 'Laws can be changed,' said Fudge savagely. 'Of course, they can,' said Dumbledore, inclining his head. 'And you certainly seem to be making many changes, Cornelius. Why, in the few short weeks since I was asked to leave the Wizengamot [the court of the Ministry of Magic], it has already become the practice to hold a full criminal trial to deal with a simple matter of underage magic!'

There are not even rules that govern the process of elections. There is no evidence of elections taking place for any key public office position. For example, we are told that Rufus Scrimgeour is appointed (not elected) as a Minister of Magic, but we do not know by who (Rowling, 2005, p. 27). Ministry officials often take advantage of their powers by drafting laws and regulations that promote their and/or their friends' interests. Further, the implementation and enforcement of the laws are often done selectively. For example, Harry has committed in the past a worse crime, blowing up Ms. Marjorie Dursley, but no charges were filed against him because, in that period, Harry was still the darling of the Minister of Magic (Rowling, 1999b, p. 28). There are even cases where wizards are accused and sentenced without any trial or court hearing. Since there is no constitution, the authoritarian Ministry can arbitrarily manipulate the existing laws and regulations to fit its needs. For example, at a certain point, the Ministry decides that all halfblood and mudblood wizards must be questioned to make sure that they did not 'steal' magic powers (Rowling, 2007, p. 136) (Note: Wizards with one muggle-parent are called 'halfblood' (Rowling, 2003, p. 584). Wizards with two muggle parents are called 'mudbloods' (Rowling, 1999a, p.96).). The wizards that fail the test are imprisoned.

Day-to-day policing is done by the wizards and witches that work at the Ministry. Some of them are aurors— wizards whose job is to locate and apprehend 'dark wizards'—wizards that practice various types of 'dark arts' and are known for their dislike of muggles. There are also divisions in the Ministry that handle specific types of offenses and crimes. Examples include the Office of the Improper Use of Magic, the Office of the Regulation and Control of Magical Creatures, the Office of the Magical Law Enforcement, etc. Some of the offenses they deal with include buying and selling stolen caldrons (Rowling, 2003, pp. 18, 80), car theft (Rowling, 2003, p. 377), sale of fake amulets that supposedly protect against werewolves (Rowling, 2005, p. 73) and various kinds of fake protections that supposedly guard against You-Know-Who and Death Eaters (Rowling, 2005, p. 56), etc. Thus, like muggles, the Potterians seem to have their share of thieves, crooks and other common criminals.

Further, in the Potterian legal system, we see very little of what might resemble private laws such as tort laws or contract laws. For example, although the Hogwarts students are often injured (e.g. during the Quidditch games), there is no mention of any kind of lawsuit in this context.

The word 'contract' is mentioned only three times in the 7-volume set. The first time it is mentioned by Dumbledore explains the Triwizard Tournament rules and warns the students: 'Once a champion has been selected by the Goblet of Fire, he or she is obliged to see the tournament through to the end. The placing of your name in the goblet constitutes a binding, magical contract' (Rowling, 2000, pp. 166, 179). However, we are not told how this contract is enforced.

The second time contract is mentioned is when Hermione Granger organizes a study group of students that take lessons (without the school's permission) from Harry on how to defend against Voldemort's Death Eaters. At the meeting where they decide on the formation of the group, they agree to keep the group's existence secret by adding their name to a list of the attendees: 'So if you sign, you're agreeing not to tell Umbridge or anybody else what we're up to.' 'There was an odd feeling in the group now. It was as though they had just signed some kind of contract' (Rowling, 2003, p. 259).

The third time the term 'contract' is mentioned when we are told that the house-elf Hokey's contract prohibits her from saying what she thinks about her mistress Hepzibah Smith's looks (Rowling, 2005, 285–286): '"How do I look?" Said Hepzibah, turning her head to admire the various angles of her face in the mirror. 'Lovely, madam,' squeaked Harry. Harry could only assume that it was down in Hokey's contract that she must lie through her teeth when asked this question because Hepzibah Smith looked a long way from lovely in his opinion'. As Gava & Paterson (2010) note, however, this does not constitute a contract in a legal sense because it is not signed voluntarily: house elves are enslaved by their masters and thus they must follow their orders. In addition, there is a mention of an 'unbreakable vow' that Snape made with Draco Malfoy's mom that he would protect Draco (Rowling, 2005, p. 214). The vow, however, more resembles a promise than a contract.

'Private property' is mentioned in the story only once, when Mr. Gaunt tells a ministry official that he cannot just show up unannounced on his private property (Rowling, 2005, p. 132, Demsetz, 1967). But there is indirect evidence that Potterian wizards respect private property, at least in the context of inheritance, which seems to be rather common in their economy. For example, Harry inherits a large amount of money from his parents (Rowling, 1998, p. 48, Rowling, 1999a, p. 30). He also inherits Sirius Black's house along with everything in the house, including the house-elf Kreacher (Rowling, 2005, pp. 33–34). There are numerous other cases of inheritance transfer from generation to generation.

The *government*, however, does not seem to be committed to respecting the Potterians' inheritance wills. *The Decree for Justifiable Confiscation* gives the Ministry the power to confiscate 'within 31 days' anything that is inherited. This particular law was created to prevent wizards from passing on Dark Artifacts, and it can be applied only if there is 'powerful evidence that the deceased's possessions are illegal' (Rowling, 2007, p. 80). However, the Ministry of Magic applies to the law its own interpretation, by arbitrarily and selectively using it to confiscate valuable artifacts. For example, the Minister of Magic Rufus Scrimgeour relies on this law to harass and interrogate Harry and his friends when he informs them about the inheritance Dumbledore has left for them (Rowling, 2007, pp. 80–85). Moreover, the Minister refuses to hand the sword of Godric Gryffindor to Harry, against the will that Dumbledore has left, arguing that 'the sword of Godric Gryffindor is an important historical artifact' (Rowling, 2007, p. 85). As another example, the High Inquisitor and Head Mistress of Hogwarts, Ms. Dolores Umbridge uses her absolute powers and authority to arbitrarily confiscate Harry Potter's and Ron Weasley's broomsticks without any hearing or other due process (Rowling, 2003, p. 311).

The Potterian government also interferes at will in the operation of private businesses. Goblins, for example, complain at one point that the Ministry interferes with Gringotts' affairs, making it unsafe for its employees (Rowling, 2007, p. 296).

According to Hayek (2006), rule of law ' ... means that government in all its actions is bound by rules fixed and announced beforehand—rules which make it possible to foresee with fair certainty how the authority will use its coercive powers in given circumstances ... the essential point, that the discretion left to the executive organs wielding coercive power should be reduced as much as possible, is clear enough ... under the Rule of Law, the government is prevented from stultifying individual efforts by ad hoc action ... It is the Rule of Law, in the sense of the rule of formal law, the absence of legal privileges of particular people designated by authority, which safeguards that equality before the law which is the opposite of arbitrary government' (pp. 75–76, 82).

Clearly, Potterians' legal system is far from a rule of law, as characterized by Hayek (2006). Rather, it seems closer to a rule of government in Nietzschean spirit (Nietzsche 1997, Hillman, 2009). Since there are no lawyers, no independent judiciary and no court system, rule of law is doubtful. Moreover, given the low ethical and moral standards of many of the Ministry's officials, the Potterians' fate, freedom and welfare seem to depend entirely on the goodwill of the government's officials. The Potterian wizards are not equal before the law. The Ministry of Magic has no regard for the rule of law, has no ethical restraints, uses its monopoly power selectively by applying its own interpretation to the laws and by denying the rights of the wizards that it dislikes. Since the government does not treat or view them equally, the wizards cannot trust their government or its intentions. They are terrified by the government's powerful bureaucrats and officials, and by the ease at which they selectively interpret, manipulate or even retroactively change the laws and regulations to harass citizens they dislike, while protecting and helping their, usually wealthy and well-connected, friends.

## MARKET STRUCTURE: MONOPOLIES, OLIGARCHIES, AND OTHER PATHOLOGIES

The number of consumers in the Potterian economy is relatively small. Moreover, the population size has remained stable for generations. Hogwarts was founded around 990 A.D. and its size has not changed since, suggesting that the number of wizard children remained stable for at least that long. In addition, most wizards live in London or near Hogsmeade, a village in northern England. Consequently, two shopping centers, one in Hogsmeade and the other in Diagon Alley in London, seem to satisfy the shopping needs of the English wizards.

Most businesses operating at these shopping centers have been around for a long period. The low business turnover is the result of several factors. First, the Potterian government's regulations give few existing well-connected businesses monopoly power. By applying these regulations selectively, public officials block competition. Consequently, new businesses open rarely. For example, at one point an entrepreneur, Ali Bashir, wants to import flying carpets that would substitute for flying brooms. The proposal is supported by a senior official, Mr. Barty Crouch, who has a personal interest in the importation of carpets. However, his initiative is blocked because the Ministry defines carpets as Muggle Artifacts, i.e. as objects that are 'too similar to objects used by muggles' (Rowling, 2000, p. 59). Although the argument is weak, the regulations cannot be changed due to technicalities. The same official, however, does not prohibit his sons from selling muggle products (Rowling, 2005, p.118).

In another instance, a public official, Percy Weasley, is asked to formulate and pass a law that will prohibit the importation of caldrons (Rowling, 2000, pp. 36–37), and thus protect the interests of a local producer. The public ends up paying higher prices because of these import restrictions.

Second, the lack of financial markets further reduces business turnover. Most wizards do not have the capital needed to open a business and are therefore forced to solicit the help of wealthy wizards. That is how Borgin and Burkes opened a shop in the 19th century, and that is how the Weasley twins managed to open a shop about 100 years afterward. There is no mention in the books of any other shop opening in Diagon Alley during that period.

A third factor contributing to the low business turnover is the monopoly on information. The *Daily Prophet* has a monopoly, and its reporting is systematically biased in favor of wealthy wizards (Rowling, 2000, p. 213, Rowling, 2003, pp. 72, 423) (There is another publication, *The Quibbler* (Rowling, 2003, p. 144), which is as reliable as the modern-day tabloids that publish articles such as 'I Was Kidnapped by Aliens!' 'I Saw Bigfoot in My Back Yard!' etc.). Also, it selectively prevents information about new products from reaching the public. When the Weasley twins open their shop, the paper does not publish their advertisements, nor share with the public information about the shop, despite the shop's success. (For example, it seems that the wizards that read the *Daily Prophet*, including Harry Potter—a devoted reader of the newspaper, know nothing about the twins' joke shop (Rowling, 2000, p. 36).

These barriers and restrictions lead to high entry costs, limit the number of sellers and give the existing firms market power. For example, the UK wizards have only one wand-maker, the Ollivanders (Rowling, 1998, p. 53). While it is hard to assess the effect of the market concentration on markups, the extent of price rigidity and the frequent use of convenient prices suggest that the Potterian retailers have substantial profit margins. These margins would likely fall if the markets were to open to imports (Rowling, 2000, pp. 36–37).

The low business turnover also limits product variety and innovations. For example, the same textbooks have been used for over 20 years. Similarly, children eat the same candies and collect the same cards as did their parents (Rowling, 2005, p. 31). Thus, the lack of competition limits the choice of the Potterians, although there is demand for new products. For example, after Zonko's Joke Shop goes bankrupt, the Weasleys' Wizard Wheezes has no competition, although there is a demand for the new jokes and tricks that the Weasleys sell (See Rowling (2000, p. 36), Rowling (2003, p. 403), Rowling (2005, pp. 76, 78, 158–159) and Tirole (1988).).

Thus, the Potterian economy is a small, monopoly-dominated economy with a large corrupt public sector and an inefficient credit market. Potterian markets lack competition because of the government's selective use of its regulatory powers in many spheres of life. This leads to low business turnover, a limited variety of goods and services, and thus to a limited choice, welfare loss and inefficient outcomes. The monopoly on information worsens the state of affairs further because the wizards have no access to useful information that could help them improve their private lives (Note: Compare this with the effect of the internet, which allows consumers to compare the prices of similar goods from thousands of sellers almost effortlessly, on the demand for goods at shopping malls.).

The inefficient and selective regulatory interventions of the Potterian public officials suggest that the world-view the books promote is in line with the economic models of public choice. At the same time, however, the actions of these officials and their outcomes for the Potterian economy and the Potterian public resemble the communist model. The *Daily Prophet* reminds us of the Soviet *Pravda*, which had a monopoly on information in the former Soviet Union and served as a propaganda machine for the communist government. More generally, the limited choice of the Potterians very much resembles the limited choice in the former USSR and other Soviet-bloc communist countries.

## INCOME DISTRIBUTION, INCOME INEQUALITY, AND SOCIAL IMMOBILITY

The wizards' society is composed of a low class, a large middle class and a small elite. The middle-class wizards earn enough to live comfortably but not enough to save. They, therefore, work almost their entire lives. Wealthy wizards enjoy a luxurious lifestyle and own almost all the assets and capital (Rowling, 1999a, p. 19).

Social mobility between the high class and the other classes is mostly downward. There is just one prominent character that makes it from the bottom to the top, Voldemort, but several characters, including Sirius Black, Andromeda Tonks (Black) and the Gaunts, that moved down the social ladder (Note: Voldemort comes from a well-established family that was stripped of its assets (Rowling, 2005, p. 138).). Mobility between low- and middle-class wizards is more common. For example, Hermione starts as a mudblood wizard and ends as the Minister of Magic (Source: https://harrypotter.fandom.com/wiki/Hermione_Granger, accessed August 5, 2021.). Most of the Weasley brothers have also improved their economic conditions.

One reason for the limited social mobility is the insufficient free enterprise due to negative social images of businesspeople in the Potterian society. The Weasleys' mother does not encourage her children's business aspirations because she believes that public sector jobs bring a better reputation (Rowling, 2000, p. 36, Rowling, 2003, p. 79).

Another reason for the low social mobility is that wealthy wizards view themselves as superior, especially to wizards with muggle predecessors (Rowling, 2003, pp. 84, 584, Rowling, 2005, p. 417), who are considered a threat because of their different culture (Rowling, 2000, p. 66). Muggle-born wizards are subject to constant harassment, ridicule, abuse and discrimination, and are derogatorily called 'mudbloods' (Rowling, 1999a, pp. 72–74, Rowling, 2005, p. 74). Wealthy wizards favor policies that limit the influence of middle-class wizards. Most of the supporters of Voldemort, for example, come from wealthy families, and their goal is to enslave the middle class, especially the mudbloods.

The middle-class wizards are unlikely to move up the social ladder through marriage as they rarely socialize with wealthy wizards. Wealthy wizards are often disinherited by their families if they associate themselves with middle-class wizards (Rowling, 1999a, p. 96, Rowling, 2003, p. 103), which along with intermarriages further block upward mobility (Rowling, 2003, p. 85). Thus, the biases of the elite against lower-class wizards minimize the opportunities for upward social mobility, resulting in social mobility being mostly downward (Dearden et al., 1997) and is expected to persist (Mulligan, 1999).

The social structure of the wizard society leads to a struggle between middle- and higher-class wizards. The upper-class wizards use their influence and even violence, to repress the middle class, while the middle-class wizards scorn the upper-class wizards. The upper-class wizards even try to control the thoughts and beliefs of the middle-class wizards by influencing the curriculum used in the school.

Although this seems like a Marxian-style social struggle, it is a struggle that is fought in a society with central planning where the government owns and regulates production. In the Marxian model, free markets are usually considered the cause of social struggles because the owners of the capital ('capitalists') can collect riches only by robbing the middle class ('proletariat'). Also in Marxian models, central planning is the solution, because it takes the power away from the upper classes and transfers it to the government officials who represent the proletariat, and who make the decisions for all. Yet in the Potterian model, the government is controlled by the elite, and the elite use the government to repress the middle class (Murphy et al., 1993).

The outcome is that in the Potterian model, the class struggle focuses on controlling the government. The Marxian model predicts that once the middle class enjoys a relatively high standard of living and good education, there will be a class struggle that ends in the collapse of the class system and greater equality. In the Potterian economy, however, even after a class struggle that ends in the defeat of the high class, the income differences remain large. Thus, when it comes to inequality outcomes in a society with a large government, the Potterian model is more consistent with the public choice model than with the Marxian one.

## INTERNATIONAL TRADE AND MIGRATION

Wizards can travel long distances easily and cheaply. There are hardly any restrictions on international travel. Nevertheless, there is little international trade, which is partly due to regulations protecting local producers. Trade and travel are limited also by linguistic barriers as most wizards are monolingual. In addition, there are cultural barriers, as the wizards do not know much about other people's customs, traditions, etc. (Rowling, 2000, pp. 163–164, 363).

These barriers reduce the demand for foreign goods, which may explain the absence of foreign or ethnic restaurants in Potterian London and Hogsmeade. Wizards also know very little about the quality of foreign goods. For example, when a leading English wand-maker disappears, wizards do not know where to find another wand-maker although there are several quality wand producers in other countries (Rowling, 2005, p. 70).

In addition, prejudices against foreigners and non-wizards eliminate interactions with muggles almost completely. Indeed, one of the main responsibilities of the Ministry of Magic is to ensure that muggles are not aware of wizards' existence (Rowling, 1998, p. 42). For example, Arthur Weasley's job in The Office of Misuse of Muggle Artifacts is to prevent any interaction between muggles and wizards (Rowling, 1999a, p. 20). Thus, although trade in muggle goods could be beneficial, it takes place only under special circumstances and only for specific goods (Rowling, 2005, p. 77). Another barrier to trade is the absence of immigrant networks (Rauch, 2001). In the Potterian economy, immigrants are rare. For example, there are only three characters with substantive roles, which have a foreign background, or their names suggest so. These are Viktor Krum, a Bulgarian wizard, Fleur Delacour, who came from France and Cho Chang. Just one of them Cho Chang, studies at Hogwarts. Thus, whereas almost 20% of the UK primary school pupils are exposed at home to a language other than English, there are very few, if any, such students in the only school for wizards in the UK

(Source: the UK Department of Education's Statistical Frist Release, 'Schools, pupils, and their characteristics: January 2015.').

These social norms and prejudices translate into formal trade barriers. For example, there is no protest when a junior official uses quality as a pretext to block the importation of considerably cheaper foreign goods, although the official himself admits that they are of almost the same quality as the locally produced goods (Rowling, 2000, pp. 36–37, 59).

However, some types of foreign work are so profitable that they are tolerated. Wizards use 'house elves', a special kind of humanoids, to do manual and dirty work, perhaps analogous to foreign workers. Although the elves provide many useful services, they work in terrible conditions for almost no pay. They do not own even their clothing. They are nevertheless diligent, work without a break for many years, accept slavery conditions and are extremely afraid of being sacked because unemployment is almost certain death for them. These considerations make the use of elves so profitable, that despite the wizards' prejudices against any type of humanoids, they employ elves in large numbers (Rowling, 2000 pp. 64, 80, 89).

In sum, wizards could benefit by trading with muggles. They could also import muggle-produced goods or adapt their designs and thus increase the variety of goods at a relatively low cost. However, trade is limited by cultural norms and prejudices that make most wizards view muggle-made goods as inferior (Rowling, 1999a, p. 25, Rowling, 2005, p. 77). Further, cultural differences prevent most trade between wizards from different countries, and between wizards and other humanoids. It, therefore, seems that unlike most models of trade, transaction opportunities in the Potterian economy are limited by culture. Thus, much of the potential trade in the Potterian economy is blocked by protectionism (Grossman and Helpman, 1994) and cultural prejudices (Bala and Long, 2005).

In summary, whereas in economic models, investments depend on borrowing cost and rate of return, in the Potterian model, they are determined primarily by prejudices and social norms. This is consistent with studies that show that social norms are a barrier to trade, except that in the Potterian economy, the effect of social norms on trade is perhaps an order of magnitude higher in comparison to the effects reported in empirical studies (See McCallum (1995), Nunn & Wantchekon (2011) and Clerides et al. (2015).).

## WAR ECONOMICS

In Rowling (2005, 2007), the wizards fight a war against 'Death Eaters', a group of wizards that either belongs to the upper class or aspire to associate themselves with the upper class. Their goal is to take control over the wizard world, thereby enslaving the middle class and driving out those wizards whose ancestors were not wizards. They are extremely committed and are willing to use any form of violence, including suicide attacks, to achieve their goal.

The signs of danger appear two years before the first acts of violence, but the government ignores them (Rowling, 2000) (Note: Compare this to the recent criticisms of the US government concerning its slow and delayed responses to ISIS terror www.nytimes.com/2014/09/29/world/middleeast/president-obama.html, accessed June 6, 2022), flood water (www.msnbc.msn.com/id/9614737/, accessed June 6, 2022), bird flu (www.msnbc.msn.com/id/9661312/, accessed June 6, 2022) and Ebola outbreak (www.nytimes.com/2014/09/26/world/africa/obama-warns-of-slow-response-to-ebola-crisis.html?_r=0, accessed June 6, 2022).). This is consistent with Tuchman's (1990) theory of march of folly: the government not only ignores signs of warning but also silences anybody who expresses opinions other than the official ones.

When violence breaks out, however, the government has a sudden need to obtain military equipment. As more resources are spent on supplying military needs, common people find it hard to obtain the goods they need for daily life. Indeed, one of Harry Potter's teachers, Professor Slughorn, complains that due to the war, prices are sky-high and that it is difficult to obtain even the most elementary products (Rowling, 2005, p. 43).

The government also fails to prepare adequately and, consequently, it has to buy costly equipment that would be unnecessary if it had provided its staff with proper training. Additional cost born by the public because of the government inefficiency is the price paid for goods sold by swindlers that offer a false sense of security (Rowling, 2005, pp. 73, 78) (Note: Similar security needs lead some modern governments to spend millions of dollars on useless bomb detectors sold by swindlers. Source: www.guardian.co.uk/uk/2013/apr/23/somerset-business-guilty-fake-bombs, accessed June 6, 2022. After the crash of the Russian Metrojet flight 9268 in Egypt on October 31, 2015, it was reported that many Sharm el-Sheikh hotels use fake bomb detectors. A November 10, 2015 report of CNN described them as 'magic wand' detectors. Source: http://edition.cnn.com/2015/11/10/middleeast/egypt-sharm-fake-bomb-detectors/, accessed June 6, 2022.).

In addition, Terror generates fear, which influences people's moods by making everything less enjoyable (Note: The effects of fear might be long lasting. Indeed, many wizards remember and react irrationally when they hear the name Voldemort even though for many years, he was thought to be dead (Huddy et al., 2002).). Indeed, the usually crowded 'Leaky Cauldron' bar is empty because even the most loyal consumers seem to have lost their appetite (Rowling, 2005, p. 72). In Rowling (2005, p. 72), the bar owner notices Hagrid, who is well known for his fondness for alcohol. From the barman's reaction, it is clear that Hagrid is one of his last loyal customers.

These events demonstrate that influential events such as wars can have long-term effects on people's preferences. For example, as predicted by recent studies, the war brings about a significant drop in the demand for dining at restaurants and bars (Gould and Klor, 2010; Becker and Rubinstein, 2011). In Israel, during the Second Intifada in 2000–2002, there were Palestinian terror attacks in public places. The consumers responded by eating more at home/office. Restaurants responded by offering delivery services. The share of restaurants offering delivery services, however, did not decrease after the cessation of violence, suggesting that the taste for deliveries remained.

When the Potterians see the inability of their government to respond, they form the 'Order of the Phoenix,' whose goal is to fight the terrorists. Their efforts, however, do not fully compensate for the lack of government action because of the high private costs of fighting terrorists. Indeed, those who fight the terrorists become primary targets.

The members of the Order of the Phoenix are therefore isolated and gain little public support. Without public support, their efforts can at best only slow down the progress of the terrorists, but they cannot prevent them from gaining almost full control over most of the institutions of the Potterian economy (Rowling, 2003, p. 88).

The Potterian model is therefore consistent with the economists' insight that governments should provide basic public goods such as security because the private sector cannot provide them efficiently (Samuelson, 1954). In the Potterian model, however, the government also controls the production of many goods that are not public, suggesting public-choice type distributive outcomes (Tullock, 1959).

The war underscores the Potterian government's inefficiency in responding in a timely fashion to the signs of danger. Furthermore, although the war might seem like a class struggle, its outcome contradicts the Marxian interpretation of a class struggle. In the Potterian model, as in the Marxian model, the upper class struggles to secure its status. In the Marxian model, however, once the middle class becomes aware of the struggle, it fights back and ultimately all classes are eliminated and a new economic order is established.

In addition, in the Marxist model, class struggle is an outcome of capitalistic motives, where the high class robes the middle class of its share in production. The solution in the Marxist model is central planning by the government. In the Potterian model, however, there is already central planning. The government, however, is controlled by the upper class who uses it as means to control the middle class.

The Potterians' war ends when more middle-class wizards join their comrades in the fight (Rowling, 2007, p. 511). The outcome, however, is not a new social order. Although the middle-class wins, the high-class wizards still preserve their high-class status (Rowling, 2007,

p. 605) and their hold over the government, which controls the economy.

## INNOVATIONS AND TECHNOLOGICAL PROGRESS

According to Snir and Levi (2010), the Potterian economy is not growing because there is no growth in the labor force, there is no accumulation of physical or human capital, and there is no evidence of any kind of technological progress.

The Potterians' broomstick industry is an exemption. A careful reading of the Harry Potter books suggests that the Potterian broomstick industry has been experiencing significant technological progress over time. For example, we are told that there are several makes and models of broomsticks. These include Cleansweep-5 (Rowling, 1999a, p. 71), Cleansweep-6 (Rowling, 2003, p. 143), Cleansweep-7 (Rowling, 1998, p. 98) and Cleansweep-11 (Rowling, 2003, p. 202), Nimbus-2000 (Rowling, 1998, p. 108) and Nimbus-2001 (Rowling, 1999a, p. 71), Comet-260 (Rowling, 1998, p. 107) and Comet-290 (Rowling, 2003, p. 128), the Shooting Star (Rowling, 1999a, p. 30), the Bluebottle (Rowling, 2000, pp. 62–63), the Silver Arrow (Rowling, 1999b, p. 162) and the top-of-the-line Firebolt (Rowling, 1999b, p. 32).

There is also evidence that over time there were non-trivial improvements in the broomsticks' production, especially in their quality. For example, some of the early models were simple and basic, but more recent ones are more advanced. For example, the Shooting Star, the model owned by the Hogwarts School, is a basic broomstick, relatively cheap, 'very slow and jerky' (Rowling, 1999a, p. 30, Rowling, 1999b, p. 121), similar to the Bluebottle family series with an anti-burglar buzzer (Rowling, 2000, pp. 62–63). Nimbus-2000, on the other hand, is a top-of-the-line broomstick from the time of its release (Rowling, 1999a, p. 30), until Nimbus-2001 is released. The Firebolt, which is the latest addition, is perhaps 'the Ferrari' of the broomsticks. It is the best, fastest and most aerodynamically efficient model, offering a smooth action and fine control. It is a dream broomstick. Indeed, Harry repeatedly visits 'Quality Quidditch Supplies' to look at and admire the prototype Firebolt model that the store displays, and is considering spending all his savings to buy it (Rowling, 1999b, p. 32).

The way the Potterian broomstick industry functions and develops is comparable to the modern auto industry, particularly because new broomstick models are released almost every year. The newer model broomsticks use better materials (e.g. type of wood such as Spanish Oak), have better precision, offer better balance, are faster, etc. The Firebolt, for example, can accelerate from 0 to 150 mph in 10 seconds, which is at least twice as fast as the Cleansweep. Comet-260 looks ' … like a joke next to the Firebolt' (Rowling, 1999b, p. 162). Thus, each model is of better quality than the older one, which is indicative of technological improvements in the broomstick industry, a process resembling a Schumpeterian mechanism of creative destruction.

An interesting question that follows from these observations is how much of the broomstick technological innovations translate into economy-wide innovations. Our reading of the books suggests that broomsticks are primarily used for leisure-related activities, especially for the Quidditch games. For day-to-day purposes, Potterians use other means of transportation including apparition—a spell that allows the caster to move from one spot to another instantaneously (Rowling, 2000, pp. 43–44), magic-propelled boats (Rowling, 1998, p. 42), flying horseless carriages (Rowling, 2000, p. 109), the Floo-network that allows immediate transportation to any place in the network (Rowling, 2000, p. 29), Knight-Bus (Rowling, 1999b, p. 21), a magical train, the Hogwarts Express (Rowling, 1998, p. 60), the Ministry of Magic cars (Rowling, 1999b, p. 45) and Portkeys—objects that transport to a pre-determined location when touched (Rowling, 2000, p. 46).

In other words, the technological progress in the production of broomsticks contributes to the Potterians' leisure-related activities, the game of Quidditch being the primary example. This conclusion is also consistent with the observation that most stores in Diagon Alley sell the same old stuff over centuries, except for the Weasley twins' Joke Shop, which sells some new models of games and toys, which are also an example of leisure goods.

Thus, in general, we do not see any evidence of the innovations in the broomstick industry spreading to the general economy. This is consistent with the finding that the Potterian economy is stagnating; there is hardly any growth in the labor force, there is no accumulation of knowledge (human capital) and we see no evidence of large-scale capital investments. Consequently, we conclude that very little (if any) of the broomstick technology innovations spread to other industries, or more generally, to the wider economy.

The lack of technological progress is also consistent with the common image of centrally planned economies. In the Potterian economy, with the government either producing many goods or regulating their production, government officials do not benefit from the introduction of new products and therefore they have little incentive to introduce them. On the contrary, they might have incentives to erect barriers to such innovations if they see technological progress as a threat to their employment, which is a common view among labor unions. Perhaps it is not surprising, therefore, that the only industry in which technology is progressing is not a part of the public sector.

## POTTERIAN EDUCATION SYSTEM: INVESTMENT IN HUMAN CAPITAL

Harry Potter books revolve around the life at Hogwarts School of Magic. It is fitting, therefore, to end with a short discussion of the Potterian education system (Note: See Snir & Levy (2010) for a discussion of the Potterian education system in the context of Solow growth model.). The wizards' education system is publicly financed. It gives the students basic training and ensures that all wizards graduate with knowledge that allows them to find a job. The school is also one of the only institutions where the rules treat all members of society equally. It is also the only institution where middle- and high-class wizards interact on a day-to-day basis.

The wizards value education highly, and thus the public image of teachers is positive, so much so that they are even willing to sacrifice future income opportunities just to become teachers. For example, Gilderoy Lockhart, the celebrity author of seven books (Rowling, 1999a, pp. 28–29, 38), becomes a professor at Hogwarts (Rowling, 1999a, p. 64). The value of education is further underscored by the high status of the school headmaster. The current headmaster, for example, is considered by many to be the greatest wizard of his time (Rowling, 1998, p. 66) (Note: See the number of Hogwarts' former headmasters that are included in the list of the most famous wizards: www.hp-lexicon.org/wizards/card_wizards.html, accessed June 6, 2022.).

The Potterian education system is not problem-free, however. Hogwarts is under the supervision of the Ministry of Magic and, thus, subject to the influence of politicians. The government can interfere with the school curriculum at will (Rowling, 2003, p. 229). Another weakness of Hogwarts is that its curriculum has not been revised for hundreds of years. Consequently, the graduates do not know more than their predecessors. The subjects and classes they take are those that were taken by the students of previous generations (Rowling, 1999a, p. 161). Similarly, the textbooks they use were also used by their parents and even grandparents. For example, the used copy of *Advanced Potion-Making* that Harry Potter borrowed from Professor Slughorn (Rowling, 2005, pp. 123, 125) was also used by Professor Snape's mom, Eileen Prince (Rowling, 2005, p. 417), and by the Half-Blood Prince—Professor Snape (Rowling, 2005, p. 126), who was a classmate of Harry's parents (Rowling, 2005, p. 360). The old-fashioned curriculum does not encourage innovative and creative thinking (Snir and Levy, 2010). The Potterian students focus primarily on practical skills but learn very little theory, literature, arts or philosophy. Most lack creative skills and cannot think originally or independently (Rowling, 2000, p. 437).

Potterians do not have elementary schools or institutes of higher education like universities. Upon graduating from Hogwarts, wizards choose a profession and usually stick to it (Rowling, 1999a, pp. 161–162). Since there is no incentive to study further, Potterians' stock of knowledge does not increase beyond what is acquired in the school. Their education system, therefore, resembles the Marxian system. Education is compulsory and all students receive the same education. Yet even in this seemingly equal education system, there are inequalities. Upon arrival at Hogwarts, the students are divided into four 'houses' based, to a large extent, on their ancestral history. Most high-class wizards end up in one particular house, whereas the middle-class wizards are scattered between the other three houses.

The students, school experience depends largely on the house to which they belong. They, therefore, graduate with a strong association with the students from their own house, and thus the high-class wizards leave the school with different values, norms and associates than the wizards that belonged to other houses. Thus, the system that should have fostered equality of values ends up instead fomenting social division, tension and struggle.

Given the central role of education in the Potterian economy, we could consider the role of the Hogwarts School as a mechanism that contributes to the Potterians' economic growth via endogenous accumulation of human capital (Uzawa, 1965; Lucas Jr., 1988). In Lucas' model, human capital accumulation is modeled in the same way as physical capital accumulation is modeled in the Solow model. However, there is no knowledge accumulation in the Potterian economy. Thus, the stock of human capital is essentially fixed. Moreover, the contribution of education to the wizards' productivity after they graduate and start working, most of them for the Potterian government, is unclear. Since there is no evidence of significant physical capital investment in the Potterian economy, the contribution of the wizards' schooling to the productivity of physical capital is also doubtful. Along with the evidence that the Potterian economy is not growing (Snir and Levy, 2010), this suggests that the endogenous human capital accumulation mechanism does not play a role in the Potterian economy. As Long (2005) notes, 'If only "He-Who-Must-Not-Be-Named" understood neo-classical endogenous growth theory, it might have been a different story.'

## CONCLUSION: SUMMARY, POSSIBLE OBJECTIONS AND CAVEATS

This paper is motivated, first by the evidence that much of the general knowledge about economic issues is transmitted and learned through various print and electronic media, and second, by the recent findings in psychology and neuroscience about the powerful influence that literary works, particularly fiction, have on the human mind. Building on these findings, we analyze the Potterian economy to assess the economic ideas and insights that these books convey to the readers, i.e. the principles of Potterian economics.

We find that the Potterian model is not in line with a single economic model. Rather, it is a mix of ideas from various models. These include Marxian notions of class struggle and equality, public choice aspects of inefficient and corrupt government, sticky prices in the Keynesian spirit, full employment of the type found in Classical models, etc.

Because the Potterian economics mixes ideas from different models and worldviews, many elements of the Potterian model are mutually inconsistent and contradictory, and thus perhaps even confusing. For example, the Potterian economic model is critical of market-based systems, yet it belittles government. The government is corrupt, yet it has public support. Many welfare-improving and mutually beneficial transactions do not take place, and there are no credit markets because of biases and prejudices, yet the books are often viewed and described as rejecting stereotypes. Potterian money is made of precious metals, yet its purchasing power is unrelated to its commodity value. The Potterian wizards value education, yet their school curriculum excludes theoretical subjects, their knowledge does not exceed much of their parents' knowledge and they do not have institutes of high education, such as universities or colleges.

Moreover, the Potterian model misses many deep and fundamental aspects of economic analysis. For example, the Potterian bank does not serve as an intermediary between savers and investors, the Potterian money lacks some key attributes (e.g. divisibility, portability, and homogeneity) that are essential for it to serve as an efficient medium of exchange or store of value, arbitrage opportunities are not exploited and efficiency-improving transactions go unnoticed.

In addition, the Potterian model lacks discussion of important aspects of real-world economies and societies. One important theme that is absent from the world of Harry Potter is religion. There are only two occasions where religion-related acts take place. One is the Christmas dinner party, which is held annually at the Hogwarts (e.g. Rowling, 1998, pp. 131–132). The second is the inscription Harry finds on his parents' gravestone: 'The last enemy that shall be destroyed is death' (Rowling, 2007, pp. 216–217), which comes from the *New Testament, 1 Corinthians* 15:26, King James Version.

Taxes of any kind are absent as well. The word 'tax' is mentioned in the seven-book series only once, in Rowling (2000, p. 3), in the context of the ownership of Riddle House: 'The wealthy man who owned the Riddle House these days neither lived there nor put it to any use; they said in the village that he kept it for "tax reasons," though nobody was very clear what these might be.' Thus, if the wizards are subject to some taxes like ordinary muggles are, the author does not give us any information about it. It could be that taxes in the Potterian economy come in the form of constraints and limitations, for example, limitations on what the Potterian wizards can and cannot do. Such limitations could be interpreted as a tax. That, however, still leaves unanswered the question: why do Potterians choose this kind of 'limitation-tax' over income tax, which is more standard? The puzzle is particularly interesting given that the Potterian government, that is the Ministry of Magic, is quite powerful, and it is the main employer in the Potterian economy. Therefore, it would have no difficulty collecting such taxes.

According to Ross (2015), '"In a July 2015 story, a journalist estimated it would cost more than $43,000 a year to go to Hogwarts, including all the Diagon Alley necessities like wands, robes, books and so on." "My friends and I are having a super-intense debate about the cost of tuition at Hogwarts. Thanks." "Rowling quickly shut down the idea on Twitter, clarifying that magical education is, in fact, free": "There's no tuition fee! The Ministry of Magic covers the cost of all magical education!"' (Source: A. Ross, 'The 50 Most Important Things We've Learned from J.K. Rowling,' *Time*, July 30, 2015.)

Thus, according to J.K. Rowling, there are no tuition fees at Hogwarts. But then the absence of taxes is particularly puzzling because, in any economy, taxes would be essential for funding public expenditures, such as public education. How is 'free education' funded? Who pays for it? The lack of taxes cannot be explained by the medieval structure of the Potterian society either, since taxes are mentioned numerous times in the *Bible*, in both the *Old Testament* (e.g. *Exodus* 30:11–16) and the *New Testament* (e.g. *Romans* 13:1–7). It is well known also that taxes were collected in Ancient Greece (e.g. Sosin, 2014), as well as in the Roman Empire (e.g. Hopkins, 1980). The absence of taxes in the Potterian economy is therefore mystifying, difficult to understand and a significant distortion.

It follows that a naïve reader of Harry Potter would get a distorted view of economics. Consider some of the lessons that the principles of Potterian economics teach: governments are corrupt, wasteful and not trustworthy; markets are not fair because transactions are zero-sum; political process is not transparent; government is big and controls much of the production; governments cannot provide security, the most basic public good; in the marketplace, the wealthy have advantage over the poor; markets encourage crony capitalism; capitalists want to enslave the proletariat; bankers are goblins; businessmen are deceptive and devious; wealthy people are mean and unethical; there is monopoly on information; power is concentrated; wealth is used to influence the government; domestic producers should be protected from foreign competition even if they are inefficient; international trade is not good; ignorance about foreigners, foreign cultures and foreign languages is the norm; lack of institutions offering financial intermediary services is the norm; no interest payments are expected on deposits; commodity money, coin denominations and attributes, and their exchange rates, all lead to higher transaction costs; paper checks are non-existent; arbitrage opportunities are not exploited; monopolies are common; economic growth is not important; economic stagnation is not something that should bother us; technological progress is not essential; innovative and creative thinking is rare; prices are expected to remain unchanged; capital investment is non-existent; human capital does not accumulate; public employees should expect life-time job-security regardless of how inefficient and unproductive they are; the default employer is the public sector, not the private sector; downward social mobility is the norm; there are no taxes yet education is free; high education is not critical; curriculum can remain unchanged for centuries; only practical skills are valued, theoretical knowledge is of no value; there is a constant class struggle; and this is only a partial list.

Many of these lessons seem shallow and uninformed characterizations of markets, market institutions and arrangements and market participants and their motives and goals. In light of the evidence about the potential influence of popular fiction on readers' views and opinions, these can influence and shape the public's understanding of economic issues and matters.

Perhaps more importantly, however, these distorted views of the economy can potentially contribute to the public's biases, misconceptions and more generally to their economic illiteracy, as documented by Caplan (2007), Rubin (2003) and others. Because most of the public never take college- or university-level courses in economics, they are susceptible to these kinds of subtle influences (Salonikov et al., 2022). It is likely, therefore, that the public that is exposed to such views and sentiments about economics, economic institutions, economic decisions and their outcomes, can be persuaded more easily by populist arguments against certain policies, against businessmen, against bankers and other beneficial service providers, against authorities (e.g. the central bank), etc.

Moreover, the confusing insights the Harry Potter books convey about economics and economic matters by mixing various conflicting and mutually inconsistent messages, observations and arguments can lead to a process where a naïve citizen might lose faith in both market institutions as well as in government authority. The results might be apathy and indifference, which should be troubling.

These findings imply that popular literature along with other information intermediaries may be contributing to the public's lack of understanding of many economic issues, as some studies of economic and financial literacy have documented recently. But we should note that some of the biases we identify, many of them stereotypical, are not new, as they have been around for centuries (Harap, 2003). For example, the negative portrayal of bankers has been a recurring theme in popular literature for a long time. Similarly, populist views and criticisms of governments, class struggles and other social ills and institutions are not new.

These observations suggest that in addition to directly influencing the public views and opinions, Harry Potter books could be reinforcing the existing norms and beliefs. This reinforcement mechanism might be playing an important role in transmitting the biases and other social treats through generations, via the process of cultural transmission of values (Bisin and Verdier, 2000, 2008; Necker and Voskort, 2014). Moreover, the formation and propagation of these biases and norms may be taking place from the period of early youth, because this is the age group in which fiction's influence is likely to be particularly strong and long-lasting. Further, an entire generation of children and youngsters grew up along with Harry Potter, closely following his adventures for over 20 years since the first book in the series was published in 1998. Therefore, given the books extraordinarily large worldwide audience, and given that a significant proportion of the readers are kids, these biased and distorted views about many fundamental economic issues could potentially persist, and even spread further. A folk economic interpretation of the Potterian model, therefore, suggests that popular intermediaries might be playing an important role in spreading biases and ignorance on important economic issues.

We shall note two possible objections to our interpretation. First, one could argue that because many of the Potterian economic principles coincide with popular folk economics, we should not expect them to be consistent with professional economists' models, views or insights. For example, Harry Potter books are liked by many adults that read them, but they were primarily written for children. That might explain, one could argue, why the Potterian economy is so simple, with no capital markets, no taxes, etc. Potterian principles of economics pay attention to the Potterians' unequal income distribution, emphasizing the affluence of the upper-class wizards and the poverty of the lower-class wizards, perhaps because children can relate to poor vs rich, but not to deeper notions of incentives and efficiency. These simplifying assumptions are necessary to make the books children-friendly.

Following this line of argument, other aspects of the Potterian world can be explained by the fact that the books belong to a fantasy genre, and the events they describe are set in medieval times. Indeed, wizards, goblins, dragons and other similar creatures are typically associated with medieval times. That can explain also the use of commodity money, why goblins are unfriendly, etc.

Similarly, the stagnation of the Potterian economy is necessary for making the story credible, as it takes place during the medieval period, a period in which the world was stagnating because there was no increase in capital stock or population, and there was little technological progress. The corrupt and inefficient government could be a literary device similar to those used in comic books such as 'Superman' or 'Batman': corrupt characters are required to have a need for a hero in the form of Harry Potter and his sidekicks.

The second type of argument could be that Potterian economics merely captures the author's subjective opinions and views, which reflect her personal experiences and biases about economic issues, and the financial consequences of her life experience as a divorced, unemployed and welfare-dependent single parent, followed by phenomenal success (Source:

https://www.businessinsider.com/the-rags-to-riches-story-of-jk-rowling-2015-5, accessed June 6, 2022.).

These observations and arguments are important. However, the books' broad, near-universal success suggests that in addition to telling a story, which seems to appeal to a large and diverse audience, the author may have also been able to capture the popular beliefs and sentiments of hundreds of millions of people across the globe. In other words, people from a broad and varied range of backgrounds, values, cultures, societies, religions and places, irrespective of age and gender, all seem to be able to relate to the Potterian story and the sentiments it conveys, consciously and perhaps subconsciously.

Under this reasoning, we argue, Potterian economics can still inform us about the formation of folk economics, even if from the author's point of view, it merely reflects her personal opinion based on her private life experiences, or even if the entire project was intended just for children's bedtime reading. For similar reasons, the value of studying Potterian economics is not diminished even if the author is 'expressive' (Hillman, 2010b), i.e. she is expressing not necessarily her own beliefs about economic matters, but rather her beliefs about her readers' beliefs.

Thus, regardless of the motivation of the author for writing these books, and regardless of whether the books reflect her private opinions, or her beliefs about her readers' opinions, interpreting the Potterian economic model as a reflection of folk economics suggests that popular intermediaries could potentially play an important role in spreading biases, distorted opinions and ignorance on important economic issues. Thus, rather than dismissing the 'mishmash' of mutually inconsistent and contradictory economic ideas found in the Harry Potter books, we suggest taking them seriously to try and understand their sources and persistence.

As Gary Becker is quoted in the August 21, 2011, *New York Times* article (p. SR5), writers do not always understand the intricacies of how markets work and function. It is not easy to write an adventure story that contains all social and market institutions and get it fully right. It is similar to writing a formal economic model: it requires a solid and deep understanding of economic ideas. But this is not limited to economics, however, as it applies to other fields of knowledge as well. Therefore, it is not surprising that some of the most successful fiction authors, it turns out, have spent an enormous amount of time on research and study (Francis, 2012; Popoola, 2015). For example, according to Freedman (2005, p. xiii), Isaac Asimov was the best physicist among non-physicists, and the best zoologist among non-zoologists, hinting at the seriousness in which Asimov did his research before writing his books and essays. Similarly, it took J.R.R. Tolkien 17 years to complete *The Lord of the Rings* (Source: middle-earth.xenite.org/2011/09/16/how-long-did-it-take-j-r-r-tolkien-to-write-the-lord-of-the-rings/, accessed June 6, 2022.).

As a caveat, we should note that the interpretation of Potterian economics, which we lay out in this paper, is based on our understanding of the economics of Harry Potter. Many of the ideas and the interpretations we offer seem non-controversial and consistent with the interpretations of other scholars and writers cited above, most of them outside the field of economics. However, on some aspects of the Potterian world, there is less consensus and thus they are subject to different interpretations.

For example, in an essay published in June 2004 in the French daily *Le Monde*, Ilias Yocaris, a professor of literary theory and French literature, argued that 'the fantastic universe of Harry Potter is a capitalist universe.' In July 2004, however, the same newspaper published an article by Isabelle Smadja, a professor of philosophy, in which she argues exactly the opposite: 'Far from being a capitalist lackey, Harry Potter is, in fact, the first fictional hero of the anti-globalist, anti-free market, pro-Third World, "Seattle" generation … [Smadja] suggests that Yocaris has been confused by the fact that the Potter books are, themselves, such a global, commercial and marketing success. Examination of the text suggests that they are, in fact, a "ferocious critique of consumer society and the world of free enterprise."' (Source: http://www.nzherald.co.nz/lifestyle/news/article.cfm?c_id=6&objectid=3576571, accessed June 6, 2022. *New York Times* translated Ilias Yocaris' article to English and published it on its editorial page in July 18, 2004, available at http://www.nytimes.com/2004/07/18/opinion/harry-potter-market-wiz.html, accessed June 6, 2022.)

We agree that some of the aspects of the Potterian economy may be subject to a different interpretation than ours. We believe, however, that one main reason for these discrepancies is precisely what we argued above—principles of Potterian economics are a mix of ideas from different models, not always consistent with each other, quite often mutually orthogonal and contradictory. A story that contains a mix of ideas and worldviews that are inconsistent with each other will likely lead to different interpretations, depending on readers' points of view.

Before we end, consider the following thought experiment. Suppose that we took the same approach as we employed in this paper, and used it to analyze the economies of Sweden, or China, or Zimbabwe, or Lichtenstein, or Myanmar, or Georgia, or any other set of countries. These real-world economies will unlikely be consistent with any specific professional economic model. Rather, like the Potterian economic model, they will likely be consistent with some parts of some of the models.

For example, the Potterian economy is one where international trade is severely restricted by protectionism, there is hardly any migration, the economy is in permanent stagnation, innovations and technological progress are limited to the broomstick industry without economy-wide spillovers, the stock of human capital is not increasing and investments are non-existent. These are not consistent with a model where decision-makers seek to maximize social welfare. Yet these are also characteristics of certain real economies.

We believe this point highlights a key advantage of this paper. It shows why the Potterian economy resonates with people. Despite its shortcomings and inaccuracies, it is consistent with folk economics, which while perhaps problematic for human flourishing in a Smithian sense, it captures and reflects people's views on many economic and social issues. It is stable equilibrium, just not one that will get us to today's level of prosperity in the developed world, even with Wizarding powers.

Future studies should analyze the economic content of other literary texts to better understand how popular opinions about the economy are influenced by fiction and media, and how they form and change over time (Cowen, 2008; Miller and Watts, 2011). For example, the Potterian model points to a change in the image of the government as reflected in fiction. For example, Tolkien's *Lord of the Rings*, published in the 1950s (Tolkien, 1993), depicts the government as efficient and benevolent. In contrast to this idealistic and perhaps naïve view, the Harry Potter books portray the government officials as corrupt, dishonest, incompetent and unkind, implying that today's public is perhaps more realistic. Interestingly, this change occurred during the period when economists moved away from studying normative models of the government to studying models that emphasize the role of private and group-specific incentives (Hillman, 2009).

As another example, Adhia (2013) provides evidence of ideological change in India. Since the 1980s, he notes, the popular sentiments in India have evolved from condemning profits as anti-social to accepting—and even applauding—business success. Applying the method of content analysis to Hindi films, he finds that rich merchant characters have changed from being depicted as villains to being depicted as heroes (Note: The methodology we have employed in this paper is similar in spirit to content analysis, a widely used empirical methodology employed in other social sciences and humanities. See Levy et al. (2002) for an example and a brief survey.). These developments point to changes that modern societies experience and suggest fruitful avenues for future research.

## SUPPLEMENTARY DATA

Supplementary data are available at *Oxford Open Economics* online.

## Funding

We have not received any grant from funding agencies in the public, commercial or not-for-profit sectors.

## Conflict of Interest

None declared.

## Data Availability

Data availability does not apply in this case. The online supplementary appendix contains a detailed reference to all economic themes, topics and issues we have identified in the Harry Potter books, along with the quotations of the relevant texts from the books, and their exact locations.

## Acknowledgments

We thank an anonymous reviewer for helpful comments and the editor Amrita Dhillon for guidance. Many colleagues and students have commented on earlier versions of this paper. In particular, we would like to thank Zurab Abramishvili, Markus Brückner, Roger Congleton and Pierre-Guillaume Méon, the discussants at the Silvaplana Workshop on Political Economy. In addition, we thank Anat Alexandron, Sima Amram, Paul Anglin, Katrina Babb, Anindya Banerjee, Bob Barsky, Yoram Bauman, Mark Bergen, Sabrina Artinger, Helen Casey, Allan Chen, Ong EeCheng, Raphael Franck, Max Gillman, Allen Goodman, Danielle Gurevitch, Yuval Heller, Arye Hillman, Daniel Houser, Miriam Krausz, Christoph Kuzmics, Frank Lechner, Steven Levitt, Sarit Levy, Lorence Maimony, Asher Meir, Hugo Mialon, Igal Milchtaich, Lavinia Moldovan, Shmuel Nitzan, Chryssa Papathanassiou, Ron Peretz, Paul Rubin, Adi Schnytzer, Matthew Shapiro, Ainit Snir, the late Heinrich Ursprung, Timothy Wong, and the seminar participants at Bar-Ilan University, at the Silvaplana Workshop on Political Economy, at the International School of Economics Tbilisi (ISET), at the National University of Singapore, at the American Economic Association Conference on Teaching and Research in Economic Education, and at the American Economic Association's annual conference, for comments, suggestions and conversations about the economics of magic. Avihai Levy and Eliav Livneh, two part-time wizards, provided research assistance. Earlier versions of this paper were circulated under the titles 'Abracadabra! Social Norms and Public Perceptions through Harry Potter's Looking Glasses' and 'Popular Perceptions and Political Economy in the Contrived World of Harry Potter.' All errors are ours.

# Online Supplementary Appendix

# Potterian Economics

The tables below list the economic ideas, principles, concepts, observations, and events that we identify in the Potterian economic model and the corresponding quotes from the Harry Potter books. Many of the economic ideas appear in the Harry Potter books more than once. For example, various make and model broomsticks which are listed in Appendix E, and which we discuss in the paper in section 12 in the context of technological innovations, are mentioned in the books numerous times. To construct these tables, we tried to identify the episodes and the locations in the story where these ideas appear for the first time. We shall note that the list is not exhaustive and there are many more economic ideas in the world of Harry Potter.

## Contents

**Appendix A. Economic Ideas[1]**

| Economic Idea | Quote | Reference |
| --- | --- | --- |
| Commodity money denominations | "The gold ones are Galleons," he explained. "Seventeen silver Sickles to a Galleon and twenty-nine Knuts to a Sickle, it's easy enough." | Rowling, 1998, p. 49 |
| Foreign currency | "You're not the first one who's had trouble with money," said Mr. Roberts, scrutinizing Mr. Weasley closely." | Rowling, 2000, p. 50 |
| Coin values are independent of their values as a commodity | "I had two try and pay me with great gold coins the size of hubcaps ten minutes ago." | Rowling, 2000, p. 50 |
| Commodity money - heavy and cumbersome | "Ron purchased a dancing shamrock hat and a large green rosette, he also bought a small figure of Viktor Krum, the Bulgarian Seeker…"Wow, look at these!" said Harry…"Omnioculars," said the saleswizard eagerly…"Bargain - ten Galleons each."…"Three pairs," said Harry firmly to the wizard…Their money bags considerably lighter, they went back to the tents. | Rowling, 2000, pp. 60–61 |
| Counterfeit money | "Well, let's check how yeh've done!" said Hagrid. "Count yer coins! An' there's no point tryin' ter steal any, Goyle," he added, his beetle-black eyes narrowed. "It's leprechaun gold. Vanishes after a few hours." | Rowling, 2000, p. 350 |
| | "I know that, Harry, but if she wakes up and the locket's gone – I need to duplicate it – *Geminio!* There…That should fool her…" | Rowling, 2007, p. 173 |
| | "They have added Germino…Curse!" said Griphook. "Everything you touch will…multiply, but the copies are worthless – and if you continue to handle the treasure, you will eventually be crushed to death by the weight of expanding gold!" | Rowling, 2007, p. 356 |
| Constraints on Converting metals into gold | "Sorcerer's Stone…will transform any metal into pure gold…but the only Stone currently in existence belongs to Mr. Nicolas Flamel…who celebrated his six hundred and sixty-fifth birthday last year." | Rowling, 1998, p. 143 |
| Opportunity cost of using precious metals for making coins | "And now Wormtail was whimpering. He pulled a long, thin, shining silver dagger from inside his cloak." | Rowling, 2000, p. 413 |
| | "Harry walked up the worn stone steps, staring at the newly materialized door. Its black paint was shabby and scratched. The silver doorknocker was in the form of a | Rowling, 2003, p. 45 |

---

[1] This appendix contains a list of the ideas found in Potterian economics which are discussed in the body of the paper.

| | twisted serpent. There was no keyhole or letterbox.” | |
|---|---|---|
| | “Dumbledore gave a great sniff as he took a golden watch from his pocket and examined it.” | Rowling, 1998, p. 8 |
| Transaction cost of withdrawing money from the bank | “An’ I’ve also got a letter here from Professor Dumbledore,”… “Very well,” he said…“I will have someone take you down…” Griphook whistled and a small cart came…hurtled through a maze of twisting passages…left, right, right, left, middle fork, right, left,…The rattling cart seemed to know its own way…they plunged even deeper “…I think I’m gonna be sick.”…Hagrid…had to lean against the wall to stop his knees from trembling…Inside were mounds of gold coins…silver…bronze Knuts…He turned to Griphook. “…can we go more slowly?” “One speed only,” said Griphook. They were going even deeper now and gathering speed. The air became colder and colder…One wild cart ride later they stood…outside Gringotts.” | Rowling, 1998, pp. 47–49 |
| The amount of money that Harry withdraws from the bank completely fills his bag | Once Harry had refilled his money bag with gold Galleons, silver Sickles, and bronze Knuts from his vault at Gringotts, he had to exercise a lot of self-control not to spend the whole lot at once. | Rowling, 1999b, p. 31 |
| Cash (Clower) constraint | “I haven’t got any money — and you heard Uncle Vernon last night…he won’t pay for me to go and learn magic.” | Rowling, 1998, p. 41 |
| Wizards’ difficulties in handling foreign currency | “They had reached the station. There was a train to London in five minutes’ time. Hagrid, who didn’t understand “Muggle money,” as he called it, gave the bills to Harry so he could buy their tickets.” | Rowling, 1998, p. 43 |
| | “Help me, Harry,” he muttered, pulling a roll of Muggle money... “This one’s a - a - a ten? Ah yes, I see the little number on it now…So this is a five?” “A twenty,” Harry corrected him... “Ah yes, so it is…I don’t know, these little bits of paper…” “You foreign?” said Mr. Roberts as Mr. Weasley returned with the correct notes. “Foreign?” repeated Mr. Weasley, puzzled. “You’re not the first one who’s had trouble with money” | Rowling, 2000, p. 50 |
| | “Stored in an underground vault at Gringotts in London was a small fortune that his parents had left him. Of course, it was only in the Wizarding world that he had money; you couldn’t use Galleons, Sickles, and Knuts in Muggle shops.” | Rowling, 1999a, p. 30 |
| Just-below (or psychological) prices | "A plump woman outside an Apothecary was shaking her head as they passed, saying, "Dragon liver, sixteen Sickles an ounce, they're mad…"" | Rowling, 1998, p. 46 |
| Sticky prices I – Daily Prophet costs 1 Knut for 7 years | "Hermione, however, had to move her orange juice aside quickly to make way for a large damp barn owl bearing a sodden *Daily Prophet* in its beak. "What are you still getting that for?" said Harry irritably, thinking of Seamus as Hermione placed a Knut in the leather pouch on the owl's leg and it took off again." | Rowling, 2003, p. 167 |
| Sticky prices II – Floo Powder's price is the same for over 100 years | "No shortage of Floo powder has ever been reported, nor does anybody know anyone who makes it. Its price has remained constant for one hundred years: two Sickles a scoop. Every wizard household carries a stock of Floo powder, usually conveniently located in a box or vase on the mantelpiece." | [pottermore head.tumb lr.com/pos t/1021758 02190/floo -powder](pottermorehead.tumblr.com/post/102175802190/floo-powder), accessed on May, 31, 2015 |
| Gringotts bank – a monopoly run by goblins | "They didn' keep their gold in the house, boy! Nah, first stop fer us is Gringotts. Wizards' bank." "Wizards have *banks*?" "Just the one. Gringotts. Run by goblins." | Rowling, 1998, p. 41 |
| Goblins – greedy bankers | "They passed a group of goblins who were cackling over a sack of gold that they had undoubtedly won betting on the match, and who seemed quite unperturbed by the trouble at the campsite." | Rowling, 2000, p. 81 |
| Counterfeit money – even school boys can do that | "Hermione soon devised a very clever method of communicating the time and date of the next meeting… She gave each of the members of the D.A. a fake Galleon …"You see the numerals around the edge of the coins? … On real Galleons that's just a serial number referring to the goblin who cast the coin. On these fake coins, though, the numbers will change to reflect the time and date of the next meeting. The coins will grow hot when the date changes." | Rowling, 2003, p. 297 |
| Introduction of counterfeit gold Galleons – makes people extra cautious | "Yeah…I prefer your way," said Harry, grinning, as he slipped his Galleon into his pocket. "I suppose the only danger with these is that we might accidentally spend them." "Fat chance," said Ron, who was examining his own fake Galleon with a slightly mournful air, "I haven't got any real Galleons to confuse it with." | Rowling, 2003, p. 298 |
| Withdrawing money from the bank – the Gringotts | "Morning," said Hagrid to a free goblin. "We've come ter take some money outta Mr. Harry Potter's safe." "You have his key, sir?"…"Got it," said Hagrid at last, holding up a tiny golden key. The goblin looked at it closely. "That seems to be in order." | Rowling, 1998, p. 47 |
| Services offered at | "When the cart stopped…Griphook unlocked the door… Inside were mounds of gold coins. Columns of silver. | Rowling, 1998, pp. |

| Gringotts - safe keeping | Heaps of little bronze Knuts. "All yours," smiled Hagrid …"The gold ones are Galleons," he explained. "Seventeen silver Sickles to a Galleon and twentynine Knuts to a Sickle, it's easy enough. Right, that should be enough fer a couple o' terms, we'll keep the rest safe for yeh." | 48–49 |
|---|---|---|
| | "Harry. Gringotts is the safest place in the world fer anything yeh want ter keep safe." | Rowling, 1998, p. 41 |
| Services offered at Gringotts – exchange of wizard money for precious stones | "A pair of goblins bowed them through the silver doors and they were in a vast marble hall. About a hundred more goblins were sitting on high stools behind a long counter… examining precious stones through eyeglasses." | Rowling 1998, p. 47 |
| Services offered at Gringotts – exchange of wizard money for Muggle-money | "You be careful, Arthur," said Mrs. Weasley sharply as they were bowed into the bank by a goblin at the door. "…Hermione's parents…were standing nervously at the counter that ran all along the great marble hall…"But you're *Muggles*!" said Mr. Weasley delightedly…"What's that you've got there? Oh, you're changing Muggle money. Molly, look!" He pointed excitedly at the ten-pound notes in Mr. Granger's hand." | Rowling, 1999a, p. 37 |
| Wizards face difficulties in using Muggle money | "There was a train to London in five minutes' time. Hagrid, who didn't understand "Muggle money," as he called it, gave the bills to Harry so he could buy their tickets." | Rowling, 1998, p. 43 |
| | "Help me, Harry," he muttered, pulling a roll of Muggle money... "This one's a - a - a ten? Ah yes, I see the little number on it now…So this is a five?" "A twenty," Harry corrected him... "Ah yes, so it is…I don't know, these little bits of paper…" | Rowling, 2000, p. 50 |
| Wizards' and Muggles' lack of interaction | "But what does a Ministry of Magic *do*?" "Well, their main job is to keep it from the Muggles that there's still witches an' wizards up an' down the country." | Rowling, 1998, p. 42 |
| | "Hey, Harry,"…"have you heard?...Sirius Black's been sighted." "Where?" said Harry and Ron quickly. "Not too far from here," said Seamus…"It was a Muggle who saw him…'Course, she didn't really understand. The Muggles think he's just an ordinary criminal, don't they?" | Rowling, 1999b, p. 80 |
| | "You foreign?" said Mr. Roberts as Mr. Weasley returned with the correct notes. "Foreign?" repeated Mr. Weasley, puzzled. "You're not the first one who's had trouble with | Rowling, 2000, p. 50 |

| | money" | |
|---|---|---|
| Wizards need to borrow from illegal usurers or friends | "Turns out he's [Ludo Bagman] in big trouble with the goblins. Borrowed loads of gold off them. A gang of them cornered him in the woods after the World Cup and took all the gold he had, and it still wasn't enough to cover all his debts. They followed him all the way to Hogwarts to keep an eye on him." | Rowling, 2000, p. 471 |
| | "Harry, you help yourself to anything you want…No charge." "I can't do that!" said Harry, who had already pulled out his money bag to pay…"You don't pay here," said Fred firmly, waving away Harry's gold…"You gave us our start-up loan, we haven't forgotten," said George sternly." | Rowling, 2005, p. 78 |
| Fred and George Weasley borrow from Harry Potter | "Harry…had forced the Weasley twins to take the thousand Galleons prize money he had won in the Triwizard Tournament to help them realize their ambition to open a joke shop." | Rowling, 2003, p. 79 |
| Wizards borrow from Goblins | "Turns out he's [Ludo Bagman] in big trouble with the goblins. Borrowed loads of gold off them. A gang of them cornered him in the woods after the World Cup and took all the gold he had, and it still wasn't enough to cover all his debts. They followed him all the way to Hogwarts to keep an eye on him." | Rowling, 2000, p. 471 |
| Weasley twins consider gambling to obtain funds needed for opening their Joke-Shop | "We'll bet thirty-seven Galleons, fifteen Sickles, three Knuts," said Fred as he and George quickly pooled all their money, "that Ireland wins - but Viktor Krum gets the Snitch. Oh and we'll throw in a fake wand."…"Boys," said Mr. Weasley under his breath, "I don't want you betting…That's all your savings" | Rowling, 2000, p. 57 |
| Government depends on donations from wealthy individuals | "Mr. Weasley and Mr. Malfoy looked at each other… Malfoy's cold gray eyes swept over Mr. Weasley, and then up and down the row. "Good lord, Arthur," he said softly. "What did you have to sell to get seats in the Top Box? Surely your house wouldn't have fetched this much?" Fudge, who wasn't listening, said, "Lucius has just given a very generous contribution to St. Mungo's Hospital for Magical Maladies and Injuries, Arthur. He's here as my guest." | Rowling, 2000, p. 66 |
| Gringotts' employees offer private usury services | "Turns out he's [Ludo Bagman] in big trouble with the goblins. Borrowed loads of gold off them. A gang of them cornered him in the woods after the World Cup and took all the gold he had, and it still wasn't enough to cover all his debts. They followed him all the way to Hogwarts to keep an eye on him." | Rowling, 2000, p. 471 |
| Wizards that make windfall gains spend them immediately | "The clipping had clearly come out of the wizarding newspaper, the *Daily Prophet*…Harry picked up the clipping…and read:…*Arthur Weasley…has won the annual Daily Prophet Grand Prize Galleon Draw. A delighted Mr. Weasley told the Daily Prophet, "We will be spending the gold on a summer holiday in Egypt."* | Rowling, 1999b, p. 5 |
| Wizards view financial service providers as immoral | "Wizards have *banks*?" "Just the one. Gringotts. Run by goblins." Harry dropped the bit of sausage he was holding. "*Goblins*?" "Yeah — so yeh'd be mad ter try an' rob it, I'll tell yeh that. Never mess with goblins, Harry. | Rowling, 1998, p. 41 |
| | "They passed a group of goblins who were cackling over a sack of gold that they had undoubtedly won betting on the match, and who seemed quite unperturbed by the trouble at the campsite." | Rowling, 2000, p. 81 |
| | "Absolute nightmare," said Bagman to Harry in an undertone, noticing Harry watching the goblins too. "Their English isn't too good…it's like being back with all the Bulgarians at the Quidditch World Cup…but at least they used sign language another human could recognize." | Rowling, 2000, p. 287 |
| | "The goblins play as dirty as him. They say you drew with Diggory, and Bagman was betting you'd win outright. So Bagman had to run for it. He did run for it right after the third task." | Rowling, 2000, p. 471 |
| Goblins' inferior image inhibits most forms of interaction between wizards and Goblins | "If there was a wizard of whom I would believe that they did not seek personal gain," said Griphook finally, "it would be you, Harry Potter. Goblins … are not used to… the respect that you have shown this night. Not from wand-carriers." | Rowling, 2007, p. 323 |
| | "Then I have to say this," Bill went on. "If you have struck any kind of bargain with Griphook, and most particularly if that bargain involves treasure, you must be exceptionally careful. Goblin notions of ownership, payment, and repayment are not the same as human ones." …However, there is a belief among some goblins, and those at Gringotts are perhaps most prone to it, that wizards cannot be trusted in matters of gold and treasure, that they have no respect for goblin ownership." | Rowling, 2007, p. 342 |
| Rent-seeking - Wealthy individuals often fund public goods | "I'll leave a note for Dumbledore when I drop you off, he ought to know Malfoys been talking to Fudge again." "What private business have they got together, anyway?" "Gold, I expect," said Mr. Weasley angrily. "Malfoy's been giving generously to all sorts of things for years… | Rowling, 2003, p. 116 |

| and thus exert influence on public officials and public policy | gets him in with the right people…then he can ask favors …delay laws he doesn't want passed…oh, he's very well-connected, Lucius Malfoy." | |
|---|---|---|
| Newspaper reports portray the officials positively, who reciprocate by making decisions favoring the reporters and wealthy wizards | "Rita…said…"All right, Fudge is leaning on the *Prophet*, but it comes to the same thing. They won't print a story that shows Harry in a good light. Nobody wants to read it. It's against the public mood." | Rowling 2003, p. 423 |
| The Minister of Magic is appointed, not elected | "Newly appointed Minister of Magic, Rufus Scrimgeour, spoke today of the tough new measures taken by his Ministry to ensure the safety of students returning to Hogwarts School of Witchcraft and Wizardry this autumn." | Rowling 2005, p. 27 |
| Government – Ministry of Magic | "Hagrid read his newspaper, the *Daily Prophet*. "Ministry o' Magic messin' things up as usual," Hagrid muttered, turning the page. "There's a Ministry of Magic?" Harry asked..."'Course," said Hagrid. "They wanted Dumbledore fer Minister, o' course, but he'd never leave Hogwarts, so old Cornelius Fudge got the job. Bungler if ever there was one. So he pelts Dumbledore with owls every morning, askin' fer advice." | Rowling, 1998, p. 42 |
| Inept public employees | "You wouldn't believe how many people, even people who work at the Ministry, can't do a decent Shield Charm" | Rowling, 2005, p. 78 |
| Bribery | "It was Umbridge's lie that brought the blood surging into Harry's brain and obliterated his sense of caution – that the locket she had taken as a bribe from a petty criminal was being used to bolster her own pure-blood credentials." | Rowling, 2007, p. 172 |
| | "When you stripped this house of all the valuables you could find," Harry began again, "you took a bunch of stuff from the kitchen cupboard. There was a locket there." Harry's mouth was suddenly dry: He could sense Ron and Hermione's tension and excitement too. "What did you do with it?" "Why?" asked Mundungus. "Is it valuable?" "You've still got it!" cried Hermione. "No, he hasn't," said Ron shrewdly. "He's wondering whether he should have asked more money for it." "More?" said Mundungus. | Rowling, 2007, p. 144 |
| | "That wouldn't have been effing difficult…bleedin' gave it away, di'n' I? No choice." "What do you mean?" "I was selling in Diagon Alley and she come up to me and asks if I've got a license for trading in magical artifacts. Bleedin' snoop. She was gonna fine me, but she took a fancy to the locket an' told me she'd take it and let me off that time, and to fink meself lucky." "Who was this woman?" asked Harry. "I dunno, some Ministry hag." | |
| Nepotism is common | "What does your dad do at the Ministry of Magic, anyway?" "He works in the most boring department," said Ron. "The Misuse of Muggle Artifacts Office." "The *what*?" "It's all to do with bewitching things that are Muggle-made, you know, in case they end up back in a Muggle shop or house. Like, last year, some old witch died and her tea set was sold to an antiques shop. This Muggle woman bought it, took it home, and tried to serve her friends tea in it. It was a nightmare — Dad was working overtime for weeks." | Rowling, 1999a, p. 20 |
| | "You are sweet," beamed Mrs. Weasley…"Yes, Rufus Scrimgeour has set up several new offices in response to the present situation, and Arthur's heading the Office for the Detection and Confiscation of Counterfeit Defensive Spells and Protective Objects." | Rowling, 2005, p. 56 |
| | "I've been promoted," Percy said before Harry could even ask, and from his tone, he might have been announcing his election as supreme ruler of the universe. "I'm now Mr. Crouch's personal assistant, and I'm here representing him." | Rowling, 2000, p. 268 |
| | "In a surprise move last night the Ministry of Magic passed new legislation… "'The Minister has been growing uneasy about goings-on at Hogwarts for some time,' said junior Assistant to the Minister, Percy Weasley." | Rowling, 2003, p. 229 |
| | "D'you know what you want to do after Hogwarts?" Harry asked the other two... "Not really," said Ron slowly. "Except…well…" He looked slightly sheepish. "What?" Harry urged him. "'Well, it'd be cool to be an Auror [at the Ministry of Magic],' " said Ron in an off-hand voice. "Yeah, it would," said Harry fervently. "But they're, like, the elite," said Ron." | Rowling, 2003, p. 170 |
| | "Good training for when we're all Aurors," said Ron excitedly, attempting the Impediment Curse on a wasp that had buzzed into the room and making it stop dead in | Rowling, 2000, p. 392 |

| | midair." | |
|---|---|---|
| Rent-seeking: Wealthy wizards that fund public officials' office expenditures have access to the officials and influence their decisions | "Malfoy's been giving generously to all sorts of things for years… gets him in with the right people…then he can ask favors …delay laws he doesn't want passed…oh, he's very well-connected, Lucius Malfoy." | Rowling, 2003, p. 116 |
| Mr. Malfoy knows in advance about the Ministry's planned raid | "Mr. Malfoy, what a pleasure to see you again," said Mr. Borgin in a voice as oily as his hair. "Delighted — and young Master Malfoy, too — charmed. How may I be of assistance? I must show you, just in today, and very reasonably priced —" "I'm not buying today, Mr. Borgin, but selling," said Mr. Malfoy. "Selling?" The smile faded slightly from Mr. Borgin's face. "You have heard, of course, that the Ministry is conducting more raids," said Mr. Malfoy, taking a roll of parchment from his inside pocket and unraveling it for Mr. Borgin to read. "I have a few — ah — items at home that might embarrass me, if the Ministry were to call…" Mr. Borgin fixed a pair of pince-nez to his nose and looked down the list. "The Ministry wouldn't presume to trouble you, sir, surely?" Mr. Malfoy's lip curled. "I have not been visited yet. The name Malfoy still commands a certain respect, yet the Ministry grows ever more meddlesome. There are rumors about a new Muggle Protection Act — no doubt that flea-bitten, Muggle-loving fool Arthur Weasley is behind it —" Harry felt a hot surge of anger. "— and as you see, certain of these poisons might make it appear —" "I understand, sir, of course," said Mr. Borgin. "Let me see…"I am in something of a hurry, Borgin, I have important business elsewhere today —" …They started to haggle. "Done," said Mr. Malfoy at the counter. "Come, Draco —"… "Good day to you, Mr. Borgin. I'll expect you at the manor tomorrow to pick up the goods." | Rowling, 1999a, pp. 33–34 |
| Junior public officials make efforts to please their superiors | "Percy hurried forward with his hand outstretched. Apparently his disapproval of the way Ludo Bagman ran his department did not prevent him from wanting to make a good impression."<br><br>"Mr. Crouch!" said Percy breathlessly, sunk into a kind of halfbow that made him look like a hunchback. "Would you like a cup of tea?" "Oh," said Mr. Crouch, looking | Rowling, 2000, p. 56<br><br>Rowling, 2000, pp. 58–59 |

| | over at Percy in mild surprise. "Yes - thank you, Weatherby." Fred and George choked into their own cups. Percy, very pink around the ears, busied himself with the kettle." | |
| --- | --- | --- |
| | "I just can't justify taking more time off at the moment," he told them. "Mr. Crouch is really starting to rely on me." "Yeah, you know what, Percy?" said George seriously. "I reckon he'll know your name soon." | Rowling, 2000, p. 104 |
| | "Are you sure you wouldn't like to stay at Hogwarts tonight, Barty?" "No, Dumbledore, I must get back to the Ministry," said Mr. Crouch. "It is a very busy, very difficult time at the moment…I've left young Weatherby in charge…Very enthusiastic…a little overenthusiastic, if truth be told…" | Rowling, 2000, p. 182 |
| | "What do they think they're doing, annoying senior Ministry members?" Percy hissed, watching Fred and George suspiciously. "No respect…" Ludo Bagman shook off Fred and George fairly quickly, however, and, spotting Harry, waved and came over to their table. "I hope my brothers weren't bothering you, Mr. Bagman?" said Percy at once." | Rowling, 2000, p. 273. |
| Many offices are overstaffed with low productivity workers | "You realize Bertha Jorkins has been missing for over a month now? Went on holiday to Albania and never came back?" | Rowling, 2000, p. 40 |
| Hogwarts' Healer, Madam Pomfrey cures them all | "How exactly did it happen, Harry?" Harry retold the story "…and then I got the bezoar down his throat and his breathing eased up a bit, Slughorn ran for help, McGonagall and Madam Pomfrey turned up, and they brought Ron up here. They reckon he'll be all right. Madam Pomfrey says he'll have to stay here a week or so…keep taking essence of rue…" | Rowling, 2005, p. 263 |
| Private Property and inheritance | "Ministry, is it?" said the older man, looking down at Ogden. "Correct!" said Ogden angrily, dabbing his face. "And you, I take it, are Mr. Gaunt?" "S'right," said Gaunt. "Got you in the face, did he?" "Yes, he did!" snapped Ogden. "Should've made your presence known, shouldn't you?" said Gaunt aggressively. "This is private property. Can't just walk in here and not expect my son to defend himself." | Rowling, 2005, p. 132 |
| | "When the cart stopped at last beside a small door in the passage wall, Hagrid got out and had to lean against the | Rowling, 1998, p. |

| | | 48 |
|---|---|---|
| | wall to stop his knees from trembling. Griphook unlocked the door. A lot of green smoke came billowing out, and as it cleared, Harry gasped. Inside were mounds of gold coins. Columns of silver. Heaps of little bronze Knuts. "All yours," smiled Hagrid. All Harry's — it was incredible. The Dursleys couldn't have known about this or they'd have had it from him faster than blinking. How often had they complained how much Harry cost them to keep? And all the time there had been a small fortune belonging to him, buried deep under London." | |
| | "Stored in an underground vault at Gringotts in London was a small fortune that his parents had left him. Of course, it was only in the Wizarding world that he had money; you couldn't use Galleons, Sickles, and Knuts in Muggle shops." | Rowling, 1999a, p. 30 |
| | "You see," Dumbledore said, turning back to Harry and again speaking as though Uncle Vernon had not uttered, "if you have indeed inherited the house, you have also inherited —"…"As you can see, Harry," said Dumbledore loudly, over Kreacher's continued croaks of "wont, won't, won't," "Kreacher is showing a certain reluctance to pass into your ownership."… "Give him an order," said Dumbledore. "If he has passed into your ownership, he will have to obey."…"Well, that simplifies matters," said Dumbledore cheerfully. "It means that Sirius knew what he was doing. You are the rightful owner of number twelve, Grimmauld Place and of Kreacher." | Rowling, 2005, pp. 33–34 |
| | "Can't the Order control Mundungus?" Harry demanded of the other two in a furious whisper. "Can't they at least stop him stealing everything that's not fixed down when he's at headquarters?" "Shh!" said Hermione desperately, looking around to make sure nobody was listening; there were a couple of warlocks sitting close by who were staring at Harry with great interest, and Zabini was lolling against a pillar not far away. "Harry, I'd be annoyed too, I know it's your things he's stealing—" "Harry gagged on his butterbeer; he had momentarily forgotten that he owned number twelve, Grimmauld Place. "Yeah, it's my stuff!" he said. "No wonder he wasn't pleased to see me! Well, I'm going to tell Dumbledore what's going on, he's the only one who scares Mundungus." | Rowling, 2005, p. 161 |
| Inheritance confiscation attempts by the government | "I have some questions for the three of you, and I think it will be best if we do it individually…"We're not going anywhere," said Harry, while Hermione nodded vigorously. "You can speak to us together, or not at | Rowling, 2007, pp. 80-85 |

all."…"I am here, as I'm sure you know, because of Albus Dumbledore's will." Harry, Ron, and Hermione looked at one another. "A surprise, apparently! You were not aware then that Dumbledore had left you anything?" "A-all of us?" said Ron, "Me and Hermione too?" "Yes, all of –" But Harry interrupted. "Dumbledore died over a month ago. Why has it taken this long to give us what he left us?" "Isn't it obvious?" said Hermione, before Scrimgeour could answer. "They wanted to examine whatever he's left us. You had no right to do that!" she said, and her voice trembled slightly. "I had every right," said Scrimgeour dismissively. "The Decree for Justifiable Confiscation gives the Ministry the power the confiscate the contents of a will –" "That law was created to stop wizards passing on Dark artifacts," said Hermione, "and the Ministry is supposed to have powerful evidence that the deceased's possessions are illegal before seizing them! …Harry spoke: "So why have you decided to let us have our things now? Can't think of a pretext to keep them?" "No, it'll be because thirty-one days are up," said Hermione at once. "They can't keep the objects longer than that unless they can prove they're dangerous. Right?" "… how do you account for the fact that he remembered you in his will? He made exceptionally few personal bequests. … Why do you think you were singled out?" "I…dunno," said Ron…Scrimgeour… removed a scroll of parchment which he unrolled and read aloud. "'*The Last Will and Testament of Albus Percival Wulfric Brian Dumbledore*'…Yes, here we are…'*To Ronald Bilius Weasley, I leave my Deluminator, in the hope that he will remember me when he uses it.*'" Scrimgeour took from the bag an object … leaned forward and passed the Deluminator to Ron, who took it and turned it over in the fingers looking stunned. "That is a valuable object," said Scrimgeour, watching Ron. … To what use did he think you would put to the Deluminator, Mr. Weasley?" "Put out lights, I s'pose," mumbled Ron. "What else could I do with it?" Evidently Scrimgeour had no suggestions. After squinting at Ron for a moment or tow, he turned back to Dumbledore's will. "'*To Miss Hermione Jean Granger, I leave my copy of* The Tales of Beedle the Bard*, in the hope that she will find it entertaining and instructive*.'" Scrimgeour now pulled out of the bag a small book … Harry saw that the title was in runes; he had never learned to read them. As he looked, a tear splashed onto the embossed symbols. "Why do you think Dumbledore left you that book, Miss Granger?" asked Scrimgeour. "He…he knew I liked books," said Hermione in a thick voice, mopping her eyes with her sleeve. "But why that particular book?" "I don't know. He must have thought I'd enjoy it." "Did you ever discuss codes, or any means of passing secret messages, with Dumbledore?" "No, I didn't," said Hermione,…Scrimgeour turned back to the will. "'*To Harry James Potter*,'" he read, and Harry's insides contracted with a sudden excitement, "'*I leave the Snitch he caught in his first Quidditch match at Hogwarts, as a reminder of the rewards of perseverance and skill*.'" As Scrimgeour pulled out the tiny, walnut-sized golden ball, …"Why did Dumbledore leave you this Snitch?" asked Scrimgeour. "No idea," said Harry…"What …could it be?" "I'm asking the questions," said Scrimgeour,."I notice that your birthday cake is in the shape of a Snitch," Scrimgeour said to Harry…"I don't think there's anything hidden in the icing," said Scrimgeour, "but a Snitch would be a very good hiding place for a small object…"Because Snitches have flesh memories," she said…"Correct," said Scrimgeour…"It occurs to me that Dumbledore, who had prodigious magical skill, whatever his other faults, might have enchanted this Snitch so that it will open only for you."…"Take it," said Scrimgeour quietly…"That's all, then, is it?" asked Hermione, making to raise herself off the sofa. "Not quite," said Scrimgeour, who looked bad tempered now. "Dumbledore left you a second bequest, Potter." "What is it?" asked Harry…"The sword of Godric Gryffindor," he said…"So where is it?" Harry asked suspiciously. "Unfortunately," said Scrimgeour, "that sword was not Dumbledore's to give away. The sword of Godric Gryffindor is an important historical artifact, and as such, belongs –" "It belongs to Harry!" said Hermione hotly. "It chose him, he was the one who found it, it came to him out of the Sorting Hat –" "According to reliable historical sources, the sword may present itself to any worthy Gryffindor," said Scrimgeour. "That does not make it the exclusive property of Mr. Potter, whatever Dumbledore may have decided." …"You go too far!" shouted Scrimgeour, standing up: Harry jumped to his feet too. …"No! D'you want to give him an excuse to arrest us?" "Remembered you're not at school, have you?" said Scrimgeour breathing hard into Harry's face. "Remembered that I am not Dumbledore, who forgave your insolence and insubordination? You may wear that scar like a crown, Potter, but it is not up to a seventeen-year-old boy to tell me how to do my job! It's time you learned some respect!"…"I don't like your methods, Minister," said Harry.

| Contract | "Finally, I wish to impress upon any of you wishing to compete that this tournament is not to be entered into lightly. Once a champion has been selected by the Goblet of Fire, he or she is obliged to see the tournament through to the end. The placing of your name in the goblet constitutes a binding, magical contract. There can be no change of heart once you have become a champion. Please be very sure, therefore, that you are wholeheartedly prepared to play before you drop your name into the goblet. Now, I think it is time for bed. Good night to you all." | Rowling, 2000, p. 166 |
|---|---|---|
| | "Empty threat, Karkaroff," growled a voice from near the door. "You can't leave your champion now. He's got to compete. They've all got to compete. Binding magical contract, like Dumbledore said. Convenient, eh?" | Rowling, 2000, p. 179 |
| | "I – I think everybody should write their name down, just so we know who was here. But I also think," she took a deep breath, "that we all ought to agree not to shout about what we're doing. So if you sign, you're agreeing not to tell Umbridge or anybody else what we're up to."…When the last person – Zacharias – had signed, Hermione took the parchment back and slipped it carefully into her bag. There was an odd feeling in the group now. It was as though they had just signed some kind of contract." | Rowling, 2003, p. 259 |
| | "How do I look?" said Hepzibah, turning her head to admire the various angles of her face in the mirror. "Lovely, madam," squeaked Hokey. Harry could only assume that it was down in Hokey's contract that she must lie through her teeth when asked this question, because Hepzibah Smith looked a long way from lovely in his opinion. | Rowling, 2005, pp. 285–286 |
| Unbreakable Vow | "So Snape was offering to help him?"…"Yes, Snape was offering to help him!" said Harry. "He said he'd promised Malfoy's mother to protect him, that he'd made an Unbreakable Oath or something —" "An Unbreakable Vow?" said Ron, looking stunned. "Nah, he can't have…Are you sure?" "Yes, I'm sure," said Harry. "Why, what does it mean?" "Well, you can't break an Unbreakable Vow…" "I'd worked that much out for myself, funnily enough. What happens if you break it, then?" "You die," said Ron simply. "Fred and George tried to get me to make one when I was about five. I nearly did too, I was holding hands with Fred and everything when Dad found us. He went mental," said Ron. | Rowling, 2005, p. 214 |

| The Ministry decides to change the time of Harry's court hearing but informs him on this change at the last minute | A cold male voice rang across the courtroom. "You're late." "Sorry," said Harry nervously "I — I didn't know the time had been changed." "That is not the Wizengamot's [the court of the Ministry of Magic] fault," said the voice. "An owl was sent to you this morning. Take your seat." | Rowling, 2003, p. 103 |
|---|---|---|
| According to the Minister of Magic, laws can be changed | "The Ministry does not have the power to expel Hogwarts students, Cornelius, as I reminded you on the night of the second of August," said Dumbledore. "Nor does it have the right to confiscate wands until charges have been successfully proven…In your admirable haste to ensure that the law is upheld, you appear, inadvertently I am sure, to have overlooked a few laws yourself." "Laws can be changed," said Fudge savagely. "Of course they can," said Dumbledore, inclining his head. "And you certainly seem to be making many changes, Cornelius. Why, in the few short weeks since I was asked to leave the Wizengamot [the court of the Ministry of Magic], it has already become the practice to hold a full criminal trial to deal with a simple matter of underage magic!" | Rowling, 2003, p. 112 |
| Hogwarts' Head Mistress and High Inquisitor amasses absolute power and authority | I contacted the Minister at once, and he quite agreed with me that the High Inquisitor has to have the power to strip pupils of privileges, or she – that is to say, I – would have less authority than common teachers!…I was reading out our amendment…hem, hem…'High Inquisitor will henceforth have supreme authority over all punishments, sanctions and removal of privileges pertaining to the students of Hogwarts, and the power to alter such punishments, sanctions and removals of privileges as may have been ordered by other staff members. Signed, Cornelius Fudge, Minister of Magic, Order of Merlin First Class, etc., etc.'" She rolled up the parchment and put it back into her handbag, still smiling. "So…I really think I will have to ban these two from playing Quidditch ever again," she said, looking from Harry to George and back again. Harry felt the Snitch fluttering madly in his hand. "Ban us?" he said, and his voice sounded strangely distant. "From playing …ever again?" "Yes, Mr. Potter, I think a lifelong ban ought to do the trick," said Umbridge, her smile widening still further as she watched him struggle to comprehend what she had said. "You and Mr. Weasley here. And I think, to be safe, this young man's twin ought to be stopped, too – if his teammates had not restrained him, I feel sure he would have attacked young Mr. Malfoy as well. I will want their broomsticks | Rowling, 2003, pp. 310–311. |

| | confiscated, of course; I shall keep them safely in my office, to make sure there is no infringement of my ban. | |
|---|---|---|
| Selective enforcement of the laws by the Ministry of Magic | "So all that remains," said Fudge, now buttering himself a second crumpet, "is to decide where you're going to spend the last two weeks of your vacation. I suggest you take a room here at the Leaky Cauldron and…" "Hang on," blurted Harry. "What about my punishment?" Fudge blinked. "Punishment?" "I broke the law!" Harry said. "The Decree for the Restriction of Underage Wizardry!" "Oh, my dear boy, we're not going to punish you for a little thing like that!" cried Fudge, waving his crumpet impatiently. "It was an accident! We don't send people to Azkaban just for blowing up their aunts!" | Rowling, 1999b, p. 28 |
| Half-blood and mud-blood wizards must be questioned to make sure that they did not "steal" magic powers | "*Muggle-born Register!*" she read aloud. "'*The Ministry of Magic is undertaking a survey of so-called "Muggle-borns" to better understand how they came to possess magical secrets.* "'*Recent research undertaken by the Department of Mysteries reveals that magic can only be passed from person to person when Wizards reproduce. Where no proven Wizarding ancestry exists, therefore, the so-called Muggle-born is likely to have obtained magical power by theft or force.* "'*The Ministry is determined to root out such usurpers of magical power, and to this end has issued an invitation to every so-called Muggle-born to present themselves for interview by the newly appointed Muggle-born Registration Commission.*'" "People won't let this happen," said Ron. "It *is* happening, Ron," said Lupin. "Muggle-borns are being rounded up as we speak." "But how are they supposed to have 'stolen' magic?" said Ron. "It's mental, if you could steal magic there wouldn't be any Squibs, would there?" "I know," said Lupin. "Nevertheless, unless you can prove that you have at least one close Wizarding relative, you are now deemed to have obtained your magical power illegally and must suffer the punishment." | Rowling, 2007, p. 136 |
| Criminal activities | "And you off buying stolen cauldrons! Didn't I tell you not to go? Didn't I!" "I – well, I –" Mundungus looked deeply uncomfortable. "It — it was a very good business opportunity, see –" | Rowling, 2003, p. 18 |
| | "Mundungus!" said Hermione. "What's he brought all those cauldrons for?" "Probably looking for a safe place to keep them," said Harry. "Isn't that what he was doing the night he was supposed to be tailing me? Picking up dodgy cauldrons?" "Yeah, you're right!" said Fred, as the front door opened; Mundungus heaved his cauldrons through it and disappeared from view. "Blimey, Mum | Rowling, 2003, p. 80 |
| | won't like that…" | |
| | Once they had eaten their Christmas lunch, the Weasleys, Harry and Hermione were planning to pay Mr. Weasley another visit, escorted by Mad-Eye and Lupin. Mundungus turned up in time for Christmas pudding and trifle, having managed to 'borrow' a car for the occasion, as the Underground did not run on Christmas Day. The car, which Harry doubted very much had been taken with the knowledge or consent of it's owner, had had a similar Enlarging Spell put upon it as the Weasley's old Ford Anglia; | Rowling, 2003, p. 377 |
| | On the other hand, a number of shabby-looking stalls had sprung up along the street. The nearest one, which had been erected outside Flourish and Blotts, under a striped, stained awning, had a cardboard sign pinned to its front: **AMULETS - Effective Against Werewolves, Dementors, and Inferi!** A seedy-looking little wizard was rattling armfuls of silver symbols on chains at passersby. "One for your little girl, madam?" he called at Mrs. Weasley as they passed, leering at Ginny. "Protect her pretty neck?" "If I were on duty…" said Mr. Weasley, glaring angrily at the amulet seller. "Yes, but don't go arresting anyone now, dear, we're in a hurry," said Mrs. Weasley, nervously consulting a list. | Rowling, 2005, p. 73 |
| | "Well, you see, in all the panic about You-Know-Who, odd things have been cropping up for sale everywhere, things that are supposed to guard against You-Know-Who and the Death Eaters. You can imagine the kind of thing…so-called protective potions that are really gravy with a bit of bubotuber pus added, or instructions for defensive jinxes that actually make your ears fall off…Well, in the main the perpetrators are just people like Mundungus Fletcher, who've never done an honest day's work in their lives and are taking advantage of how frightened everybody is." | Rowling, 2005, p. 56 |
| New businesses open rarely | "It's this joke shop idea they've got," said Ron. "I thought they were only saying it to annoy Mum, but they really mean it, they want to start one. They've only got a year left at Hogwarts, they keep going on about how it's time to think about their future, and Dad can't help them, and they need gold to get started." | Rowling, 2000, pp. 366–367 |
| Attempts by an entrepreneur to import flying carpets are | "He wants a word with you about your embargo on flying carpets." Mr. Weasley heaved a deep sigh…"Carpets are defined as a Muggle Artifact by the Registry of Proscribed Charmable Objects"…"Well, they'll never replace brooms | Rowling, 2000, p. 59 |

| blocked | in Britain, will they?" said Bagman. "Ali thinks there's a niche in the market for a family vehicle," said Mr. Crouch. "I remember my grandfather had an Axminster that could seat twelve - but that was before carpets were banned, of course." | |
|---|---|---|
| *The Squibbler* – unreliable tabloids | "Of course not," said Hermione scathingly, before Harry could answer. "*The Quibbler*'s rubbish, everyone knows that." | Rowling, 2003, p. 144 |
| *Daily Prophet* publishes biased information | "Harry…was talking more than he'd talked in days - about how no one believed he hadn't entered the tournament of his own free will, how Rita Skeeter had lied about him in the *Daily Prophet*" | Rowling, 2000, p. 213 |
| | "They're trying to discredit him," said Lupin. "Didn't you see the *Daily Prophet* last week? They reported that he'd been voted out of the Chairmanship of the International Confederation of Wizards because he's getting old and losing his grip, but it's not true" | Rowling 2003, p. 72 |
| Harry Potter knows nothing about the Weasleys' Wizard Wheezes – the Weasley twins' joke shop | "What are Weasleys' Wizard Wheezes?" Harry asked… Ron and Ginny both laughed, although Hermione didn't. "Mum found this stack of order forms when she was cleaning Fred and George's room," said Ron quietly. "Great long price lists for stuff they've invented. Joke stuff, you know. Fake wands and trick sweets,…I never knew they'd been inventing all that"…"and, you know, they were planning to sell it at Hogwarts to make some money" | Rowling 2000, p. 36 |
| The wizards have only one wand-maker | "Don' mention it," said Hagrid gruffly. "Don' expect you've had a lotta presents from them Dursleys. Just Ollivanders left now — only place fer wands, Ollivanders, and yeh gotta have the best wand." | Rowling, 1998, p. 53 |
| Opening the market to caldron imports is expected to drive the prices down | "What are you working on?" said Harry. "A report for the Department of International Magical Cooperation," said Percy smugly. "We're trying to standardize cauldron thickness. Some of these foreign imports are just a shade too thin - leakages have been increasing at a rate of almost three percent a year"… "unless some sort of international law is imposed we might well find the market flooded with flimsy, shallow bottomed products that seriously endanger -" | Rowling, 2000, pp. 36–37 |
| Kids eat the same candies and collect the same cards as their parents | "Madam Rosmertas finest oak-matured mead," said Dumbledore, raising his glass to Harry, who caught hold of his own and sipped. He had never tasted anything like it before, but enjoyed it immensely...Harry could not suppress a suspicion that Dumbledore was rather enjoying himself. | Rowling, 2005, p. 31 |
| Zonko's Joke | "The walk into Hogsmeade was not enjoyable…More | Rowling, |
| Shop goes out of business | than once Harry wondered whether they might not have had a better time in the warm common room, and when they finally reached Hogsmeade and saw that Zonko's Joke Shop had been boarded up, Harry took it as confirmation that this trip was not destined to be fun." | 2005, pp. 158–159 |
| Lack of competition limits the Potterians' choice although there's demand for new products | "Mum found this stack of order forms when she was cleaning Fred and George's room," said Ron quietly. "Great long price lists for stuff they've invented. Joke stuff, you know. Fake wands and trick sweets, loads of stuff. It was brilliant, I never knew they'd been inventing all that…" | Rowling, 2000, p. 36 |
| | "But the common room was packed and full of shrieks of laughter and excitement; Fred and George were demonstrating their latest bit of joke shop merchandise." | Rowling, 2003, p. 403 |
| | "…they headed farther along the street in search of Weasleys' Wizard Wheezes, the joke shop run by Fred and George."…"And he and Harry led the way into the shop. It was packed with customers; Harry could not get near the shelves." | Rowling, 2005, p. 76 |
| | "We've just developed this more serious line," said Fred…"Well, we thought Shield Hats were a bit of a laugh, you know, challenge your mate to jinx you while wearing it and watch his face when the jinx just bounces off. But the Ministry bought five hundred for all its support staff! And we're still getting massive orders!" "So we've expanded into a range of Shield Cloaks, Shield Gloves…" "And then we thought we'd get into the whole area of Defense Against the Dark Arts, because it's such a money spinner," continued George enthusiastically. "This is cool. Look, Instant Darkness Powder… Handy if you want to make a quick escape." "And our Decoy Detonators…" | Rowling, 2005, p. 78 |
| Wizards view muggle-made goods as inferior | "Why would anyone bother making door keys shrink?" said George. "Just Muggle-baiting," sighed Mr. Weasley. "Sell them a key that keeps shrinking to nothing so they can never find it when they need it. Of course, it's very hard to convict anyone because no Muggle would admit their key keeps shrinking — they'll insist they just keep losing it." | Rowling 1999a, p. 25 |
| | "Harry left Hermione dabbing her black eye with paste and followed Fred toward the back of the shop, where he saw a stand of card and rope tricks. "Muggle magic tricks!" said Fred happily, pointing them out. "For freaks like Dad, you know, who love Muggle stuff. It's not a big | Rowling, 2005, p. 77 |

| | earner…" | |
|---|---|---|
| Wealthy wizards enjoy a luxurious life style, and own almost all the assets and capital | "Judging by the fact that Draco Malfoy usually had the best of everything, his family was rolling in wizard gold; he could just see Malfoy strutting around a large manor house." | Rowling, 1999a, p. 19 |
| Voldemort comes from a well-established family that was stripped of its assets | "That old man was —?" "Voldemort's grandfather, yes," said Dumbledore. "Marvolo, his son, Morfin, and his daughter, Merope, were the last of the Gaunts, a very ancient Wizarding family noted for a vein of instability and violence that flourished through the generations due to their habit of marrying their own cousins. Lack of sense coupled with a great liking for grandeur meant that the family gold was squandered several generations before Marvolo was born. He, as you saw, was left in squalor and Poverty." | Rowling, 2005, pp. 138 |
| Businessmen's negative image | "Mum wants them to go into the Ministry of Magic like Dad, and they told her all they want to do is open a joke shop." | Rowling, 2000, p. 36 |
| | "Mrs. Weasley…did not think running a joke shop was a suitable career for two of her sons." | Rowling, 2003, p. 79 |
| Wizards with muggle predecessors, are considered by wealthy wizards a threat because of their different culture | "Harry knew exactly what was making Mr. Malfoy's lip curl like that. The Malfoys prided themselves on being purebloods; in other words, they considered anyone of Muggle descent, like Hermione, second-class. However, under the gaze of the Minister of Magic, Mr. Malfoy didn't dare say anything." | Rowling, 2000, p. 66 |
| Wealthy pure-blood wizards consider themselves superior to *mud-blood* wizards | "I would have thought you'd be ashamed that a girl of no wizard family beat you in every exam," snapped Mr. Malfoy." | Rowling, 1999a, p. 34 |
| | "Malfoy called her 'Mudblood,' Hagrid"…Hagrid looked outraged…"But I don't know what it means. I could tell it was really rude, of course." "It's about the most insulting thing he could think of," gasped Ron…"Mudblood's a really foul name for someone who is Muggle-born — you know, non-magic parents. There are some wizards — like Malfoy's family — who think they're better than everyone else because they're what people call pure- | Rowling, 1999a, pp. 72–74 |

| | blood." | |
|---|---|---|
| | "If you're wondering what the smell is, Mother, a Mudblood just walked in," said Draco Malfoy." | Rowling, 2005, p. 74 |
| | "You're lying, filthy Mudblood, and I know it! You have been inside my vault at Gringotts! Tell the truth, *tell the truth*!" | Rowling, 2007, p. 308 |
| Wealthy pure-blood wizards consider themselves superior to *half-blood* wizards | "Shut your mouth!" Bellatrix shrieked. "You dare speak his name with your unworthy lips, you dare besmirch it with your half-blood's tongue, you dare" | Rowling, 2003, p. 584 |
| | "…they thought Voldemort had the right idea, they were all for the purification of the wizarding race, getting rid of Muggle-borns and having pure-bloods in charge. They weren't alone, either, there were quite a few people, before Voldemort showed his true colors, who thought he had the right idea about things" | Rowling, 2003, p. 84 |
| | "He'd play up the pure-blood side so he could get in with Lucius Malfoy and the rest of them…he's just like Voldemort. Pure-blood mother, Muggle father…ashamed of his parentage" | Rowling, 2005, p. 417 |
| Wealthy wizards associated with middle class wizards are often disinherited by their families | "The tapestry looked immensely old…the golden thread with which it was embroidered still glinted brightly enough to show them a sprawling family tree dating back (as far as Harry could tell) to the Middle Ages. Large words at the very top of the tapestry read: "The Noble and Most Ancient House of Black Toujours pur." "You're not on here!" said Harry, after scanning the bottom of the tree closely. "I used to be there," said Sirius, pointing at a small, round, charred hole in the tapestry, rather like a cigarette burn. "My sweet old mother blasted me off after …I'd had enough." "Where did you go?" asked Harry, staring at him. "Your dad's place," said Sirius…"when I was seventeen I got a place of my own. My Uncle Alphard had left me a decent bit of gold – he's been wiped off here, too, that's probably why" | Rowling, 2003, pp. 83–84 |
| Intermarriages further block upward mobility | "The pure-blood families are all interrelated," said Sirius. "If you're only going to let your sons and daughters marry pure-bloods your choice is very limited; there are hardly any of us left. Molly and I are cousins by marriage and Arthur's something like my second cousin once removed." | Rowling, 2003, p. 85 |
| Cultural | "There was a greater variety of dishes in front of them | Rowling, |

| barriers – wizards lack basic knowledge about other people's customs and traditions | than Harry had ever seen, including several that were definitely foreign. "What's that?" said Ron, pointing at a large dish..."Bouillabaisse," said Hermione…"It's French," said Hermione, "I had it on holiday summer before last. It's very nice." "I'll take your word for it," said Ron, helping himself to black pudding." | 2000, pp. 162–163 |
|---|---|---|
| | "Excuse me, are you wanting ze bouillabaisse?" It was the girl from Beauxbatons…Ron…stared up at her, opened his mouth to reply, but nothing came out except a faint gurgling noise. "Yeah, have it," said Harry, pushing the dish toward the girl. "You 'ave finished wiz it?" "Yeah," Ron said breathlessly. "Yeah, it was excellent."...Ron was still goggling at the girl as though he had never seen one before." | Rowling, 2000, p. 163 |
| | "When the second course arrived they noticed a number of unfamiliar desserts too. Ron examined an odd sort of pale blancmange closely, then moved it carefully a few inches to his right, so that it would be clearly visible from the Ravenclaw table." | Rowling, 2000, p. 164 |
| | "I'll be havin' a few words with her, an' all," said Hagrid grimly, stomping up the stairs. "The less you lot 'ave ter do with these foreigners, the happier yeh'll be. Yeh can trust any of 'em." | Rowling, 2000, p. 363 |
| When a leading English wand-maker disappears, wizards can't find another wand-maker | "Talking of Diagon Alley," said Mr. Weasley, "looks like Ollivander's gone too." "The wandmaker?" said Ginny, looking startled. "That's the one. Shop's empty. No sign of a struggle. No one knows whether he left voluntarily or was kidnapped." "But what'll people do for wands?" | Rowling, 2005, p. 70 |
| Trade in Muggle goods takes place only under very special circumstances and only for very specific goods | "But what does a Ministry of Magic *do*?" "Well, their main job is to keep it from the Muggles that there's still witches an' wizards up an' down the country." | Rowling, 1998, p. 42 |
| | "What does your dad do at the Ministry of Magic, anyway?" "He works in the most boring department," said Ron. "The Misuse of Muggle Artifacts Office." "The *what*?" "It's all to do with bewitching things that are Muggle-made, you know, in case they end up back in a Muggle shop or house. Like, last year, some old witch died and her tea set was sold to an antiques shop. This Muggle woman bought it, took it home, and tried to serve her friends tea in it. It was a nightmare — Dad was working overtime for weeks." | Rowling, 1999a, p. 20 |
| | "Harry…followed Fred toward the back of the shop, where he saw a stand of card and rope tricks. "Muggle magic tricks!" said Fred happily, pointing them out. "For freaks like Dad, you know, who love Muggle stuff. It's not a big earner…" | Rowling, 2005, p. 77 |
| A junior official uses quality as a pretext to block an importation of considerably cheaper foreign goods | "What are you working on?" said Harry. "A report for the Department of International Magical Cooperation," said Percy smugly. "We're trying to standardize cauldron thickness. Some of these foreign imports are just a shade too thin - leakages have been increasing at a rate of almost three percent a year"… "unless some sort of international law is imposed we might well find the market flooded with flimsy, shallow bottomed products that seriously endanger -" | Rowling, 2000, pp. 36–37 |
| | "He wants a word with you about your embargo on flying carpets." Mr. Weasley heaved a deep sigh…"Carpets are defined as a Muggle Artifact by the Registry of Proscribed Charmable Objects"…"Well, they'll never replace brooms in Britain, will they?" said Bagman. "Ali thinks there's a niche in the market for a family vehicle," said Mr. Crouch. "I remember my grandfather had an Axminster that could seat twelve - but that was before carpets were banned, of course." | Rowling, 2000, p. 59 |
| Senior wizard officials are driven by ego-rents for power | "Fudge… [the current minister of magic] is frightened of Dumbledore?" said Harry incredulousy. "Frightened of what he is up to… Fudge thinks… Dumbledore wants to be minister of magic… Deep down Fudge knows Dumbledore is much cleverer than he is… but it seems that he's become fond of power." | Rowling, 2003, p. 89 |
| | "A well… said Thicknesse [the presiding minister of magic]. "If you ask me, the blood traitors are as bad as the mudbloods." | Rowling, 2007, p. 247 |
| Despite the wizards' prejudices against any type of humanoids, they are willing to employ elves in large numbers, who work hard under terrible | "He is wanting paying for his work, sir." "Paying?" said Harry blankly. "Well - why shouldn't he be paid?" Winky looked quite horrified…"House-elves is not paid, sir!"… "No, no, no. I says to Dobby, I says, go find yourself a nice family and settle down, Dobby. He is getting up to all sorts of high jinks, sir, what is unbecoming to a house-elf… "Well, it's about time he had a bit of fun," said Harry. "House-elves is not supposed to have fun, Harry Potter," said Winky firmly, from behind her hands. "House-elves does what they is told. I is not liking heights at all, Harry Potter"…"but my master sends me to the Top Box and I comes, sir."…"Why's he sent you up here, if he | Rowling, 2000, p. 64 |

| conditions and almost without pay | knows you don't like heights?" said Harry, frowning. "Master - master wants me to save him a seat, Harry Potter. He is very busy,"…"Winky does what she is told. Winky is a good house-elf." | |
|---|---|---|
| | "You know, house-elves get a very raw deal!" said Hermione indignantly. "It's slavery, that's what it is! That Mr. Crouch made her go up to the top of the stadium, and she was terrified, and he's got her bewitched so she can't even run when they start trampling tents! Why doesn't anyone do something about it?" "Well, the elves are happy, aren't they?" Ron said. "You heard old Winky back at the match… 'House-elves is not supposed to have fun'…that's what she likes, being bossed around…" | Rowling, 2000, p. 80 |
| | "You may rest assured that she will be punished," Mr. Crouch added coldly. "M-m-master…" Winky stammered, looking up at Mr. Crouch, her eyes brimming with tears. "M-m-master, p-p-please…" Mr. Crouch stared back, his face somehow sharpened, each line upon it more deeply etched. There was no pity in his gaze. "Winky has behaved tonight in a manner I would not have believed possible," he said slowly. "I told her to remain in the tent. I told her to stay there while I went to sort out the trouble. And I find that she disobeyed me. This means clothes." "No!" shrieked Winky, prostrating herself at Mr. Crouch's feet. "No, master! Not clothes, not clothes!" | Rowling, 2000, p. 89 |
| The first signs of danger appear two years earlier but the government ignores them | "But he cannot now give testimony, Cornelius," said Dumbledore.…"He cannot give evidence about why he killed those people." "Why he killed them? Well, that's no mystery, is it?" blustered Fudge. "He was a raving lunatic! From what Minerva and Severus have told me, he seems to have thought he was doing it all on You Know-Who's instructions!" "Lord Voldemort was giving him instructions, Cornelius," Dumbledore said. "Those peoples deaths were mere by-products of a plan to restore Voldemort to full strength again. The plan succeeded. Voldemort has been restored to his body." Fudge…began to sputter, still goggling at Dumbledore. "You-Know-Who …returned? Preposterous. Come now, Dumbledore …" "As Minerva and Severus have doubtless told you," said Dumbledore, "we heard Barty Crouch confess. Under the influence of Veritaserum, he told us how he was smuggled out of Azkaban, and how Voldemort…went to free him from his father and used him to capture Harry. The plan worked, I tell you. Crouch has helped Voldemort to return." "See here, Dumbledore…you - you can't seriously believe that You-Know-Who's back? Come | Rowling, 2000, pp. 453–454 |

| | now, come now…certainly, Crouch may have believed himself to be acting upon You-Know-Who's orders - but to take the word of a lunatic like that, Dumbledore…" | |
|---|---|---|
| Professor Slughorn complains that due to the war prices are sky-high | "There was a final plunk from the piano,…My last bottle, and prices are sky-high at the moment." | Rowling, 2005, p. 43 |
| Cost born by the public for the government inefficiency - the price paid for goods sold by swindlers that offer a false sense of security | "…a number of shabby-looking stalls had sprung up along the street. The nearest one…had a cardboard sign pinned to its front: ***AMULETS, Effective Against Werewolves, Dementors, and Inferi!*** A seedy-looking little wizard was rattling armfuls of silver symbols on chains at passersby. "One for your little girl, madam?" he called at Mrs. Weasley as they passed, leering at Ginny. "Protect her pretty neck?" "If I were on duty…" said Mr. Weasley, glaring angrily at the amulet seller. "Yes, but don't go arresting anyone now, dear, we're in a hurry," said Mrs. Weasley…" | Rowling, 2005, p. 73 |
| | "You wouldn't believe how many people, even people who work at the Ministry, can't do a decent Shield Charm," said George. "'Course, they didn't have you teaching them, Harry." "…the Ministry bought five hundred for all its support staff! And we're still getting massive orders!" "So we've expanded into a range of Shield Cloaks, Shield Gloves…" "…"And then we thought we'd get into the whole area of Defense Against the Dark Arts, because it's such a money spinner," continued George… "This is cool. Look, Instant Darkness Powder, we're importing it from Peru. Handy if you want to make a quick escape." | Rowling, 2005, p. 78 |
| The usually crowded "Leaky Cauldron" bar is empty because even the most loyal consumers seem to have lost their appetite | "The Leaky Cauldron was, for the first time in Harry's memory, completely empty. Only Tom the landlord, wizened and toothless, remained of the old crowd." | Rowling, 2005, p. 72 |
| | "The bar of the Leaky Cauldron was nearly deserted. Tom, the stooped and toothless landlord, was polishing glasses behind the bar counter; a couple of warlocks having a muttered conversation in the far corner glanced at Hermione and drew back into the shadows. "Madam Lestrange," murmured Tom, and as Hermione paused he inclined his head subserviently." | Rowling, 2007, p. 347 |
| The bar owner notices Hagrid. From the | "The Leaky Cauldron was, for the first time in Harry's memory, completely empty. Only Tom the landlord, wizened and toothless, remained of the old crowd. He | Rowling, 2005, p. 72 |
| barman's reaction it is clear that Hagrid is one of his last loyal customers | looked up hopefully as they entered, but before he could speak, Hagrid said importantly, "Jus' passin' through today, Tom, sure yeh understand, Hogwarts business, yeh know." | |
| Makes and Models of the broomsticks used by the Potterian wizards | See Table A5 in this appendix. | |
| Wealthy wizards are even willing to give up on some of their income just to become teachers | "There was also a list of the new books he'd need for the coming year. *SECOND-YEAR STUDENTS WILL REQUIRE:*<br>*Standard Book of Spells Grade 2 by Miranda Goshawk*<br>*Break with a Banshee by Gilderoy Lockhart*<br>*Gadding with Ghouls by Gilderoy Lockhart*<br>*Holidays with Hags by Gilderoy Lockhart*<br>*43 Travels with Trolls by Gilderoy Lockhart*<br>*Voyages with Vampires by Gilderoy Lockhart*<br>*Wanderings with Werewolves by Gilderoy Lockhart*<br>*Year with the Yeti by Gilderoy Lockhart*<br>Fred, who had finished his own list, peered over at Harry's. "You've been told to get all Lockhart's books, too!" he said." | Rowling, 1999a, pp. 28–29 |
| | An hour later, they headed for Flourish and Blotts…bookshop…a large crowd jostling outside the doors, trying to get in. The reason for this was proclaimed by a large banner stretched across the upper windows: *GILDEROY LOCKHART will be signing copies of his autobiography MAGICAL ME today 12:30P.M.to 4:30P.M.* "We can actually meet him!" Hermione squealed. "I mean, he's written almost the whole booklist!"…A harassed looking wizard stood at the door, saying, "Calmly, please, ladies…Don't push, there…mind the books, now…" | Rowling, 1999a, p. 38 |
| | "They had reached Lockhart's classroom…When the whole class was seated, Lockhart cleared his throat loudly and silence fell. He reached forward, picked up Neville Longbottom's copy of *Travels with Trolls*, and held it up to show his own, winking portrait on the front. "Me," he said, pointing at it and winking as well. "Gilderoy Lockhart... "I see you've all bought a complete set of my books — well done." | Rowling, 1999a, p. 64 |

| The current headmaster is considered by many to be the greatest wizard of his time | Harry unwrapped his Chocolate Frog and picked up the card…Harry turned over his card and read:<br>*ALBUS DUMBLEDORE*<br>*CURRENTLY HEADMASTER OF HOGWARTS*<br>*Considered by many the greatest wizard of modern times, Dumbledore is particularly famous for his defeat of the dark wizard Grindelwald in 1945, for the discovery of the twelve uses of dragon's blood, and his work on alchemy with his partner, Nicolas Flamel. Professor Dumbledore enjoys chamber music and tenpin bowling.* | Rowling, 1998, p. 66 |
|---|---|---|
| The government can interfere with the school curriculum and governance | *Daily Prophet*: "In a surprise move last night the Ministry of Magic passed new legislation giving itself an unprecedented level of control at Hogwarts School " "…the passing of Educational Decree Number Twenty-three…creates the new position of Hogwarts High Inquisitor…'The Inquisitor will have powers to inspect her fellow educators and make sure that they are coming up to scratch." | Rowling, 2003, p. 229 |
| The Hogwarts' graduates don't know more than their predecessors | "The time had come to choose their subjects for the third year…"I just want to give up Potions," said Harry. "We can't," said Ron gloomily. "We keep all our old subjects, or I'd've ditched Defense Against the Dark Arts." | Rowling, 1999a, p. 161 |
| | "Harry bent swiftly over the tattered book Slughorn had lent him. To his annoyance he saw that the previous owner had scribbled all over the pages, so that the margins were as black as the printed portions. Bending low to decipher the ingredients (even here, the previous owner had made annotations and crossed things out)…" | Rowling, 2005, p. 123 |
| | "I just tried a few of the tips written in the margins, honestly, Ginny, there's nothing funny-" | Rowling, 2005, p. 125 |
| | "Harry bent low to retrieve the book, and as he did so, he saw…scribbled along the bottom of the back cover in the same small, cramped handwriting…***This book is the property of the Half Blood Prince.***" | Rowling, 2005, p. 126 |
| | "It's just that I was right about Eileen Prince once owning the book. You see…she was Snape's mother!"… "I was going through the rest of the old *Prophets* and there was a tiny announcement about Eileen Prince marrying a man called Tobias Snape, and then later an announcement saying that she'd given birth…"Snape must have been proud of being 'half a Prince', you see? Tobias Snape was a Muggle from what it said in the *Prophet*" "Yeah, that fits," said Harry. "He'd play up the pure-blood side so he | Rowling, 2005, p. 417 |

| | could get in with Lucius Malfoy and the rest of hem… Pure-blood mother, Muggle father…ashamed of his parentage, trying to make himself feared using the Dark Arts, gave himself an impressive new name…the Half-Blood Prince - how could Dumbledore have missed —?" | |
|---|---|---|
| | "…that the parents he would destroy in his murderous quest were people that Professor Snape knew, that they were your mother and father —" | Rowling, 2005, p. 360 |
| Potterian students lack creative skills and cannot think originally, which diminishes entrepreneurial spirit | "It hasn't been easy, Harry, guiding you through these tasks without arousing suspicion. I have had to use every ounce of cunning I possess, so that my hand would not be detectable in your success. Dumbledore would have been very suspicious if you had managed everything too easily. As long as you got into that maze, preferably with a decent head start - then, I knew, I would have a chance of getting rid of the other champions and leaving your way clear. But I also had to contend with your stupidity. The second task…that was when I was most afraid we would fail. I was keeping watch on you, Potter. I knew you hadn't worked out the egg's clue, so I had to give you another hint" | Rowling, 2000, p. 434 |
| After Graduation from Hogwarts, wizards choose a profession | The time had come to choose their subjects for the third year…Percy Weasley was eager to share his experience. "Depends where you want to go, Harry," he said. "It's never too early to think about the future, so I'd recommend Divination. People say Muggle Studies is a soft option, but I personally think wizards should have a thorough understanding of the non-magical community, particularly if they're thinking of working in close contact with them — look at my father, he has to deal with Muggle business all the time. My brother Charlie was always more of an outdoor type, so he went for Care of Magical Creatures. Play to your strengths, Harry." | Rowling, 1999a, pp. 161–162 |
| Goblins can easily distinguish between the fake and real Galleons | "The long counter was manned by goblins sitting on high stools serving the first customers of the day. Hermione, Ron, and Travers headed toward an old goblin who was examining a thick gold coin through an eyeglass…The goblin tossed the coin he was holding aside, said to nobody in particular, "Leprechaun," and then greeted Travers." | Rowling, 2007, p. 351 |
| Inflation of home prices | "This house!" shrieked Uncle Vernon, the vein his forehead starting to pulse. "*Our* house! House prices are skyrocketing around here! | Rowling, 2007, p. 20 |
| Inflation | "Yes, dragon," repeated the wizard conversationally. "My last bottle, and prices are sky-high at the moment. Still, it might be reusable." | Rowling, 2005, p. 43 |

## Appendix B. Additional Economic Ideas[2]

| Additional Economic Ideas | Quote | Reference |
|---|---|---|
| Rational ignorance-inattention | "Haven't…you been getting the *Daily Prophet*!" Hermione asked nervously. "Yeah, I have!" said Harry. "Have you – er – been reading it thoroughly?" Hermione asked…"Not cover to cover," said Harry defensively. | Rowling, 2003, p. 55 |
| Barter exchange | "Ron had taken out a lumpy package and unwrapped it. There were four sandwiches inside. He pulled one of them apart and said, "She always forgets I don't like corned beef." "Swap you for one of these," said Harry, holding up a pasty." | Rowling, 1998, p. 66 |
| | "Though the goblins of Gringotts will consider it base treachery, I have decided to help you – for payment." …"I want the sword. The sword of Godric Gryffindor."… The sword is the price of my hire, take it or leave it!" | Rowling, 2007, pp. 334–335 |
| | "A large sign had been affixed to the Gryffindor noticeboard; so large it covered everything else on it – the lists of secondhand spellbooks for sale…the offers to barter certain Chocolate Frog Cards for others" | Rowling, 2003, p. 262 |
| Barter exchange – a lack of double coincidence of wants | "I have decided to help you – for payment." …"How much do you want? I've got gold." "Not gold," said Griphook. "I have gold." | Rowling, 2007, pp. 334 |
| Implicit contract – a handshake | "I have your word, Harry Potter, that you will give me the sword of Gryffindor if I help you?" "Yes," said Harry. "Then shake," said the goblin, holding out his hand." | Rowling, 2007, p. 337 |
| Secondhand, used goods' market | "Mrs. Weasley and Ginny were going to a secondhand robe shop." | Rowling, 1999a, p. 37 |
| | "I thought they'd bring out the color of your eyes, dear," said Mrs. Weasley fondly. "Well, they're okay!" said Ron angrily, looking at Harry's robes. "Why couldn't I have some like that?" "Because…well, I had to get yours secondhand, and there wasn't a lot of choice!" said Mrs. Weasley" | Rowling, 2000, p. 101 |
| Diminishing marginal utility | "The leprechaun gold I gave you…Why didn't you tell me it disappeared?"…"I dunno…I never noticed it had gone..."Must be nice," Ron said…"To have so much | Rowling, 2000, p. 351 |

| | money you don't notice if a pocketful of Galleons goes missing." | |
|---|---|---|
| Budget constraint | "I haven't got any money — and you heard Uncle Vernon last night…he won't pay for me to go and learn magic." | Rowling, 1998, p. 41 |
| Non-tradable goods | "We need them for the Skiving Snackboxes but they're a Class C Non-Tradable Substance so we've been having a bit of trouble getting hold of them." | Rowling, 2003, p. 128 |
| Price adjustment - Haggling and negotiating over price | "In that case, perhaps we can return to my list," said Mr. Malfoy shortly. "I am in something of a hurry, Borgin, I have important business elsewhere today —" They started to haggle." | Rowling, 1999a, p. 34 |
| | "Look what Dung's got us," said George..."Venomous Tentacula seeds," said George. "Ten Galleons the lot, then Dung?" said Fred. "Wiv all the trouble I went to get 'em?" said Mundungus, his saggy, bloodshot eyes stretching even wider. "I'm sorry, lads, but I'm not taking a Knut under twenty."…Mundungus looked nervously over his shoulder…he grunted. "All right, lads, ten it is, if you'll take 'em quick." | Rowling, 2003, pp. 128–129 |
| Fund raising, donations | "All proceeds from the Fountain of Magical Brethren will be given to St. Mungo's Hospital for magical maladies and injuries." | Rowling, 2003, p. 95 |
| Consumption smoothing | "Once Harry had refilled his money bag with gold Galleons, silver Sickles, and bronze Knuts from his vault at Gringotts, he had to exercise a lot of self-control not to spend the whole lot at once." | Rowling, 1999b, p. 31 |
| Black market | "Meanwhile, a flourishing black-market trade in aids to concentration, mental agility and wakefulness had sprung up among the fifth- and seventh-years. Harry and Ron were much tempted by the bottle of Baruffio's Brain Elixir offered to them by Ravenclaw sixth-year Eddie Carmichael, who swore it was solely responsible for the nine 'Outstanding' OWLs he had gained the previous summer and was offering a whole pint for a mere twelve Galleons." | Rowling, 2003, p. 527 |
| Marketing and market research | "We're going to use it to do a bit of market research, find out exactly what the average Hogwarts student requires from a joke shop, carefully evaluate the results of our research, then produce products to fit the demand." | Rowling, 2003, p. 169 |
| Minimum wage | "Our short-term aims," said Hermione…"are to secure house-elves fair wages and working conditions." | Rowling, 2000, p. 145 |
| Representative government | "Our long-term aims include…trying to get an elf into the Department for the Regulation and Control of Magical Creatures, because they're shockingly underrepresented." | Rowling, 2000, p. 145 |

| Opening the market to caldron imports is expected to drive the prices down | "What are you working on?" said Harry. "A report for the Department of International Magical Cooperation," said Percy smugly. "We're trying to standardize cauldron thickness. Some of these foreign imports are just a shade too thin - leakages have been increasing at a rate of almost three percent a year"… "unless some sort of international law is imposed we might well find the market flooded with flimsy, shallow bottomed products that seriously endanger -" | Rowling, 2000, pp. 36–37 |
|---|---|---|
| Fine for traffic violation | "*Daily Prophet*…said: Inquiry at the Ministry of Magic - Arthur Weasley, Head of the Misuse of Muggle Artifacts Office, was today fined fifty Galleons for bewitching a Muggle car." | Rowling, 1999a, p. 142 |
| Import restrictions-regulations | "He wants a word with you about your embargo on flying carpets." Mr. Weasley heaved a deep sigh…"Carpets are defined as a Muggle Artifact by the Registry of Proscribed Charmable Objects"…"Well, they'll never replace brooms in Britain, will they?" said Bagman. "Ali thinks there's a niche in the market for a family vehicle," said Mr. Crouch. "I remember my grandfather had an Axminster that could seat twelve - but that was before carpets were banned, of course." | Rowling, 2000, p. 59 |
| Product Standardization | "We're trying to standardize cauldron thickness. Some of these foreign imports are just a shade too thin - leakages have been increasing at a rate of almost three percent a year" | Rowling, 2000, p. 36 |
| Gambling market | "Now, now, Penny, no sabotage!" said Percy heartily as she examined the Firebolt closely. "Penelope and I have got a bet on," he told the team. "Ten Galleons on the outcome of the match!" | Rowling, 1999b, p. 164 |
| Scarcity | "As Harry might have told you, the final of the Quidditch World Cup takes place this Monday night, and my husband, Arthur, has just managed to get prime tickets through his connections at the Department of Magical Games and Sports." | Rowling, 2000, p. 20 |
| Saving | "Boys," said Mr. Weasley under his breath, "I don't want you betting…That's all your savings" | Rowling, 2000, p. 57 |
| Corruption | "As Harry might have told you, the final of the Quidditch World Cup takes place this Monday night, and my husband, Arthur, has just managed to get prime tickets through his connections at the Department of Magical Games and Sports." | Rowling, 2000, p. 20 |

**Appendix C. Round Prices**[3]

| Round Prices | Quote | Reference |
|---|---|---|
| 1,500 Galleons – cursed opal necklace at Borgin and Burkes | "Is this necklace for sale?" she asked, pausing beside a glass-fronted case. "If you've got one and a half thousand Galleons," said Mr. Borgin coldly." | Rowling, 2005, p. 83 |
| 1,000 Galleons – prize for winning the Triwizard tournament | "An impartial judge will decide which students are most worthy to compete for the Triwizard Cup, the glory of their school, and a thousand Galleons personal prize money." | Rowling, 2000, pp. 121–122 |
| 1,000 Galleons – A bounty on the head of escaped Death Eaters | "A large poster had been stuck up in the window… and Harry found himself staring once more at the pictures of the ten escaped Death Eaters. The poster, *By Order of the Ministry of Magic,* offered a thousand-Galleon reward to any witch or wizard with information leading to the recapture of any of the convicts pictured" | Rowling, 2003, p. 416 |
| 700 Galleons – Prize from *Daily Prophet* drawing | "I couldn't believe it when Dad won the Daily Prophet Draw. Seven hundred galleons! Most of it's gone on this trip, but they're going to buy me a new wand for next year." | Rowling, 1999b, p. 6 |
| 500 Galleons – Goblin-made armor at Borgin and Burkes | "Mr. Burke would like to make an improved offer for the goblin-made armor," said Voldemort. "Five hundred Galleons, he feels it is a more than fair." | Rowling, 2005, p. 286 |
| 100 Galleons – Acromantula Venom, per pint | "I mean, it's almost impossible to get venom from an acromantula while it's alive…" "Slughorn seemed to be talking more to himself than Harry now. "…seems an awful waste not to collect it…might get a hundred Galleons a pint" | Rowling, 2005, p. 316 |
| 20 Galleons – Deflagration Deluxe | "Hermione, it's…twenty [Galleons] for the Deflagration Deluxe…" | Rowling, 2003, p. 472 |
| 10 Galleons – Hermione's birthday present | "I've still got ten Galleons," she said, checking her purse. "It's my birthday in September, and Mum and Dad gave me some money to get myself an early birthday present." | Rowling, 1999b, p. 36 |
| 10 Galleons – Omnioculars | "Omnioculars," said the saleswizard eagerly. "You can replay action… slow everything down…and they flash up a play-by- play breakdown if you need it. Bargain - ten Galleons each." | Rowling, 2000, p. 60 |
| 10 Galleons – | "Professor Dumbledore offered Dobby ten Galleons a | Rowling, |
| Weekly pay to house-elf | week, and weekends off," said Dobby" | 2000, p. 244 |
| 10 Galleons – Metamorph-Medals for changing your appearance, 100,000 disguises | "Some idiot's started selling Metamorph-Medals. Just sling them around your neck and you'll be able to change your appearance at will. A hundred thousand disguises, all for ten Galleons!" | Rowling, 2005, p. 58 |
| 10 Galleons – Bet on the outcome of a Quidditch match | "Now, now, Penny, no sabotage!" said Percy heartily as she examined the Firebolt closely. "Penelope and I have got a bet on," he told the team. "Ten Galleons on the outcome of the match!" | Rowling, 1999b, p. 164 |
| 10 Galleons – Slytherin's Locket | "She didn't seem to have any idea how much it was worth. Happy to get ten Galleons for it. Best bargain we ever made!" | Rowling, 2005, p. 171 |
| 10 Galleons – Unicorn hair | "Not long after this, Hagrid became tearful again and pressed the whole unicorn tail upon Slughorn, who pocketed it with cries of, "To friendship! To generosity! To ten Galleons a hair!" | Rowling, 2005, p. 320 |
| 5 Galleons – Basic Blaze box | "Hermione, it's five Galleons for your Basic Blaze box" | Rowling, 2003, p. 472 |
| 5 Galleons – Rubber chicken wand | "Bagman didn't seem to think the wand was rubbish at all…his boyish face shone with excitement as…the wand gave a loud squawk and turned into a rubber chicken… "Excellent! I haven't seen one that convincing in years! I'd pay five Galleons for that!" | Rowling, 2000, p. 57 |
| 5 Knuts per scoop – Glittery-Black Beetle Eyes | "Harry himself examined…glittery-black beetle eyes (five Knuts a scoop)." | Rowling, 1998, p. 52 |

**Appendix D. Convenient Prices**[4]

| Convenient Prices | Quote | Reference |
|---|---|---|
| 21 Galleons – Silver Unicorn Horn | "Harry himself examined silver unicorn horns at twenty-one Galleons each" | Rowling, 1998, p. 52 |
| 16 Galleons – Human skull at Borgin and Burkes | "And…what about this lovely…um…skull?" "Sixteen Galleons." | Rowling, 2005, p. 83 |
| 12 Galleons – 12-week course of Apparition lessons | "Apparition Lessons: If you are seventeen years of age…you are eligible for a twelve-week course of Apparition Lessons from a Ministry of Magic Apparition instructor…Cost: 12 Galleons." | Rowling, 2005, p. 233 |
| 9 Galleons – New copy of *Advanced Potion-Making* | "He pulled the old copy of *Advanced Potion-Making*… There sat the Prince's copy, disguised as a new book, and there sat the fresh copy from Flourish and Blotts, looking thoroughly secondhand. "I'll give Slughorn back the new one, he can't complain, it cost nine Galleons." | Rowling, 2005, p. 144 |
| 7 Galleons – New wand from Ollivanders | "Harry shivered. He wasn't sure he liked Mr. Ollivander too much. He paid seven gold Galleons for his wand, and Mr. Ollivander bowed them from his shop." | Rowling, 1998, p. 55 |
| 2 Galleons – Headless hat | Fred and George were demonstrating their latest bit of joke shop merchandise. "Headless Hats!" shouted George, as Fred waved a pointed hat decorated with a fluffy pink feather at the watching students. "Two Galleons each" | Rowling, 2003, p. 403 |
| 16 Sickles – Dragon Liver, per ounce | "A plump woman outside an Apothecary was shaking her head as they passed, saying, "Dragon liver, sixteen Sickles an ounce, they're mad…" | Rowling, 1998, p. 46 |
| 11 Sickles – Night bus to London | "Listen, how much would it be to get to London?" "Eleven Sickles," said Stan." | Rowling, 1999b, p. 22 |
| 7 Sickles – Canary Cream | "Canary Creams!" Fred shouted to the excitable crowd. "George and I invented them – seven Sickles each, a bargain!" | Rowling, 2000, p. 236 |
| 4 Sickles – Hot water and toothbrush in the color of your choice | "Listen, how much would it be to get to London?" "Eleven Sickles," said Stan, "but…for fifteen you get an 'ot-water bottle an' a toofbrush in the color of your choice." | Rowling, 1999b, p. 22 |
| 2 Sickles – Hot chocolate | "Listen, how much would it be to get to London?" "Eleven Sickles," said Stan, "…but for firteen you get 'ot | Rowling, 1999b, p. |

| on the night bus | chocolate" | 22 |
|---|---|---|
| 2 Sickles – Membership in S.P.E.W | "It's S-P-E-W. Stands for the Society for the Promotion of Elfish Welfare."…"We start by recruiting members," said Hermione happily. "I thought two Sickles to join" | Rowling, 2000, pp. 144–145 |

**Appendix E. Broomstick Makes and Models – Technological Innovations in the Potterian Economy**[5]

| Broomstick makes and models | Quote | Reference |
|---|---|---|
| Cleansweep-5 | "As for the old Cleansweeps" — he smiled nastily at Fred and George, who were both clutching Cleansweep Fives —"sweeps the board with them." | Rowling, 1999a, p. 71 |
| Cleansweep-6 | "Pausing every few pages [of *The Quibbler* magazine:], he read…"an interview with a wizard who claimed to have flown to the moon on a Cleansweep Six and brought back a bag of moon frogs to prove it." | Rowling, 2003, p. 143 |
| Cleansweep-7 | "He's just the build for a Seeker, too," said Wood… "Light —speedy—we'll have to get him a decent broom, Professor — a Nimbus Two Thousand or a Cleansweep Seven, I'd say." | Rowling, 1998, p. 98 |
| Cleansweep-11 | "He…almost walked into Ron, who was…clutching his broomstick. He gave a great leap of surprise when he saw Harry and attempted to hide his new Cleansweep Eleven behind his back." | Rowling, 2003, p. 202 |
| Nimbus-2000 | "Even Harry, who knew nothing about the different brooms, thought it looked wonderful. Sleek and shiny, with a mahogany handle, it had a long tail of neat, straight twigs and Nimbus Two Thousand written in gold near the top." | Rowling, 1998, p. 108 |
| | "They took turns riding Harry's Numbus-2000, which was easily the best broom" | Rowling, 1999a, p. 30 |
| Nimbus-2001 | "Let me show you the generous gift he's made to the Slytherin team." All seven of them held out their broomsticks. Seven highly polished, brand-new handles and seven sets of fine gold lettering spelling the words *Nimbus Two Thousand and One*… "Very latest model. Only came out last month," said Flint…"I believe it outstrips the old Two Thousand series by a considerable amount." | Rowling, 1999a, p. 71 |
| Comet-260 | "What did you say you've got at home, Malfoy, a Comet Two Sixty?" Ron grinned at Harry. "Comets look flashy, but they're not in the same league as the Nimbus." | Rowling, 1998, p. 107 |
| Comet-290 | "Ron was rhapsodizing about his new broom…"…nought to seventy in ten seconds, not bad, is it? When you think the Comet Two Ninety's only nought to sixty and that's with a decent tailwind according to Which Broomstick?" | Rowling, 2003, p. 128 |
| Shooting Star | "They took turns riding Harry's Nimbus Two Thousand, | Rowling, |

---

[5] Broomstick models and makes are discussed in the paper in the context of technological progress, in section 14.

| | which was easily the best broom; Ron's old Shooting Star was often outstripped by passing butterflies." | 1999a, p. 30 |
|---|---|---|
| | "Harry borrowed a copy of *Which Broomstick* from Wood, and decided to spend the day reading up on the different makes. He had been riding one of the school brooms at team practice, an ancient Shooting Star, which was very slow and jerky." | Rowling, 1999b, p. 121 |
| Bluebottle | "At…the field…was a gigantic blackboard…watching it, Harry saw that it was flashing advertisements across the field. *The Bluebottle: A Broom for All the Family - safe, reliable, and with Built-in Anti-Burglar Buzzer.*" | Rowling, 2000, pp. 62–63 |
| Silver Arrow | "Look at the balance on it! If the Nimbus series has a fault, it's a slight list to the tail end — you often find they develop a drag after a few years. They've updated the handle too, a bit slimmer than the Cleansweeps, reminds me of the old Silver Arrows — a pity they've stopped making them. I learned to fly on one, and a very fine old broom it was too…" | Rowling, 1999b, p. 162 |
| Firebolt | "Harry…was able to read the sign next to the broom: "THE FIREBOLT: This state-of-the-art racing broom sports a stream-lined, superfine handle of ash, treated with a diamond-hard polish and hand-numbered with its own registration number. Each individually selected birch twig in the broomtail has been honed to aerodynamic perfection, giving the Firebolt unsurpassable balance and pinpoint precision. The Firebolt has an acceleration of 150 miles an hour in ten seconds and incorporates an unbreakable braking charm. Price on request." Harry didn't like to think how much gold the Firebolt would cost." | Rowling, 1999b, p. 32 |
| | "He had never wanted anything as much in his whole life — but he had never lost a Quidditch match on his Nimbus Two Thousand, and what was the point in emptying his Gringotts vault for the Firebolt, when he had a very good broom already? Harry didn't ask for the price, but he returned, almost every day after that, just to look at the Firebolt." | Rowling, 1999b, p. 32 |
| | "On the other hand, she rides a Comet-260, which is going to look like a joke next to the Firebolt." | Rowling, 1999b, p. 162 |